\documentclass[aps,prb,twocolumn,showpacs,superscriptaddress,10pt,longbibliography]{revtex4-2}
\usepackage{amssymb}
\usepackage{graphicx}
\usepackage[colorlinks = true,linkcolor = red,citecolor = blue,urlcolor=blue]{hyperref}
\usepackage[sort&compress]{natbib}
\usepackage{scalerel}
\usepackage[normalem]{ulem}
\usepackage[dvipsnames]{xcolor}
\usepackage{bm}
\usepackage{amsmath}
\usepackage{subfigure}
\usepackage{amsfonts}
\usepackage{mathtools}
\usepackage{dsfont}
\usepackage{float}
\usepackage{lmodern}
\usepackage[many]{tcolorbox}
\usepackage{empheq}
\usepackage{soul}
\usepackage[T1]{fontenc}

\usepackage{float}

\begin{document}
\title{Bose-Einstein condensation in exotic lattice geometries}

\author{Kamil Dutkiewicz}
\affiliation{Faculty of Physics, University of Warsaw, ul. Pasteura 5, 02-093 Warszawa, Poland}
\affiliation{DIPC - Donostia International Physics Center, Paseo Manuel de Lardiz{\'a}bal 4, 20018 San Sebasti{\'a}n, Spain}
\affiliation{ICFO - Institut de Ciencies Fotoniques, The Barcelona Institute of Science and Technology, 08860 Castelldefels, Barcelona, Spain}

\author{Marcin Płodzień}
\affiliation{ICFO - Institut de Ciencies Fotoniques, The Barcelona Institute of Science and Technology, 08860 Castelldefels, Barcelona, Spain}

\author{Abel Rojo-Franc\`{a}s}
\affiliation{Departament de F\'{i}sica Qu\`{a}ntica i Astrof\'{i}sica, Facultat de F\'{i}sica, Universitat de Barcelona, E-08028 Barcelona, Spain}
\affiliation{Institut de Ci\`{e}ncies del Cosmos, Universitat de Barcelona, ICCUB, Mart\'{i} i Franqu\`{e}s 1, E-08028 Barcelona, Spain}

\author{Bruno Juli\'{a}-D\'{i}az}
\affiliation{Departament de F\'{i}sica Qu\`{a}ntica i Astrof\'{i}sica, Facultat de F\'{i}sica, Universitat de Barcelona, E-08028 Barcelona, Spain}
\affiliation{Institut de Ci\`{e}ncies del Cosmos, Universitat de Barcelona, ICCUB, Mart\'{i} i Franqu\`{e}s 1, E-08028 Barcelona, Spain}

\author{Maciej Lewenstein}
\affiliation{ICFO - Institut de Ciencies Fotoniques, The Barcelona Institute of Science and Technology, 08860 Castelldefels, Barcelona, Spain}

\author{Tobias Grass}
\affiliation{DIPC - Donostia International Physics Center, Paseo Manuel de Lardiz{\'a}bal 4, 20018 San Sebasti{\'a}n, Spain}
\affiliation{IKERBASQUE, Basque Foundation for Science, Plaza Euskadi 5, 48009 Bilbao, Spain}

\begin{abstract}
Modern quantum engineering techniques allow for synthesizing quantum systems in exotic lattice geometries, from self-similar fractal networks to negatively curved hyperbolic graphs. We demonstrate that these structures profoundly reshape Bose-Einstein condensation. 
Fractal lattices dramatically lower the condensation temperature and enhance condensation fluctuations. In a Sierpiński carpet, quasi-degeneracies in the tight-binding spectrum fragment the condensate. 
Hyperbolic lattices, on the other hand, exhibit condensation features similar to regular three-dimensional lattices, despite their embedding in only two dimensions: The critical temperature increases as the system grows, and the 
temperature-dependence of the condensate fraction follows the same power-law as for cubic lattices. We explain these similarities through the similarity of the densities of state at low energies.
When strong repulsive interactions are included, the gas enters a Mott insulating state. Using a multi-site Gutzwiller approach as well as a simple strong-coupling expansion, for the Sierpiński triangle we find a smooth interpolation between the characteristic insulating lobes of one-dimensional and two-dimensional systems. Our findings establish lattice geometry as a powerful tuning knob for quantum phase phenomena and pave the way for experimental exploration in photonic waveguide arrays and Rydberg-atom tweezer arrays.
\end{abstract}

\maketitle   

\section{Introduction}

\begin{figure*}[t]
\centering    \includegraphics[width=0.99\textwidth]{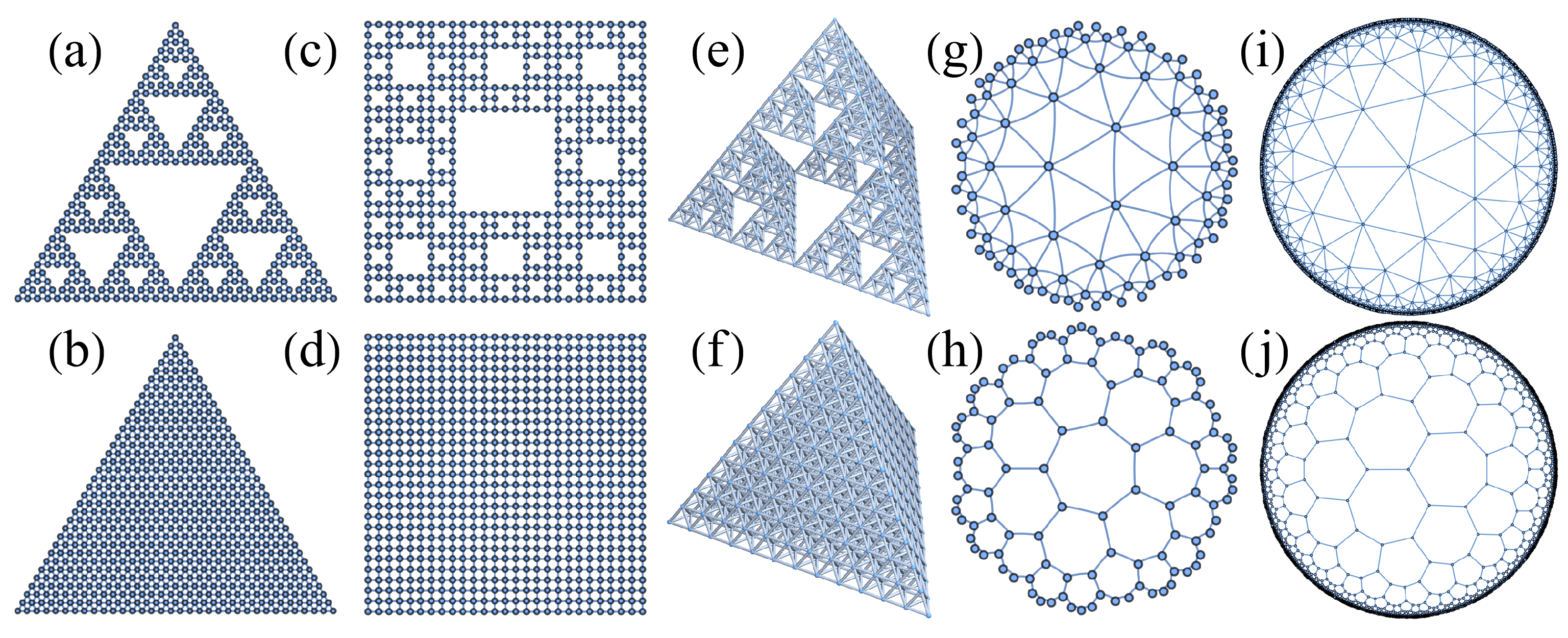}
    \caption{Illustrations of lattice geometries used in this work: (a) Sierpiński gasket with dimension $d \approx 1.585$, (b) triangular lattice with $d=2$, (c) Sierpiński carpet with $d \approx 1.893$, (d) square lattice with $d=2$, (e) Sierpiński tetrahedron with $d = 2$, and (f) tetrahedral lattice with $d=3$, 
    as well as hyperbolic lattices with:
    $\{p, q\} = \{3,7\}$, with $V = 96$ sites (g), $V=4264$ (i)   and 
    $\{p, q\} = \{7,3\}$, with $V = 112$ (h), and $V=3481$ (j) sites.
    Lattices in Fig~\ref{fig:latticeGraphs}(i,j) have a vertex as their center, while lattices in Fig~\ref{fig:latticeGraphs}(g,h) are constructed around a central $p$-polygon.
    }
    \label{fig:latticeGraphs}
\end{figure*}

In 1995, the experimental discovery of Bose-Einstein condensation in a gas of sodium \cite{Davis1995} or rubidium atoms \cite{Anderson1995} has been a ground-breaking achievement: It has confirmed the theoretical prediction made by Bose \cite{Bose1924} and Einstein \cite{Einstein1925} 70 years earlier, and at the same time, it has opened a new research field that uses cold atoms in order to scientifically explore and technologically exploit quantum phenomena. The phenomenon of Bose-Einstein condensation has not remained limited to atomic gases: Not only does the superfluid phase of a $^4$He liquid, which has been known since 1937 already, carry many characteristic signatures of a Bose-Einstein condensate (BEC), but bosonic quasiparticles can also be condensed into a BEC. Quasiparticle BECs have been realized a few years after the first atomic BECs with magnons \cite{Nikuni2000}, excitons \cite{Eisenstein2004}, or exciton-polaritons \cite{Deng2002,Kasprzak2006}. Finally, also the condensation of elementary bosons is possible: In 2010, the first BEC of photons has been achieved \cite{Klaers2010}.
The macroscopic occupation of the single-particle ground state forms the basis of Bose-Einstein condensation. It goes hand in hand with a series of fascinating properties \cite{Dalfovo1999}: The condensate exhibits long-range phase coherence, which in the presence of weak interactions turns into superfluid behavior. This can be evidenced,  for instance, through the presence of vortices \cite{Matthews1999,Madison2000}. Thermal effects and/or interactions deplete the condensate, and peculiar fluctuations of the ground state occupation have been extensively studied theoretically \cite{Gajda1997,Giorgini1998,Idziaszek1999,Meier1999,Kocharovsky2006,Idziaszek2005,Brewczyk2007,Idziaszek2009,Witkowska2009,Bienias2011,Bienias2011Attr,Kocharovsky16,Touchette15,Yukalov24,Kruk2023,Vibel2024}, and experimentally \cite{Chuu2005,Esteve2006,Maussang2010,Armijo2010,  Jacqmin2011,Chomaz2015,Gajdacz2016,Kristensen2019, Asteria_2021,Christensen2021}, for recent review, see \cite{kruk2025}.
 The existence of a condensate also depends strongly on geometric properties of the system. For instance, a uniform gas of massive bosons cannot condense at non-zero temperature in less than three spatial dimensions, but also 1D or 2D systems can show condensation in the presence of trapping potentials or in finite systems \cite{Dalfovo1999}. The present manuscript revisits the phenomenon of Bose-Einstein condensation from the point of view of exotic geometries, in particular in fractal lattices, characterized by (possibly) non-integer fractal dimensions, or in hyperbolic lattices, characterized by negative curvature.

The motivation behind this work stems from the recent progress in quantum engineering techniques which has provided us with various synthetic quantum systems in such exotic spaces~\cite{Grass2024}. This includes photonic fractal lattices~\cite{Xu2021,Biesenthal2022}, synthetic electronic lattices with fractal structure~\cite{Shang2015,Kempkes_2019_b}, 
fractal lattices of cold atoms in optical tweezers~\cite{Tian2023}, hyperbolic lattices realized with superconducting qubits \cite{Kollar2019}.
Many interesting aspects of quantum behavior in unconventional geometries have already been revealed: For instance, in quasiperiodic or fractal structures, the absence of Bloch theorem can give rise to localized or critical eigenstates \cite{Pavlyshynets2025}, as has theoretically been known since the early 1980s for 1D quasi-crystals~\cite{Aubry_1980,Kohmoto_1983}
or Sierpiński fractals~\cite{Domany_1983,Wang_1995, Manna2023, Manna_2024}.
Transport behavior and localization phenomena in fractal lattices have been theoretically studied both on the classical level \cite{Gefen1981,Alexander_1982,Rammal_1983,Gefen1983,Havlin_1987}, and in the quantum regime \cite{Darazs_2014,Kosior2017,vanVeen_2016,Rojo-Francas2024,Salvati2024}, including also topological transport behavior \cite{Brzezinska2018,Pai2019,Iliasov2020,Fremling2020, Moustaj2021, Manna2022b,Ivaki2022,Li2023,Krebbekx2023,   Moustaj2023, Stlhammar2023,Eek2024,Osseweijer2024,Canyellas2024, PhysRevResearch.2.023401, PhysRevA.105.L021302}.
Beyond the single-particle picture, also the BCS pairing behavior of fermions in fractal lattices \cite{Iliasov2024} and their Mott transition \cite{Caracanhas2017,Conte2024} has been studied. For weakly interacting bosons, the loop current behavior of mini-condensates in a  Sierpiński gasket has been analyzed~\cite{Koch2024}. 

Similarly, there has been growing interest in hyperbolic lattices as a platform for studying quantum phenomena in non-Euclidean geometries.
Recent research has explored topological states \cite{Zhang_2022,Yu_2020, Chen_2023, Tummuru_2024, Sun_2024}, fermionic systems \cite{G_tz_2024,PhysRevB.111.L121108, lenggenhager2024hyperbolicspinliquids}, quantum phase transitions \cite{Zhu_2021} and Bose-Einstein condensation \cite{Lemm_2022} in hyperbolic spaces, as well as other curved geometries \cite{Tononi_2019,Diniz2020, Tononi2020, Salasnich2022, Furutani2022, Tononi2022, sheilla2025,Tononi_2023,Nikolaieva2023,
Brito2023,Biral2024,
Tononi2024quantumvortices, Tononi2024shellshaped, Tononi_2024gastosoliton, tononi2025boseeinsteincondensationaxiallysymmetricsurfaces,tononi2025dimerproblemsphericalsurface}. 
Studies have also investigated the density of states in hyperbolic tight-binding models \cite{Mosseri_2023} and developed a framework for the crystallography of hyperbolic lattices \cite{PhysRevB.105.125118, Lenggenhager_2023}.

In this work, we focus on fractal and hyperbolic lattices, as examples of exotic geometries that extend the more commonly investigated Euclidean two-dimensional lattice types. The geometries will be introduced in detail in Sec.~\ref{sec:model}. We then consider both quantum-statistical effects of non-interacting bosons, and interaction effects in a Bose-Hubbard model.
The condensation behavior seen in these geometries is compared to the known behavior in regular lattices \cite{Ramakumar_2005}, highlighting the important effect of geometry on  captivating many-body quantum phenomena.
It is worth noting that the influence of lattice inhomogeneity on BEC \cite{Burioni_2001, PhysRevB.66.094207, Burioni_2000} and quantum properties in the Bose-Hubbard model \cite{Halu_2012, PhysRevB.70.184520, PhysRevB.70.224510} has been extensively studied.

Sec.~\ref{sec:Nonint} concentrates on a non-interacting boson gas at finite temperature. We compare various properties of the Bose-Einstein condensate in regular, fractal and hyperbolic geometries, including the scaling of condensate fraction and condensate fluctuations, off-diagonal long-range order, and dependence of critical temperature on system size.
We find that the condensate fraction in fractal lattices drops to zero at absolute temperatures much lower than in the corresponding regular lattices with the same number of sites. The finite size scaling of the critical temperature indicates that condensation in fractal lattices occurs only in finite systems, as is also the case in regular 2D lattices (cf. \cite{sym13020300, PhysRevA.9.2103}). Interestingly, this observation also holds for the Sierpiński tetrahedron lattice, which is embedded in three dimensions.
The situation is strikingly different in hyperbolic lattices -- we find that, despite being embedded in 2D, the condensate fraction scales with temperature similarly to the regular 3D lattice. Importantly, the critical temperature does not decrease monotonically with the system size, which suggests a non-zero $T_{\rm c}$ in the thermodynamic limit.
We also analyze the fluctuations above the condensate, which are found to be enhanced in the fractal geometries, but strongly suppressed in the hyperbolic lattice, as compared to regular lattices in 2D or 3D.

In Sec.~\ref{sec:inter}, we investigate the properties of interacting Bose gas at zero temperature. We consider the Bose-Hubbard model 
in graphs corresponding to lattices with different geometries (cf. \cite{Akaturk, Ghadimi, Conte2024}). We investigate the Mott insulator (MI) to superfluid (SF) phase transition as a function of the hopping amplitude $J$ and chemical potential $\mu$ \cite{van_Oosten_2001}, using both a Green function method~\cite{dosSantos2009,Bradlyn2009,Grass2011} and a multi-site Gutzwiller approach~\cite{McIntosh_2012,PhysRevA.83.053608,L_hmann_2013, luhmann2016notesclustergutzwillermethod, Fischer2011}. In contrast to the single-site Gutzwiller approach, which simplifies the many-body problem into a self-consistent local problem, the multi-site approach is able to capture also properties of the lattice geometry beyond the average coordination number $z$.
Specifically, the Gutzwiller calculation yields Mott lobes for the Sierpiński triangle that nicely illustrates the intermediate dimensionality of the system: the lobe exhibits a  kink at the tip, as also found in 1D systems \cite{Kashurnikov1996, Khner1998}, while the shoulders of the lobe remain convex, as for the lobes in regular 2D lattices \cite{Ohgoe2012}.
Based on a resummed hopping expansion, the Green function method predicts a re-entrant SF to Mott transition for the Sierpiński triangle, due to the presence of spectral gaps in the tight-binding band structure. Since the multi-site Gutzwiller method does not show such a re-entrant behavior, we tend to interpret it as an artifact of the hopping expansion.

\section{Model}
\label{sec:model}

We start with a lattice defined in the language of a mathematical graph. A lattice ${\cal L}$ is a connectivity graph  ${\cal L} = ({\cal V}, {\cal E})$ where ${\cal V} = \{1,\dots, V \}$ is a set of enumerated nodes, while ${\cal E} = \{ \langle i,j\rangle \in {\cal V}\times {\cal V}|i\ne j\}$ is a set of edges on the lattice. The set of edges defines an adjacency matrix of a graph $J_{ij}=J_{ji}$, $i,j=1,\dots,V$, which takes $1$ for any pair $\langle i,j\rangle\in{\cal E}$, and zero otherwise.
The tight-binding Hamiltonian on the lattice ${\cal L}$ reads
\begin{equation}
    \hat{H}_{\cal L}^0 = -J \sum_{ i,j \in {\cal V}} J_{ij}\hat{b}_i^\dagger \hat{b}_j  - \mu \sum_{i \in {\cal V}} \hat{n}_i,
\end{equation}
where $\hat b_i$, $\hat b_i^\dagger$, $\hat n_i$ are annihilation, creation, and number operator acting on site $i$. Here, we have chosen a grand-canonical description, in which the particle number $N=\sum_{i \in {\cal V}} \hat{n}_i$ is controlled by a chemical potential $\mu$. The parameter $J$ is the hopping amplitude. In the presence of interactions, the hopping competes with on-site repulsion $U$, and the system is described by the Bose-Hubbard Hamiltonian:
\begin{equation}
    \hat{H}_{\cal L} = \hat{H}_{\cal L}^0 + \frac{U}{2} \sum_{i \in {\cal V}} \hat{n}_i (\hat{n}_i - 1).
\end{equation}
In the present manuscript, we investigate the behavior in different lattice geometries, depicted in  Fig.~\ref{fig:latticeGraphs}, including fractal, hyperbolic, and Euclidean lattices.

Fractal lattices are constructed as repeating self-similar patterns. 
The Sierpiński gasket (triangle) is composed of 6-site triangles, the Sierpiński carpet (square) of 8-site squares without the center site, and the Sierpiński tetrahedron is made up of 10-site tetrahedrons.
An important property of fractals is their Hausdorff dimension $d$~\cite{Mandelbrot_1967}, which for fractal lattices is defined as the limit of how the number of sites $V$ scales with their linear size $L$, 
such that $\lim_{L\to\infty} V = L^d$. 
For regular lattices, their Hausdorff dimension is the same as their Euclidean dimension. In fractal lattices, $d$ can take non-integer values -- in the Sierpiński triangle the linear size doubles, while the number of sites triples in each fractal iteration, resulting in $d = \log(3)/\log(2) \approx 1.585$. In the Sierpiński carpet, the number of sites increases rapidly by a factor of $8$ in each iteration, 
resulting in $d = \log(8)/\log(3) \approx 1.893$. Interestingly, the Sierpiński tetrahedron has an integer Hausdorff dimension of $d = \log(4)/\log(2) = 2$.

Hyperbolic lattices are constructed from regular tilings of the hyperbolic plane, defined by their Schl\"{a}fli symbol $\{p,q\}$~\cite{coxeter1973}. The hyperbolic lattice is constructed with $p$-sided polygons, where the average vertex is connected by $q$ edges.
We construct our hyperbolic lattices as graphs with open boundary conditions, starting with either a $p$-sided polygon or a single vertex with $q$ $p$-sided polygons around it, see Fig. \ref{fig:latticeGraphs}. Larger lattices are constructed by adding a layer of polygons around the smaller lattice. Hyperbolic lattices that satisfy the equation $(p - 2)(q - 2) > 4$, 
such as the $\{p, q\} = \{7,3\}$ and $\{p, q\} = \{3,7\}$ lattices,
are characterized by a constant negative curvature~\cite{edmonds1982}.

\section{Non-interacting boson gas}
\label{sec:Nonint}
We first analyze the quantum-statistical behavior of non-interacting bosons in different geometries, including regular, fractal, and curved lattices. To this end, let $\epsilon_{k}/J$ denote the eigenvalues of the adjacency matrix $-J_{ij}$. The non-interacting Hamiltonian in diagonal form reads
\begin{align}
    \hat H^0= \sum_{k} (\epsilon_{k} - \mu) n_{k} 
\end{align}
where $n_k$ denotes the occupation of level $\epsilon_k$. It is given, as a function of chemical potential $\mu$ and inverse temperature $\beta=1/(k_{\rm B}T)$, by the Bose-Einstein distribution function
\begin{align}
n_{k}= \frac{1}{e^{\beta (\epsilon_{k}-\mu)}-1}.
\end{align}
 Let $k=0$ denote the ground state. If the occupation of the ground state, $n_0$, becomes macroscopic, the system is considered to be condensed. For a precise definition of such macroscopic ground state occupation, we evaluate the number of excited particles $N_{\rm ex}(\mu=\epsilon_0) = \sum_{k>0} n_{k}$ for $\mu=\epsilon_0$. Since $n_0$ (and hence also the total particle number $N=n_0+N_{\rm ex}$ diverges for $\mu=\epsilon_0$, in practice $\mu<\epsilon_0$, and $N_{\rm ex}(\mu=\epsilon_0) \equiv N_{\rm ex}^{\rm max}$ is an upper bound for the number of excited particles in the system at the given temperature, $N_{\rm ex}(\mu) < N_{\rm ex}^{\rm max}$. For some interval $ \mu_{\rm c} < \mu < \epsilon_0 $, the total number of particles $N(\mu)$ will exceed the maximum number of excited particles $N_{\rm ex}^{\rm max}$ and the system is considered to be condensed, as $n_0$ must be macroscopic. To the critical chemical potential $\mu_{\rm c}$ corresponds a critical total number of particles $N_{\rm c}= N_{\rm ex}^{\rm max}$
which depends on the chosen temperature. By inverting this function $N_{\rm c}(T)$, we obtain the critical temperature $T_{\rm c}$ below which condensation sets in for a given particle number $N_{\rm c}(T_{\rm c})$.

\subsection{Condensate fraction}

\begin{figure}[t]
\centering    \includegraphics[width=0.48\textwidth]{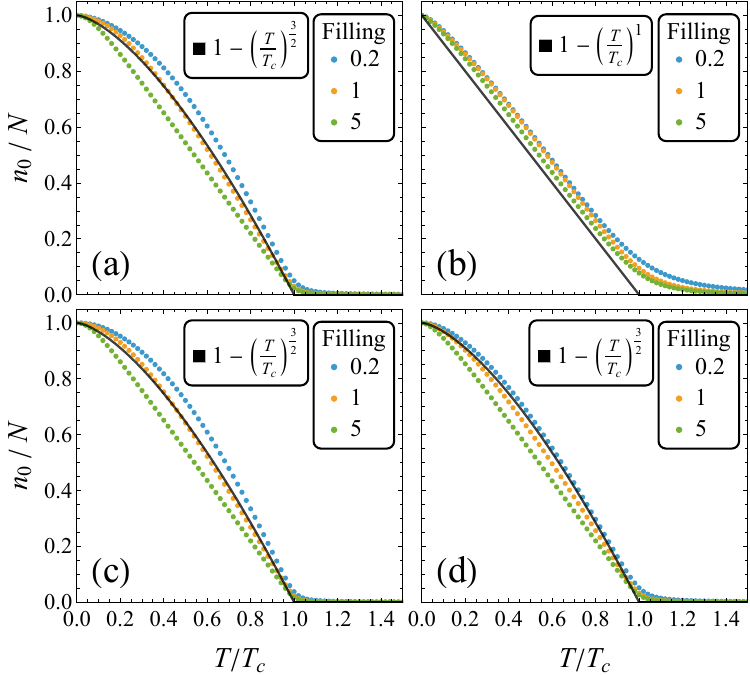}
    \caption{
    Condensate fraction in regular and hyperbolic lattice geometries. We plot the condensate fraction as a function of the temperature in units of the critical temperature $T_{\rm c}$ for (a) a three dimensional simple cubic lattice with $V = 20^3$
    sites, and (b) a two dimensional 
    square lattice with $V = 90^2$ sites, both with open boundary conditions, as well as a $\{p,q\}=\{3,7\}$ hyperbolic lattice with $V = 11173$ sites (c), and a $\{p,q\}=\{7,3\}$ hyperbolic lattice with $V = 9136$ sites (d). 
    The different colored points correspond to different lattice fillings.
    The solid black line corresponds to the theoretical curve of Eq.~(\ref{n0fit}), for the regular geometries. In the hyperbolic lattices, the condensate fraction behaves similarly to the 3D cubic lattice.
    }
    \label{fig:n0combined}
\end{figure}

\begin{figure}[t]
\centering    \includegraphics[width=0.48\textwidth]{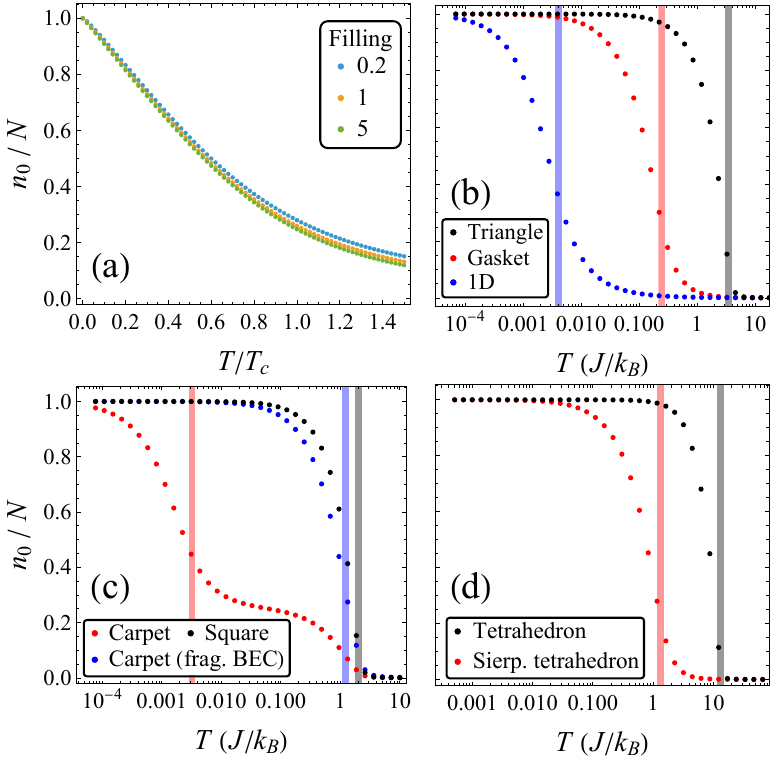}
    \caption{Condensate fraction in fractal lattices vs temperature. Panel (a) shows, on a linear scale, the condensate fraction in a Sierpiński triangle lattice ($V=3282$ sites) in units of the critical temperature. The condensate fraction drops linearly at small temperatures with a heavy tail above $T_{\rm c}$, regardless of the filling. In panels (b-d), the behavior in different fractal lattices is compared to their non-fractal counterparts with a similar number of sites: (b) Sierpiński triangle lattice ($V=3282$ sites), a standard triangular lattice ($V=3321$ sites) and a 1D lattice ($V = 3282$), (c) Sierpiński carpet ($V=4096$ sites) and square lattice ($V=4096$ sites); for the carpet, both the standard condensate and the fragmented condensate (including the four lowest eigenstates) are shown. 
    Panel (d) compares the Sierpiński tetrahedron lattice ($V=2050$ sites) to a regular tetrahedral lattice ($V=2024$ sites). In all plots, unit filling is chosen, and the condensate fraction is plotted vs a logarithmic temperature scale in units of the hopping constant $J$. The vertical lines 
    mark the critical temperatures of the fractal and reference lattices at unit filling.
    }
    \label{fig:fracN0}
\end{figure}

For any temperature, the total number of particles $N$ shall now be fixed (via $\mu$) according to a desired filling $N/V$ of the lattice, where $V$  is the number of sites in the finite lattice. According to the mentioned procedure, we then define $T_{\rm c}$, and evaluate the condensate fraction $n_0/N$ as a function of normalized temperature $T/T_{\rm c}$. Within the interval $0\leq T \leq T_{\rm c}$, The behavior of the condensate fraction can be fit to a function
\begin{align}
    \frac{n_0}{N} = 1- \left(\frac{T}{T_{\rm c}}\right)^{\alpha}.
    \label{n0fit}
\end{align}
It is known that with $\alpha=3/2$  this function describes accurately the ideal Bose gas in a 3-dimensional box, cf. Ref.~\cite{Pitaevskii-book}.

As shown in Fig.~\ref{fig:n0combined}(a), a similar behavior characterizes also the ideal gas in a 3D cubic lattice, although at small fillings a slightly slower decay of condensate fraction is observed due to the finite size effects, whereas at larger filling lattice effects cause a slightly faster decay. In a 2D square lattice, the condensate fraction decays much faster (with $\alpha\approx1$) for any filling, as shown in Fig.~\ref{fig:n0combined}(b). Remarkably, by changing from an Euclidean plane to a hyperbolic one, the behavior of the condensate fraction becomes similar to that of the cubic lattice, see Fig.~\ref{fig:n0combined}(c-d). 

Quantitatively, the condensate behavior in fractal lattices is similar to the 2D lattice, as over a broad temperature range the condensate fraction decreases linearly with $T$, see Fig.~\ref{fig:fracN0}(a) for the case of a Sierpiński triangle. However, a pronounced qualitative difference is the heavy tail of the condensate fraction in fractal lattices, with a non-zero $n_0$ even for $T \gg T_{\rm c}$. In this context, it should also be noted that the critical temperatures on fractal lattices are orders of magnitude smaller than the critical temperatures in regular lattices of comparable size. In Fig.~\ref{fig:fracN0}(b--d) we compare the behavior of the condensate fraction in a Sierpiński triangle and a regular triangular lattice (b), in a Sierpiński carpet and a regular square lattice (c), and in a Sierpiński tetrahedron and a regular tetrahedral lattice (d), with the vertical lines marking the critical temperatures in the different geometries. This comparison illustrates that very distinct temperature scales are relevant for fractal and regular lattices, and the heavy tail in the fractal lattice extends approximately in the temperature range between the critical temperature of the fractal lattice and the critical temperature of the corresponding regular lattice.

\begin{figure}[t!]
\centering    \includegraphics[width=0.48\textwidth]{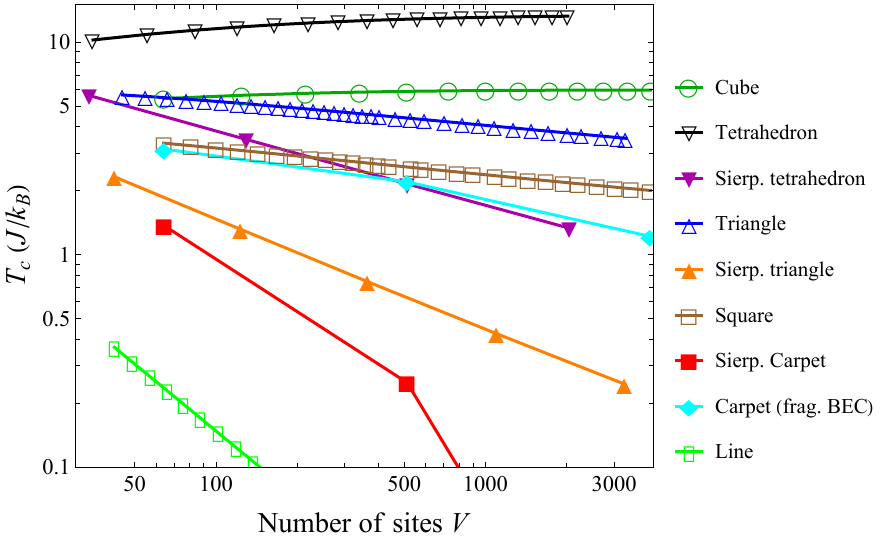}
    \caption{
    Critical temperature $T_{\rm c}$ at a unit filling as a function of the system size, on a double logarithmic scale.
    The results from a cubic, tetrahedral, triangular, square and 1D lattices are compared to results from fractal geometries: the Sierpiński gasket ($d \approx 1.585$), Sierpiński tetrahedron ($d = 2$) and Sierpiński carpet ($d \approx 1.893$).
    In the Sierpiński carpet, the condensate fragments between the lowest $4$ eigenstates, which results in rapid decay of $T_{\rm c}$. One can consider BEC through the occupancy of any of these $4$ states, which produces different results (cyan).
    In general, the $T_{\rm c}$ decreases and decays faster in lattices with smaller dimension, and does not decay in the 3D tetrahedral and cubic lattices.
    }
    \label{fig:fracTc}
\end{figure}

In Fig.~\ref{fig:fracN0}(c) we also observe an interesting behavior: The critical temperature in the carpet is several orders of magnitudes lower than the one in the square lattice. This results in a fast drop of the condensate fraction until it slows down when $n_0$ reaches $0.25$. The next drop occurs at temperatures close to the $T_{\rm c}$ of the standard geometry. This behavior can be explained by examining the spectrum of the carpet lattice: The three first excited states lie very close to the ground state (within $4\cdot 10^{-6} J$) while the next excited state is more than $2\cdot 10^{-3} J$ away from the ground state. Therefore, a condensate of the true ground state depletes quickly and fragments into the macroscopic occupation of the lowest four eigenstates, before the occupation spills towards higher branches of the spectrum. This example nicely illustrates how the fractal nature of the lattice can give rise to fractal structures in the energy spectrum, which then also manifest in the condensation behavior. One can account for this fragmentation by extending the definition of the condensate fraction to include particles occupying any of the four lowest states. As shown in Fig.~\ref{fig:fracN0}(c), the obtained temperature dependence of the condensate fraction and the corresponding critical temperature closely resemble those of the 2D square lattice, consistent with the carpet’s Hausdorff dimension $d \approx 1.893$.

\subsection{Critical temperature}

\begin{figure}[t]
\centering    \includegraphics[width=0.48\textwidth]{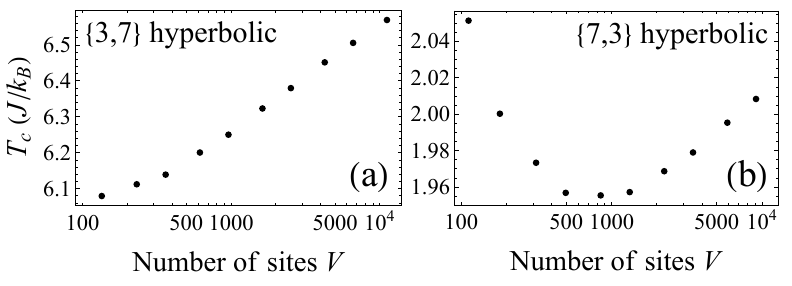}
    \caption{
    Critical temperature as a function of the system size in hyperbolic lattices. 
    The $\{p,q\}=\{3,7\}$ lattice (a) displays a small increase in critical temperatures with the system size (similarly to a cubic lattice with OBC). 
    The $\{p,q\}=\{7,3\}$ lattice (b) displays a small decrease followed by an increase in the critical temperature. 
    Critical temperature not decreasing monotonically with the number of sites suggests a finite value in the thermodynamic limit.
    }
    \label{fig:hypTc}
\end{figure}

From the discussion of the condensate fraction, we have already noted substantial differences of the critical temperature in different geometries at fixed system sizes. 
A systematic comparison of $T_{\rm c}$ as a function of system size $V$ is shown in Fig.~\ref{fig:fracTc} for different fractal lattices, as well as for a regular 1D lattice (line), regular 2D lattices (triangular and square lattice), and regular 3D lattices (cubic and tetrahedral lattices). Not surprisingly, the latter ones are the only ones in which $T_{\rm c}$ does not decrease as a function of $V$. In fact, it is well established that, in the absence of a trapping potential, the critical temperature of a 2D system vanishes in the thermodynamic limit, and remains finite only in 3 (or more) spatial dimensions. 
More remarkably, we also observe in Fig.~\ref{fig:fracTc} that $T_{\rm c}$ is strictly ordered according to the Hausdorff dimension of each structure, both regarding its absolute value at a given $V$ and regarding its slope as a function of $V$ -- with the exception of the Sierpiński carpet. The critical temperature for the carpet lattice decreases almost as rapidly as in the 1D lattice, due to condensate fragmentation discussed in the previous section. When fragmentation is taken into account, the critical temperature falls between that of the Sierpiński triangle ($d\approx1.585$) and the 2D lattices, with only slight overlap with the Sierpiński tetrahedron. We further note that the Hausdorff dimension does not \textit{uniquely} determine the behavior of $T_{\rm c}$: the triangular, square, and Sierpiński tetrahedron lattices possess the same Hausdorff dimension $d=2$, but exhibit slightly different values of $T_{\rm c}$.

Strikingly different is the behavior in hyperbolic lattices. Although these lattices can be mapped onto a 2D Poincaré disk, the critical temperatures are found to increase with the system size, for sufficiently large systems, see Fig.~\ref{fig:hypTc}. Clearly, this behavior suggests that the critical temperature does \emph{not} vanish in the thermodynamic limit. Instead, we expect that the critical temperature will saturate as in the cubic lattice, however at system sizes that are too large for our computations. In this sense, the hyperbolic lattices provide a remarkable exception to other 2D structures with vanishing $T_{\rm c}$.
In the following subsection, we will show that the different condensation behavior is explained by the very different spectral behavior in fractal, hyperbolic, or regular lattices.

\begin{figure}[t!]
\centering    \includegraphics[width=0.48\textwidth]{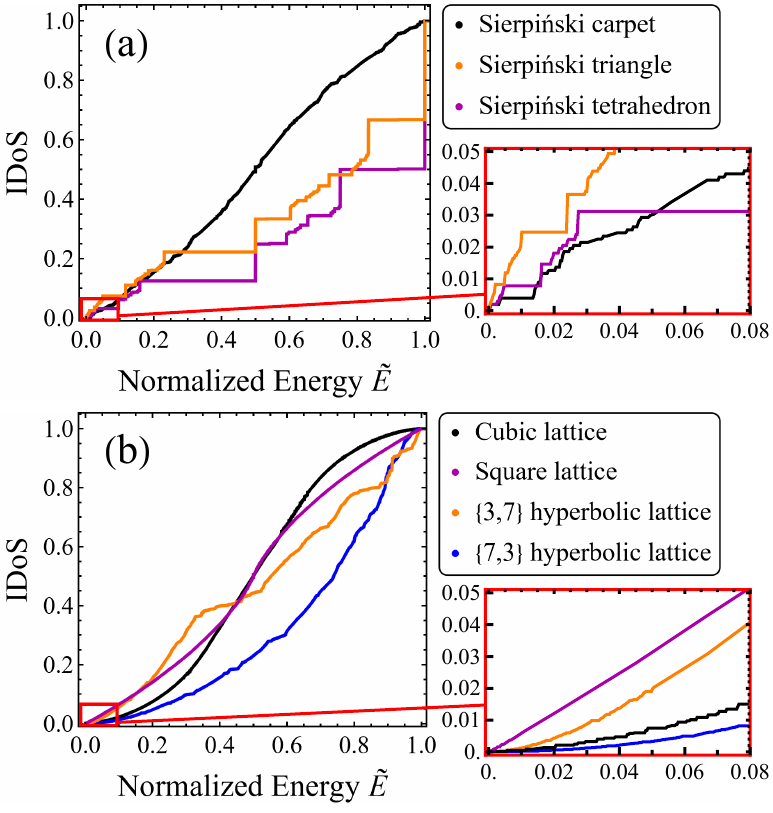}
    \caption{
    Integrated density of states (IDoS) in (a) three fractal lattices:
    Sierpiński carpet with $V=4096$ sites (black),
    Sierpiński triangle with $V=3282$ sites (orange), 
    and Sierpiński tetrahedron with $V=2050$ sites (purple), 
    as well as (b) hyperbolic and regular rectangular lattices:
    cubic lattice with $V=20^3 = 8000$ sites (black),
    rectangular lattice with $V=90^2 = 8100$ sites (purple),
    $\{p,q\}=\{3,7\}$ hyperbolic lattice with $V=11173$ sites (orange) and 
    $\{p,q\}=\{7,3\}$ hyperbolic lattices with $V=9136$ sites (blue).
    For each geometry, the energy of each state $E$ is normalized using the energy of the ground state $\epsilon_0$ and the highest energy $\epsilon^*=\max(\epsilon_{k})$, so that $\tilde{E} = (E - \epsilon_0)/(\epsilon^* - \epsilon_0)$.
    In fractal lattices, the IDoS increases rapidly near the ground state, as shown in the red frame of Fig. \ref{fig:combinedIDOS}(a).
    As can be seen in Fig. \ref{fig:combinedIDOS}(b), the IDoS in the rectangular lattice scales linearly near the ground state, while in the cubic and hyperbolic lattices it follows a power-law scaling $\rm{IDoS}(\tilde{E})\sim \tilde{E}^{\alpha}$ with $\alpha>1.5$.
    }
    \label{fig:combinedIDOS}
\end{figure}

\subsection{Density of states}

For an analysis of spectral properties it is convenient to look at the integrated density of states (IDoS) in the different geometries. In Fig.~\ref{fig:combinedIDOS}(a), we examine different fractal lattices which share similar properties, including large gaps (i.e. horizontal lines in the IDos) and huge degenerate of quasi-degenerate manifolds (i.e. vertical lines/jumps in the IDos). We note that these features are less pronounced in the Sierpiński carpet.
The IDoS of the cubic and square lattice as well as different hyperbolic lattices is shown in Fig.~\ref{fig:combinedIDOS}(b). Their behavior is much smoother, and at low energies can be well fitted by a power-law function for the cubic lattice and the hyperbolic lattices and a linear function for the square lattice. It is well established that
a linear IDoS, corresponding to  a non-zero density of states (DoS) near the ground state, prevents BEC in the thermodynamic limit \cite{Hohenberg1967,Bagnato1991}. 
From this perspective, it is not surprising that the hyperbolic lattices exhibits very similar condensation behavior as the cubic lattice, with a condensate that remains stable in the thermodynamic limit. 
We note that the DoS of hyperbolic lattices has also been studied in recent literature --
Mosseri \textit{et al.} \cite{Mosseri_2023} investigate the DoS of \{$p$,3\} lattices by computing continuous-fraction expansions of the lattice Green's functions. 
A power-law fit of the integrated density of states, $\mathrm{IDoS}(\tilde{E}) \sim \tilde{E}^{\alpha}$, to the lowest $5\%$ of the spectrum, shown in Fig.~\ref{fig:combinedIDOS}(b), yields $\alpha = 1.93$ for the ${3,7}$ hyperbolic lattice, $\alpha = 1.66$ for the ${7,3}$ hyperbolic lattice, and $\alpha = 1.65$ for the cubic lattice.
In the cubic lattice, the density of states scales approximately as the square root of the excitation energy \cite{JELITTO1969609}, corresponding to an expected $\alpha = 1.5$. The slightly higher fitted value results from fitting over a finite portion of the spectrum rather than the asymptotic low-energy regime.

Although the spectral gaps in fractal systems prevent from fitting their IDoS to smooth functions, one can nevertheless approximate their rough trend at low $E$ through a power-law fit. For all fractal structures analyzed in Fig.~\ref{fig:combinedIDOS}(a), the exponent is on the order of $1$, providing the same argument against condensation that holds in the case of the square lattice.

\subsection{Condensate fluctuations}

\begin{figure*}[t]
\centering    \includegraphics[width=0.99\textwidth]{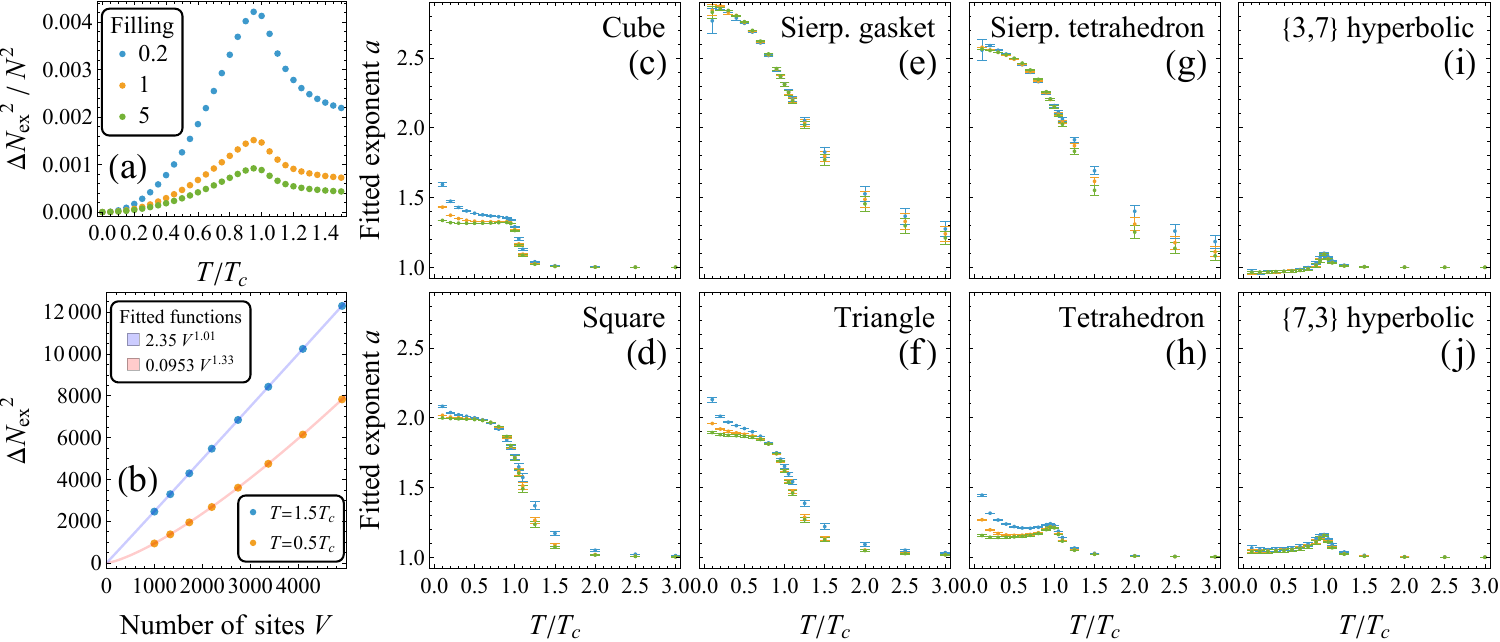}
    \caption{
    Excited particles fluctuations in various geometries.
    Panel (a) shows the fluctuations in a cubic lattice with $V=15^3$ sites 
    as a function of the temperature, peaking at the critical temperature $T_{\rm c}$. 
    Panel (b) shows the system size scaling of fluctuations in the cubic geometry. For $T=1.5 T_{\rm c}$, the fluctuations scale linearly, whereas for $T<T_{\rm c}$ the fluctuations scale anomalously.
    with an exponent $a\approx4/3$.
    In panels (c-j), we show the obtained fit parameter $a$ with error bars from the fit at various temperatures for:
    (c) cubic lattices, (d) square lattices, (e) Sierpiński triangular lattices, (f) regular triangular lattices, (g) Sierpiński tetrahedral lattices, (h) regular tetrahedral lattices, (i) $\{p,q\}=\{3,7\}$ hyperbolic lattices and (j) $\{p,q\}=\{7,3\}$ hyperbolic lattices. 
    }
    \label{fig:combinedFlucs}
\end{figure*}

Bose-Einstein condensates are not only characterized by the macroscopic occupation of the ground state, but also by the fluctuations of the occupation numbers. The fluctuations of particles out of the condensate, $\Delta N_{\rm ex}^2$, are evaluated as \cite{Pitaevskii-book}
\begin{align}
\Delta N_{\rm ex}^2 = \sum_{k>0}  n_{k}(n_{k}+1).
\end{align}
In the non-condensed regime, fluctuations of the ground state population behave normally, that is, they scale linearly with the system size,  $\Delta N_{\rm ex}^2 \propto V$, where $V$ denotes the volume of the system, or the number of sites in the lattice. In the condensed regime, fluctuations may become anomalous, subtly depending on geometric properties of the system, interactions, and even the choice of thermodynamic ensemble, see Ref.~\cite{kruk2025} for a review.
For a non-interacting continuum gas in a 3D box potential, the anomalous fluctuations are characterized by an exponent $a=4/3$, that is by the behavior $\Delta N_{\rm ex}^2 \propto V^a$, cf.~\cite{Pitaevskii-book}. 

 Here, we numerically study fluctuations in  different lattice geometries. Specifically, we have fitted the fluctuations towards a function $bV^a$, with $a,b$ fit parameters. The value of the exponent $a$ is plotted vs. temperature in Fig.~\ref{fig:combinedFlucs} for regular, fractal and hyperbolic lattices. 
 As shown in Fig.~\ref{fig:combinedFlucs}(a-c), the fluctuation scaling in a cubic lattice is similar to the 3D continuum system. Below $T_{\rm c}$, we observe an anomalous scaling of fluctuations with a relatively flat value of $a\approx 4/3$.
 With an even larger exponent $a \approx 2$, anomalous fluctuations are also observed in square lattices, see Fig.~\ref{fig:combinedFlucs}(d). The behavior in triangular lattices or tetrahedral lattices, shown in Figs.~\ref{fig:combinedFlucs}(f,h), is very similar to the square and cubic cases, respectively, suggesting that mostly the dimensionality of the lattice matters in case of regular lattices.

Significant differences can be observed for fractal and hyperbolic lattices:
\begin{itemize}
\item Both for the Sierpiński triangle and the Sierpiński tetrahedron, Fig.~\ref{fig:combinedFlucs}(e,g), the anomalous scaling of fluctuations is significantly larger than in regular 2D lattices. At very low temperatures, the exponent $a$ reaches values $a>2.5$, and it remains larger than 2 at any $T<T_{\rm c}$. In contrast to the regular lattices, where the exponent $a$ is approximately flat, the exponent $a$ becomes strongly temperature-dependent in fractal lattices.
\item Also the behavior above $T_{\rm c}$ brings the fractal in contrast to the regular lattice: While in a regular lattice $a$ drops abruptly to 1 at $T\approx T_{\rm c}$, $a$ decreases only slowly in the fractal systems, and we have $a>1$ even for temperatures far above $T_{\rm c}$. This is suggestive of a crossover regime between a condensed and non-condensed phase, in line with the heavy tail of the condensate fraction.
\item The hyperbolic systems, shown in Fig.~\ref{fig:combinedFlucs}(i,j), exhibit no or only  very weak anomalous fluctuation scaling. This makes their scaling behavior in condensed and non-condensed phases very similar, with $a\approx 1$.
\item The distinction between condensed and non-condensed regimes in the hyperbolic lattices can be obtained from the fact that $a$ exhibits a peak (at around $a\approx 1.15$) in the vicinity of the critical temperature. Such a peak is not unique to the hyperbolic lattices, and can also be observed in the case of a tetrahedral lattice.
\end{itemize}

\subsection{Long-range order}

\begin{figure}[t!]
\centering    \includegraphics[width=0.48\textwidth]{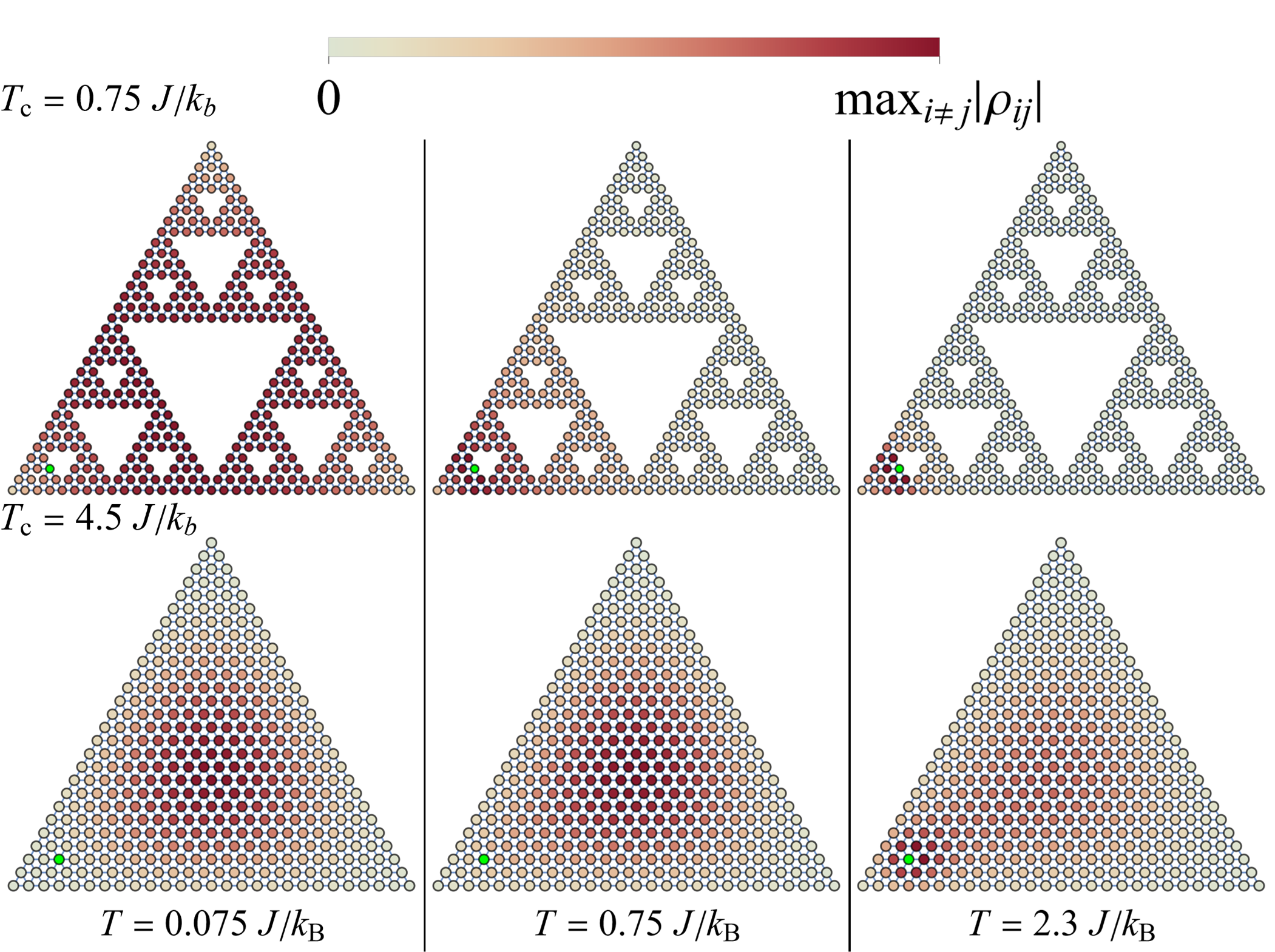}
    \caption{Thermal average of the single-particle density matrix (SPDM) elements in a Sierpiński triangle lattice with $V=366$ sites and a regular triangular lattice with $V=378$ sites. 
    The plots show $\rho_{ij}$, normalized by $\max_{i\ne j} |\rho_{ij}|$, for a fixed reference site $j$, highlighted in green.
    Each column compares the two geometries at the same absolute temperature. In the left column, the temperature $T = 0.075\,J/k_B$ is far below $T_{\rm c}$ for both fractal and regular geometry, and both structures exhibit long-range order.
    In the middle column, at $T = 0.75\,J/k_B$, the fractal lattice is at its critical temperature, and the long-range order is confined to a part of the structure. In the right column, where $T = 2.3\,J/k_B$ corresponds to half the critical temperature of the triangular lattice, long-range order is reduced in both the regular and fractal lattices.
    }
    \label{fig:fracCorr}
\end{figure}

All properties discussed so far rely exclusively on the energy spectrum of the tight-binding model in a given graph. However, as Bloch theorem does not apply to non-periodic lattice such as the fractal ones, also the eigenstates in such geometries can differ significantly from the eigenstates in regular lattices. In particular, fractal lattices also admit localized eigenstates. Here, we analyze whether this affects the spatial properties of the corresponding Bose-Einstein condensates, in particular, we study the single-particle density matrix  (SPDM) $\rho_{ij}=\langle \hat b_i^\dagger \hat b_j \rangle$, where $\langle\cdot\rangle$ shall denote thermal averaging. For regular lattices, condensation into the zero-momentum mode establishes long-range order of the SPDM, i.e. $\rho_{ij} \rightarrow \rho_0>0$ for $|i-j|\rightarrow \infty$.

As is shown in Fig.~\ref{fig:fracCorr}, long-range order is not only observed in regular lattices, but also for a condensate in the fractal lattice. 
Each panel displays the thermal average of the SPDM elements $\rho_{ij}$, calculated with respect to a single reference site (highlighted in green). In this way, the plot illustrates how correlations decay with distance from this site across the lattice.

At low temperature (left column), both the regular and fractal lattices exhibit extended coherence, with correlations remaining large throughout the system, signaling the presence of long-range order. At intermediate temperature (middle column), equal to the critical temperature of the fractal lattice, correlations remain strong only within parts of the fractal structure, while in the regular lattice they remain unchanged. Finally, at high temperature (right column), long-range order disappears also in the regular lattice, consistent with thermal depletion of the condensate.
In accordance with the lower critical temperature in the fractal lattice, the long-range order of the fractal is thus less robust against temperature as compared to the regular lattice. 

\subsection{Discussion} 
We have compared the condensation behavior of ideal bosons in different lattice geometries (Euclidean lattices, fractal lattices, hyperbolic lattices). In all fractal lattices studied (Sierpiński gasket, Sierpiński carpet, Sierpiński tetrahedron), we find a significant suppression of critical temperature as compared to regular lattice systems. However, it should be noted that the condensate fraction exhibits a heavy tail, and a significant portion of the particles remains in the ground state even above $T_{\rm c}$. This behavior is a result of fragmentation of the condensate among several states within a quasi-degenerate manifold, as seen explicitly in the case of the Sierpiński carpet, cf. Fig.~\ref{fig:fracN0}(c). In the thermodynamic limit (where no finite-size tail should appear), we do not expect condensation in the fractal lattice at finite temperature (see Fig.~\ref{fig:fracTc}(a,c)), in accordance with their low dimensionality. This is also the case for the Sierpiński tetrahedron, which is embedded in a 3D space, but with a fractal dimension of 2. More specifically, the absence of condensation in the thermodynamic limit is a consequence of the linear scaling of integrated density of states at low energies, see Fig.~\ref{fig:combinedIDOS}(a). The condensates in the fractal are characterized by highly anomalous fluctuations of condensate occupation number, $\Delta N_{\rm ex}^2 \sim V^{a}$ with $a>2.5$, see Fig.~\ref{fig:combinedFlucs}(e,g). Similarly to condensates in regular geometries, the fractal condensates exhibit off-diagonal long-range order in the one-body density matrix, see Fig.~\ref{fig:fracCorr}.

The behavior is strikingly different in the hyperbolic lattice: Even though these lattices are embedded in 2D, they show condensation behavior reminiscent of the 3D Euclidean lattice. In particular, the critical temperature does not vanish in the thermodynamic limit, see Fig.~\ref{fig:hypTc}, and the condensate fraction follows a similar power-law decay as in the cubic lattice, see Fig.~\ref{fig:n0combined}. On the other hand, anomalous fluctuations turn out to be very weak in the hyperbolic lattice, cf. Fig.~\ref{fig:combinedFlucs}(i,j).

\section{Interacting boson gas}
\label{sec:inter}

\subsection{Cluster Gutzwiller ansatz}
To numerically investigate the Mott Insulator to Superfluid (MF/SF) transition, taking into account both lattice geometry and interactions, we  employ the cluster Gutzwiller ansatz, \cite{McIntosh_2012, L_hmann_2013, luhmann2016notesclustergutzwillermethod}, adapted to exotic geometries.

\begin{figure}[t!]
\centering    
\includegraphics[width=0.450\textwidth]{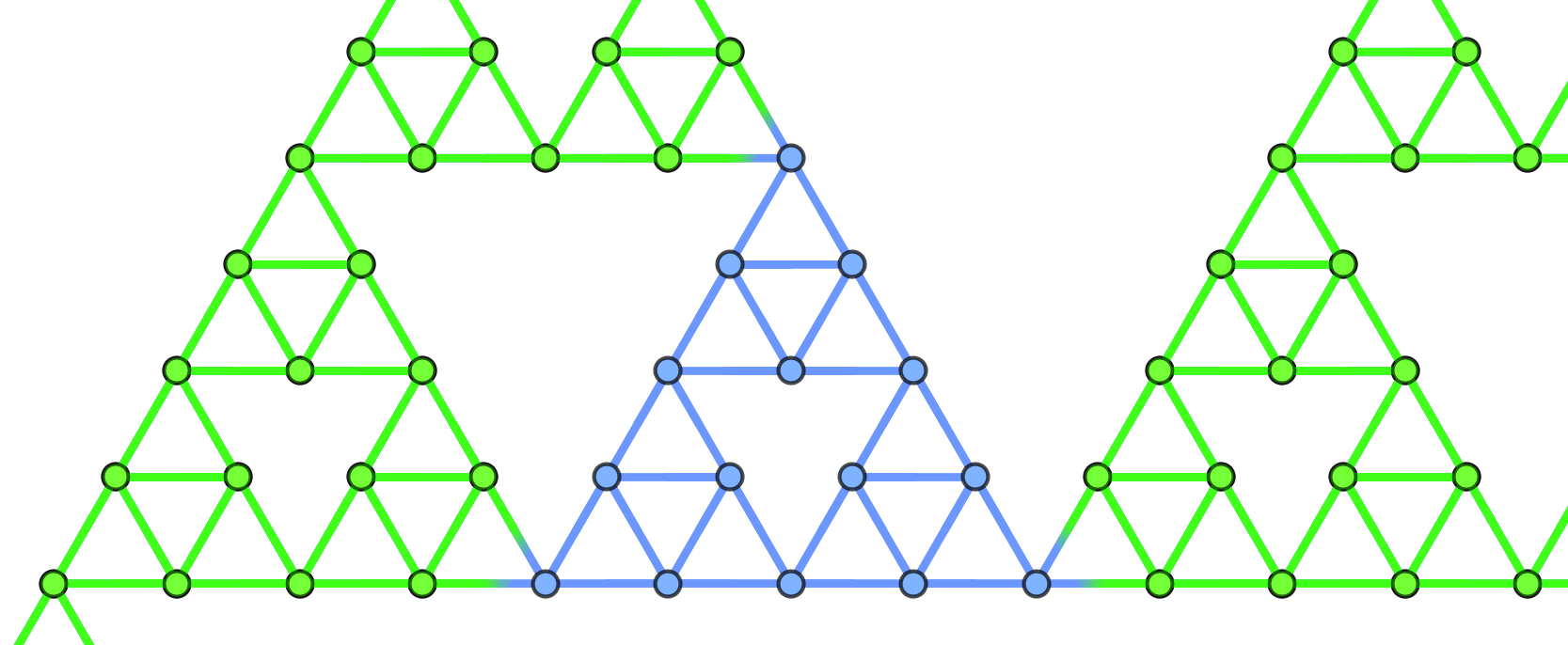}
     \caption{
     Illustration of a single cluster (blue) embedded in an infinite Sierpiński triangle lattice (green). Hopping within the cluster (between blue-blue nodes) is treated exactly, $\hat{b}^\dagger_i\hat{b}_j + h.c.$, while hopping between  neighboring clusters (between blue-green nodes), is $\phi^*\hat{b}_i + h.c.$, where $\phi$ is a mean-field order parameter.}
     \label{fig:translation}
\end{figure}

\begin{figure}[t!]
\centering
\includegraphics[width=0.99\linewidth]{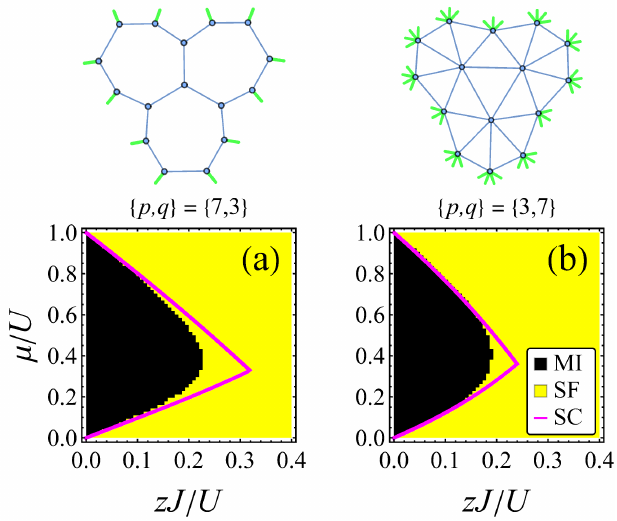}
    \caption{
    Phase diagrams of the Mott insulator and superfluid phases on 
    $\{p,q\}=\{7,3\}$ (a) and $\{p,q\}=\{3,7\}$ (b) hyperbolic lattices for clusters (with $V_{\cal C}=16$, and $V_{\cal C}=15$ sites,  respectively) shown above each plot. Green edges indicate mean-field couplings to the infinite lattice. The black regions denote the MI phase (vanishing order parameter $|\phi| = 0$), while yellow marks the SF phase where (non-zero order parameter $|\phi|$).
    The magenta line represents the strong-coupling (SC) expansion from Ref.~\cite{PhysRevB.53.2691}, calculated for the $\{7,3\}$ lattice with $V = 9136$ sites and the $\{3,7\}$ lattice with $V=11173$ sites.
    } 
    \label{fig:homo_MI_SF_hyp}
\end{figure}

\begin{figure}[t!]
\centering    
\includegraphics[width=0.99\linewidth]{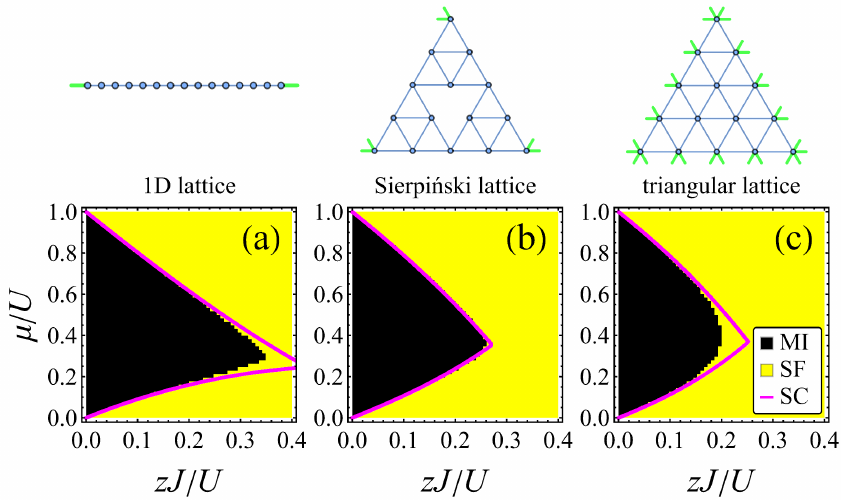}
    \caption{Same as Fig.\ref{fig:homo_MI_SF_hyp}, for lattices with dimension: $d=1$ -- 1D lattice (a), $d=\log3/\log2$ -- Sierpiński triangle (b), $d=2$  -- triangular lattice (c). The Mott lobe for the fractal lattice exhibits a shape that lies in-between the one- and two-dimensional geometries, in accordance with its fractional Hausdorff dimension.
    Each cluster has $V_{\cal C}=15$ sites.
    The magenta line represents the strong-coupling (SC) expansion from Ref.~\cite{PhysRevB.53.2691}, calculated for the 1D, Sierpiński triangle, and triangular lattices with $V=3282$, $V=3282$, and $V=3321$ sites, respectively.}
    
    \label{fig:homo_MI_SF}
\end{figure}

Let us start with the definition of a cluster. The cluster ${\cal C}$ on a lattice ${\cal L}$ is defined as a set of nodes forming a \textit{unit cell} that can tile the whole lattice while preserving the structure of the node connections, for example see Fig.~\ref{fig:translation}. As such, we define a cluster as a pair ${\cal C} = ({\cal V_{\cal C}}, {\cal E}_{\cal C})$ containing set of nodes ${\cal V}_{\cal C} = \{i, i = 1,\dots, V_{\cal C}\}$, and edges ${\cal E}_{\cal C} = \{ \langle i,j\rangle, i,j \in {\cal V}_{\cal C} \}$, with corresponding adjacency matrix $J^{\cal C}_{ij} = 1$ for any $\langle i,j\rangle\in {\cal E}$, and $0$ otherwise.
We define the Bose-Hubbard Hamiltonian on the cluster ${\cal {C}}$  
\begin{align}
    \hat{H}_{\cal C} =& -J \sum_{ i,j \in {\cal V}_{\cal C}}J^{\cal C}_{ij}\hat{b}_i^\dagger \hat{b}_j +  \frac{U}{2} \sum_{i \in {\cal V}_{\cal C}} \hat{n}_i (\hat{n}_i - 1) 
    \nonumber \\
    &- \mu \sum_{i \in {\cal V}_{\cal C}} \hat{n}_i,
\end{align}
with bosonic Hilbert space ${\cal H}$, constructed as described in Appendix \ref{app:truncation}.
Next, we define Hamiltonian describing the hopping between clusters, assuming the mean-field decoupling $\hat{b}_i^\dagger\hat{b}_j \approx \langle\hat{b}_j^\dagger\rangle\hat{b}_i\equiv \phi^*\hat{b}_i$, 
\begin{equation}
    \hat{H}_{{\partial\cal C}} = - J\sum_{i \in \cal{V}_{\partial\cal C}} \nu_i(\phi^*\hat{b}_i + \phi\hat{b}_i^\dagger),
\end{equation}
where $\cal{V}_{\partial\cal C}$ is the set of indices belonging to the boundary of cluster ${\cal C}$ (denoted as $\partial \cal C$), $\nu_i$ is the number of external couplings of site $i$ to the mean-field (denoted as green edges in Fig~\ref{fig:homo_MI_SF_hyp} and Fig~\ref{fig:homo_MI_SF}), and
$\phi = \langle GS|\hat{b}_{i_0}|GS\rangle$, where 
$i_0$ is an index of a site in the center of the cluster, and
$|GS\rangle$ is a ground state of a 
coupled-cluster Hamiltonian
\begin{equation}
    \hat{\cal{H}}_{{\cal C}} = \hat{H}_{\cal C} + \hat{H}_{\partial\cal C},
\end{equation}

The mean-field order parameter $\phi$ is a signature of the quantum phase of the system. It vanishes for Mott Insulator, and is non-zero for superfluid phase. The order parameter $\phi$ can be obtained self-consistently starting with initial small value (here we initialize it as $\phi=10^{-3}$), and iteratively updated until convergence. Since $\phi$ converges monotonically (cf. \cite{McIntosh_2012}), we can distinguish the phase after a single iteration.

We start with hyperbolic lattices. Figure \ref{fig:homo_MI_SF_hyp} presents the MI/SF phase diagram for two types of hyperbolic lattices, namely $\{p, q\} = \{7,3\}$ and $\{p, q\} = \{3,7\}$ on clusters with $V_{\cal C} = 16$ and $15$ sites respectively, where $p$ denotes number of connections to the node, and $q$ denotes the number of edges in a lattice tile.
To enable comparison across different geometries, the hopping amplitude is rescaled by the average coordination number $z$, which accounts for the total number of connections per site. 
The shapes of the lobes are similar to those in 2D Euclidean geometry. The $\{p, q\} = \{7,3\}$ lattice has a slightly taller lobe, which can be attributed to the smaller ratio of mean-field couplings to the total number of connections $\lambda$, as discussed in \cite{L_hmann_2013}.

Next, we study the MI/SF phase diagram in Sierpiński triangle fractal geometry, with Hausdorff dimension $d = \log3/\log2$, focusing on the change in the Mott-lobe shape when going from a one dimensional geometry (1D), $d=1$, to a two dimensional geometry (2D), $d=2$ on the triangular lattice.

Fig.~\ref{fig:homo_MI_SF}, presents corresponding phase diagrams, allowing to see the quantitative change of the Mott lobe for three different Hausdorff dimensions $d=1,\log3/\log2,2$.
Each cluster consists of $V_{\cal C}=15$ sites and is embedded in the surrounding lattice via the mean-field links, highlighted in green in the diagrams above the plots.
Despite differences in internal structure, all sites in each cluster have the same total number of connections, which is $z = 2$ in 1D, $z = 4$ in the Sierpiński triangle, and $z=6$ in the triangular lattice.
The rounded shape of the lobe in 2D and the sharper, more pointed shape in 1D are consistent with previous results \cite{Kashurnikov1996, Khner1998, Ohgoe2012}. 
The lobe in the Sierpiński lattice stands out by exhibiting features lying between the 1D and 2D geometries, 
reflecting the lattice’s non-integer fractal dimensionality.

The MI/SF phase boundary can be approximated using a strong-coupling expansion, as described in Ref.~\cite{PhysRevB.53.2691}. By solving Eqs.~(9) and (10) following the method described in detail in that reference, one can obtain an approximate boundary for lattices of arbitrary geometry. In Figures~\ref{fig:homo_MI_SF_hyp} and \ref{fig:homo_MI_SF}, the solid line represents the phase boundary calculated using this method for larger lattices of the corresponding geometry.
In Figure~\ref{fig:homo_MI_SF}, we see that the strong-coupling method reproduces the results for 1D and 2D lattices in a manner consistent with Ref.~\cite{PhysRevB.53.2691}, where it is more accurate for the 1D geometry. On the other hand, the cluster mean-field method becomes more accurate in higher dimensions, which leads to near-perfect agreement between the two methods for the Sierpiński triangle lattice shown in Figure~\ref{fig:homo_MI_SF}(b).
In Figure~\ref{fig:homo_MI_SF_hyp}, the strong-coupling approach yields slightly different results for the two hyperbolic geometries, and in both cases deviates from the cluster mean-field phase diagram near the tip of the Mott lobe.

\subsection{Field-theoretic description of the Mott transition}

A field-theoretic description of the Mott transition of bosons can be obtained from expanding the effective action of the Bose-Hubbard model in lowest (that is quadratic) order in the field operators. The kernel of this expansion is the inverse of the two-point Green functions. It can be obtained from the readily available local Green function through a hopping expansion. This formalism has successfully been applied to the Bose-Hubbard model in Refs.~\cite{dosSantos2009,Bradlyn2009,Grass2011}, using a time-independent description suited for studying the zero-temperature static scenario \cite{dosSantos2009}, an imaginary-time description best suited for studying equilibrium thermodynamics of the model \cite{Bradlyn2009}, or a real-time Schwinger-Keldysh description suited to capture also out-of-equilibrium behavior \cite{Grass2011}. We briefly sketch the main general steps for constructing this formalism in the appendix. Here, we quote only the main results. In particular, by derivative with respect to one of the fields, the second-order effective action yields a linear equation of motion for the field $\Psi_{j}$:
\begin{align}
\sum_j [G_{ij}(\omega)]^{-1} \Psi_{j}(\omega) = 0.
\end{align}
This expression actually holds to any order in the hopping expansion. 

To calculate the kernel of this equation, the inverse of the (retarded) Green function, $G_{ij}^{-1}(\omega)$, we take the hopping as a perturbation and carry out a first-order hopping expansion.  For brevity, let us concentrate on the zero-temperature case, where a product state of sites occupied by $n_0$ bosons, determined by the chemical potential, is the ground state of the Bose-Hubbard model in the local limit. The corresponding zero-temperature (retarded) Green function reads: $G_{ij}^{(0)}(\omega)= \delta_{ij} g(\omega)$, with 
\begin{align}
g(\omega)= & \lim_{\epsilon\rightarrow 0} \Big( \frac{n_0+1}{E_{n_0+1}-E_{n_0}-\omega-i\epsilon} - \nonumber \\ & \frac{n_0}{E_{n_0}-E_{n_0-1}-\omega-i\epsilon} \Big),
\end{align}
where $E_n=\frac{U}{2}n(n-1)-\mu n$ the energy of a site occupied by $n$ bosons. From the local Green function, we then construct the inverse Green function to first hopping order (see appendix). We obtain:
\begin{align}
[G_{ij}^{(1)}(\omega)]^{-1} = \frac{\delta_{ij}}{g(\omega)} - J_{ij}.
\end{align}
Here, $J_{ij}$ are the hopping parameters between sites $i$ and $j$, that is, the kinetic part of the Bose-Hubbard model is given by $H_{\rm kin}=-\sum_{ij}J_{ij} b_i^\dagger b_j$.

In equilibrium, we have $\Psi_j(\omega) \sim \delta(\omega)$. The equations of motion simplifies to
\begin{align}
   0=  \sum_j \left( \frac{\delta_{ij}}{g(0)} -J_{ij} \right) \Psi_j(0)  \equiv \sum_j M_{ij} \Psi_j.
\end{align}
If the matrix $M_{ij}$ is non-singular, these equations are only trivially solved by $\Psi_j=0$, corresponding to the Mott phase. However, a singular matrix $M$ admits non-trivial solutions, $\Psi_j\neq0$, indicating the transition into the superfluid phase. Note that the matrix $M_{ij}$ is diagonalized by diagonalizing the tight-binding matrix $J_{ij}$. 

For further analysis, we assume that $J_{ij}$ takes values zero or $J>0$, and the eigenvalues of $J_{ij}/J$ shall be denoted by $\lambda_k$. In this notation, the tight-binding energy spectrum is given by $\epsilon_k= -J\lambda_k$. With this, $M_{kk}=\frac{1}{g(0)} - J \lambda_k = \frac{1}{g(0)} + \epsilon_k$. Let us now focus on the parameter regime $0< \mu/U < 1$, where $n_0=1$, and $g(0)= \frac{\mu+U}{\mu(U-\mu)}>0$, and first consider a regular lattice where $J_{ij}/J$ can be diagonalized via Fourier transform, yielding an energy band with a lower band edge at $\epsilon_{\rm min}=-2dJ<0$, in the case of a $d$-dimensional hypercubic lattice. Thus, the matrix $M$ becomes singular when $\frac{1}{g(0)} - 2dJ = 0$. This means the system enters the superfluid regime for $J>\frac{\mu(U-\mu)}{4(\mu+U)}$. This expression reproduces the mean-field value of the first Mott lobe.

Now let us turn to a fractal system: The condition of a singular matrix $M$ is still 
\begin{align}
    \frac{1}{g(0)} + J \lambda_k = 0,
    \label{lobe}
\end{align} where $-J\lambda_k$ denotes an eigenstate of the tight-binding matrix. Upon increasing the hopping strength $J>0$, this condition is first met for the ground state (i.e. $k=0$) at a critical hopping strength $J_{\rm crit}= -\frac{1}{g(0)\lambda_0}$, as also the case for the regular lattice. The ground state energy of the Sierpiński gasket is identical to the one of a regular square lattice, $\epsilon_0=-4J$, and hence the two systems have identical Mott lobes. In the case of regular lattices, $\epsilon_0=-zJ$ is directly given by the coordination number $z$ of the lattice, and it immediately follows that in the case of a triangular lattice ($z=6$), the height of the lobe is suppressed. Specifically, the quantitative comparison of a Sierpiński gasket and a regular triangle shows an increased Mott phase for the gasket, and the heights at the tips of the lobes takes a ratio $3/2$.

\begin{figure}[t!]
\centering    \includegraphics[width=0.48\textwidth]{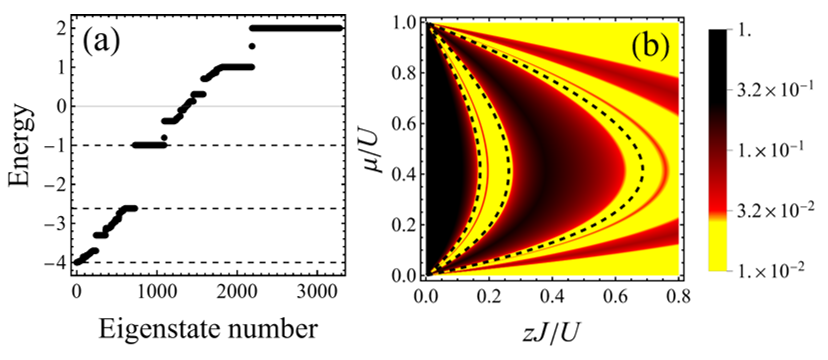}
    \caption{(a) The energy spectrum of the tight-binding model on a Sierpiński gasket ($V=3282$ sites), with the band edge energies used in panel (b) marked by the dashed lines.
    (b) The condition number of the matrix $M$ corresponding to the Sierpiński gasket with $V=3282$ sites. Yellow regions correspond to a singular matrix that admits superfluid solutions. The dashed lobe lines are produced from Eq.~(\ref{lobe}) using eigenenergies of the tight-binding
    matrix $J_{ij}$, corresponding to the ground state energy and edges of the large gap in the spectrum.
    }
    \label{fig:lobes}
\end{figure}

Apart from this quantitative change of the height of the lobe, there is also a more remarkable qualitative new feature in the fractal case: As the spectrum of the fractal lattice does not form a continuous band, but exhibits regions of vanishing density of states, the theory predicts  re-entrant Mott behavior. Specifically, the matrix $M$ will not remain singular for all  $J>J_{\rm crit}$,  but instead, for all $J$ such that pronounced band gaps $\epsilon \in J[\lambda_k,\lambda_{k+1}]$ coincide with $\epsilon=-1/g(0)$, $M$ becomes non-singular again, and the system can re-enter the Mott phase. 
This is illustrated by a plot of the condition number of $M=\frac{{\rm min}_k|m_k|}{{\rm max}_k|m_k|}$, where $m_k$ denote the eigenvalues of $M$, see Fig.~\ref{fig:lobes}(b). Major band gaps occur between eigenstates $k=3^n-1$ and $k=3^n$, for $n=6$ and $n=7$, as seen in the spectrum of the tight-binding matrix, shown in Fig.~\ref{fig:lobes}(a).

Such a re-entrance of the Mott phase is certainly a counterintuitive feature of the fractal. As such a behavior has not been observed with the Gutzwiller approach to the problem, we speculate that the re-entrance appears as an artifact of the approximation. Specifically, as shown above, the first-order hopping expansion leads to a very simple Mott condition, given entirely by the band edges of the tight-binding matrix. The re-entrant behavior appears as a consequence of this simple form, and hence, higher-order corrections, which are beyond the scope of this paper, might show different behavior.

\section{Conclusions}
Geometry and dimensionality can strongly influence the behavior of a quantum system. Many studies \cite{Darazs_2014,Kosior2017,vanVeen_2016,Rojo-Francas2024,Salvati2024,Kempkes_2019_b,Brzezinska2018,Pai2019,Iliasov2020,Fremling2020, Moustaj2021, Manna2022b,Ivaki2022,Li2023,Krebbekx2023,   Moustaj2023, Stlhammar2023,Eek2024,Osseweijer2024,Canyellas2024} have been devoted to the effect of fractal geometry on the single-particle level. In the present paper, we have explored the role of exotic geometries on quantum-statistical and quantum many-body behavior.
 Specifically, for fractal and hyperbolic lattices, we have scrutinized the phenomenon of Bose‑Einstein condensation of an ideal gas, as well as the zero‑temperature Mott‑insulator-to‑superfluid transition in the presence of strong on-site repulsion. For all fractal geometries considered, we found that the condensation temperature is dramatically lowered as compared to regular lattices of similar size, but the thermal depletion as a function of temperature has a different functional behavior, with a heavy tail above the critical temperature. Condensate fluctuations are found to be enlarged by the fractal geometries, with an anomalous scaling exponent $>2$. As in regular lattices in less than three dimensions, the critical temperature in fractal lattices decreases with system size and is expected to be zero in thermodynamically large systems. A strikingly different behavior has been seen in hyperbolic lattices: Despite the fact that these lattices can be embedded in two spatial dimensions, the critical temperature increases with system size. 
  In the presence of strong interactions, we observed changes in the shape of the Mott lobe, obtained from a cluster Gutzwiller ansatz due to fractal geometry. In accordance with the fractal dimension of the lattice, the shape appears to be intermediate to the well-known cases of 1D and 2D Mott lobes. 
  Our theoretical research is aligned with current experimental trends which have started to realize
  the topologies analyzed here in state‑of‑the‑art quantum simulators, including photonic wave‑guide arrays and tweezer‑assembled Rydberg‑atom arrays. We believe that our results will spur systematic experimental exploration of quantum matter in non‑Euclidean settings, where geometry itself becomes a tunable control parameter that can unveil genuinely new phases and critical behavior.
  
\appendix

\section{Cluster Gutzwiller method with basis truncation}
\label{app:truncation}

\begin{figure}[t!]
\centering    \includegraphics[width=0.49\textwidth]{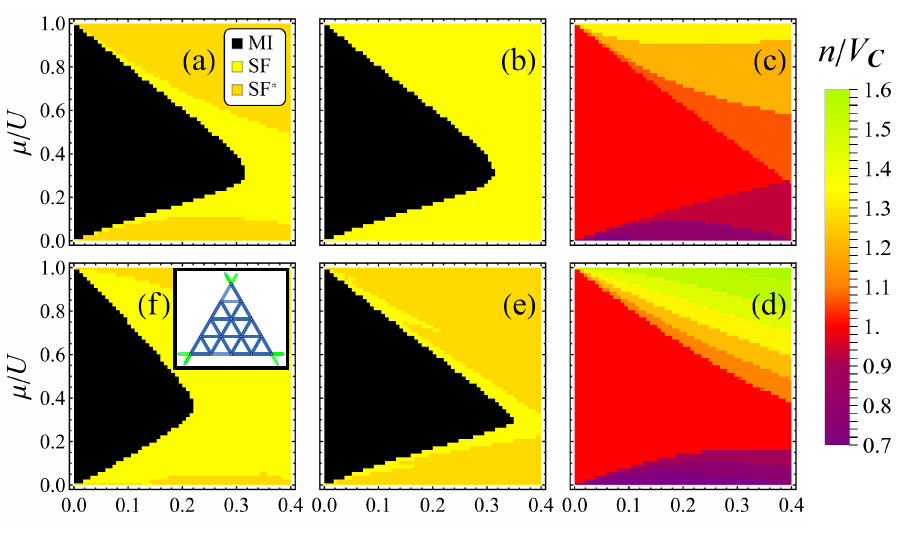}
    \caption{Panels (a,b,e,f) show obtained phase diagrams for various clusters, where the black regions correspond to the order parameter $\phi=0$, yellow regions correspond to regions where $|\phi|>0$ and orange regions correspond to regions where $\phi=0$, but  $|n/V_{\cal{C}} - 1|>1/V_{\cal{C}}$. Black regions correspond to the Mott insulator (MI) phase, while yellow and orange regions indicate the superfluid (SF) phase. Panels (c,d) show the expected number of bosons per site $n/V_{\cal{C}} \coloneqq \sum_i\langle GS|\hat{n}_i/V_{\cal C}|GS\rangle$. A 1D cluster with $V_{\cal{C}} = 9$ sites is used for panels (a-c), a 1D cluster with $V_{\cal{C}} = 15$ sites (same as in Figure \ref{fig:homo_MI_SF}(a)) is used for panels (d,e) and for panel (f), a triangular cluster connected to the bulk only through two couplings at each corner (as shown in the top right). In panel (a), the basis is constructed with $\Delta N_{\max} = 1$ and $f_{\max}=5$, while in panels (b,c) $\Delta N_{\max}=V_{\cal{C}}=9$ and $f_{\max}=V_{\cal{C}}=9$. Panels (d-f) use $\Delta N_{\max} = 5$ and $f_{\max}=5$, same as in Figures \ref{fig:homo_MI_SF_hyp} and \ref{fig:homo_MI_SF}.
    }
    \label{fig:app_Grid}
\end{figure}

In this section, we explain the details of the cluster Gutzwiller method adapted to the fractal geometry.

We start by defining a truncated grand canonical basis ${\cal H}$, which is a bosonic Hilbert space constructed as a direct sum of bosonic bases ${\cal H}_{N_i, V_{\cal C}}$ with fixed number of particles $N_i$ at given number of lattice sites $V_{\cal C}$, i.e. ${\cal H} \coloneqq \bigoplus_i {\cal H}_{N_i, V_{\cal C}}$.

We consider fluctuations of the number of bosons around the unit filling, $N_i = V_{\cal C}$. Consequently, the composite Hilbert space contains  $N_i = V_{\cal C}-\Delta N_{\max}, \dots , V_{\cal C}, \dots, V_{\cal C}+\Delta N_{\max}$, with $\Delta N_{\max} = 5$. Next, we truncate the basis, following the scheme proposed by L\"{u}hmann \cite{L_hmann_2013}, i.e.
we discard Fock states with (i) large local fluctuations, where 
the number of particles $n_i$ on any site $i$ differs significantly from $1$: $\exists_i~|n_i - 1| >3$ or (ii) large global fluctuations, where the sum over the sites $\sum_i |n_i- 1| > f_{\max}$, with $f_{\max} = 5$.
We note that this truncation is most accurate inside and near the MI phase, where the ground states have low fluctuations the of occupation operator. When the parameter $\Delta N_{\max}$ is too small, the order parameter $\phi$ may vanish even in the SF phase, when the ground state populates states with $N = V\pm \Delta N_{\max}$ particles.
We use $\Delta N_{\max} = 5$, which makes the size of the basis limited primarily by the maximum number of global fluctuations $f_{\max} = 5$.
This is sufficient for determining the Mott lobe boundaries in clusters with $V_{\cal C} = 15$ sites, although in the lower-dimensional geometries like the 1D chain and the Sierpiński triangle, we still observe regions in the SF phase where $\phi = 0$ (see Fig. \ref{fig:app_Grid} (e,f)).
We can identify such misclassified regions by checking where the number of particles per site $n/V_{\cal{C}} \coloneqq \sum_i\langle GS|\hat{n}_i/V_{\cal C}|GS\rangle$ in the ground state differs significantly from $1$: $|n/V_{\cal{C}} - 1|>1/V_{\cal{C}}$.

This is illustrated through Figures \ref{fig:app_Grid} (a-c), which show results for a 1D cluster with $V_{\cal C}=9$ sites. When $\Delta N_{\max} = 1$ and $f_{\max} = 5$ (Fig. \ref{fig:app_Grid} (a)), orange regions appear that are absent when $\Delta N_{\max} = 9$ and $f_{\max} = 9$ (Fig. \ref{fig:app_Grid} (b)), but the shape of the Mott lobe remains the same. 
Moreover, we can still distinguish between the MI and SF phases in regions where $\phi=0$ by examining the average number of particles per site $n/V_{\cal{C}}$ (Fig. \ref{fig:app_Grid} (c)). 
The same approach is illustrated for a larger 1D cluster with $V_{\cal C} = 15$ sites in Figure \ref{fig:app_Grid} (d,e) and also used in generating Figures \ref{fig:homo_MI_SF_hyp} and \ref{fig:homo_MI_SF}.

Lastly, it is worth noting that both the internal structure of a cluster and its couplings to the mean-field contribute to the shape of the Mott lobe. The phase diagram in Figure \ref{fig:app_Grid} (f) shows how a cluster with the same triangular internal structure as in Figure \ref{fig:homo_MI_SF}(c), but with couplings matching those of the Sierpiński triangle in Figure \ref{fig:homo_MI_SF}(b), more closely resembles the latter.
At the same time, the Mott lobes of 1D clusters with $V_{\cal{C}}=9$ and $V_{\cal{C}}=15$ sites differ significantly (see Fig. \ref{fig:app_Grid} (b,e)), despite being coupled to the mean-field in the same way.

\section{Effective action description of the Mott phase}
In this appendix, we derive in detail the field-theoretic description of the Mott transition. Therefore, let us start by adding a source term $H_s= \sum_i j_i(t) b_i^\dagger + \rm{h.c.}$ to the Bose-Hubbard Hamiltonian. Through this term, we obtain a partition-function-like expression that can be used as the generating functional of Green functions:
\begin{align}
{\cal Z}[j,j^*] = {\rm tr}\left( T_{\rm c} e^{-\frac{i}{\hbar} \int_{\rm c} dt H(t)} \right).
\end{align}
Here, we integrate along a time-contour c, which could just be along the imaginary time axis for an equilibrium theory, but the more general description also includes forward and backward integration along the real-time axis and is capable to describe also out-of-equilibrium behavior. In any case, $T_{\rm c}$ serves as an ordering operator along the chosen contour. 

From ${\cal Z}[j,j^*]$, a free-energy-like functional can be obtained, ${\cal F}[j,j^*]=-i{\cal Z}[j,j^*]$, and finally the effective action $\Gamma[\Psi,\Psi^*]$ via Legendre transformation from sources $j,j^*$ to fields $\Psi = \delta {\cal F}[j,j^*]/\delta j^*|_{j=j^*=0}$ and $\Psi^* = \delta {\cal F}[j,j^*]/\delta j|_{j=j^*=0}$. To this end, the sources have to expressed in terms of fields, and then the effective action is given by:  
\begin{align}
\Gamma[\Psi,\Psi^*] = &{\cal F}[j(\Psi,\Psi^*),j^*(\Psi,\Psi^*)]-
\nonumber \\ &
\sum_i \int dt \ j(\Psi,\Psi^*) \cdot \Psi + {\rm h.c.}
\end{align}
It is implied that the sources $j,j^*$ and the conjugate fields $\Psi,\Psi^*$ depend on their position on the time contour, as well as their position on lattice. 

From the definition of the fields it is seen that $\Psi_i,\Psi_i^*$ are precisely the expectation values of the bosonic operators, i.e. $\langle b_i \rangle$ and $\langle b_i^\dagger \rangle$. In the Mott phase, these expectation values vanish, and hence an expansion of the functional $\Gamma[\Psi,\Psi^*]$ to second-order in the field is sufficient. Its expression takes the following form:
\begin{align}
\Gamma[\Psi,\Psi^*] -\hbar^2 \sum_{i,j} \int dt_1 \int dt_2 \Psi_i(t_1)^* G_{ij}^{-1}(t_1,t_2) \Psi_j(t_2)
\end{align}
Here, we shall briefly comment on some details in the real-time/Keldysh formalism. As mentioned earlier, the temporal integral shall enclose a contour with forward and backward paths along the real-time axis. However, it is convenient to write the time-contour integral as a single forward integral (from $-\infty$ to $\infty$), and equip each temporal degree of freedom with an additonal path index in order to differentiate between forward and backward path. With this, the fields are doubled into two-component vector fields, and the Green function $G_{ij}(t_1,t_2)$ becomes a $2\times2$ matrix. At this point, it is very convenient to make a rotation within this path-index space, known as Keldysh rotation. This rotaion makes one element in the Green function matrix vanish, and the off-diagonal elements of this matrix are the retarded and advanced Green functions. Within this Keldysh space, the two-component fields have a  "classical" field component, $\Psi_{i,{\rm cl}}(t),\Psi_{i,{\rm cl}}(t)^*$, which is the sum of the fields on the forward and backward path, and a "quantum" field component, $\Psi_{i,{\rm q}}(t),\Psi_{i,{\rm q}}(t)^*$, which is the difference of the fields on the forward and backward path. 

Without any explicitly time-dependent potential in the Hamiltonian, the effective action is most conveniently Fourier transformed in frequency space:
\begin{align}
\Gamma[\Psi,\Psi^*] -\hbar^2 \sum_{i,j} \int d\omega \int dt_2 \Psi_i(\omega)^* [G_{ij}(\omega)]^{-1} \Psi_j(\omega).
\end{align}
The equation of motion, $\delta \Gamma / \delta \Psi_i(\omega)^*=0$  takes the form
\begin{align}
\sum_j [G_{ij}(\omega)]^{-1} \Psi_j(\omega) = 0.
\end{align}
If we now argue that the quantum component of the field should vanish, we obtain the equation of motion for the classical field which only depends on the inverse of the retarded Green function:
$\sum_j [G_{ij}^{\rm (ret)}(\omega)]^{-1} \Psi_{j,{\rm cl}}(\omega) = 0$, as also given in the main text (where, for brevity, we avoided the Keldysh notation).

In order to find the inverse Green function, $[G_{ij}^{\rm (ret)}(\omega)]^{-1}$, we use a hopping expansion in which the unperturbed Hamiltonian are all the local terms of the Bose-Hubbard model. The Green functions for the local Hamiltonian are easily obtained. For instance, the retarded Green function in this local limit, $G_{ij}^{(0,{\rm ret})}(\omega)\equiv G_i^{(0,{\rm ret})}(\omega)\delta_{ij}$ reads
\begin{align}
G_i^{(0,{\rm ret})}(\omega)= & \lim_{\epsilon\rightarrow 0} 
\sum_{n=0}^{\infty} \frac{e^{-\beta E_n}}{{\cal Z}^{(0)}}  \Big(\frac{n+1}{E_{n+1}-E_n-\omega-i\epsilon} 
\nonumber \\ & 
- \frac{n}{E_n-E_{n-1}-\omega-i\epsilon} \Big).
\end{align}
Here, $n$ is the occupation number of site $i$, and $E_n=Un(n-1)-\mu n$ the energy of the state in the atomic limit, and ${\cal Z}^{(0)} = \sum_n e^{-\beta E_n}$ the thermodynamic partition function at inverse temperature $\beta$. At zero temperature, the expression reduces to
\begin{align}
G_i^{(0,{\rm ret})}(\omega)= & \lim_{\epsilon\rightarrow 0} \Big( \frac{n_0+1}{E_{n_0+1}-E_{n_0}-\omega-i\epsilon} - 
\nonumber \\ &
\frac{n_0}{E_{n_0}-E_{n_0-1}-\omega-i\epsilon} \Big),
\end{align}
where $n_0$ is the occupation number in the ground state.

In the first order of the hopping expansion, the retarded Green function reads:
\begin{align}
G_{ij}^{(1,{\rm ret})}(\omega) =\delta_{ij} G_i^{(0,{\rm ret})}(\omega) + J_{ij} G_i^{(0,{\rm ret})}(\omega) G_j^{(0,{\rm ret})}(\omega).
\end{align}
Here, $J_{ij}$ is the hopping amplitude between sites $i$ and $j$. For the effective action description, it is necessary to invert this Green function, while keeping the expression to linear order in $J_{ij}$. One obtains
\begin{align}
[G_{ij}^{(1,{\rm ret})}(\omega)]^{-1} = [G_{i}^{(0,{\rm ret})}(\omega)]^{-1} \left(\delta_{ij} - J_{ij} G_{i}^{(0,{\rm ret})}(\omega) \right).
\end{align}
The power of the hopping expansion of the effective action is also due to the fact that this inversion automatically resums all first-order hopping terms in the free energy.

\acknowledgments T.G. acknowledges the financial support received from the IKUR Strategy under the collaboration agreement between the Ikerbasque Foundation and DIPC on behalf of the Department of Education of the Basque Government, as well as funding by the Department of Education of the Basque Government through the project PIBA\_2023\_1\_0021 (TENINT), by the Agencia Estatal de Investigación (AEI) through Proyectos de Generación de Conocimiento PID2022-142308NA-I00 (EXQUSMI), and that this work has been produced with the support of a 2023 Leonardo Grant for Researchers in Physics, BBVA Foundation. The BBVA Foundation is not responsible for the opinions, comments, and contents included in the project and/or the results derived therefrom, which are the total and absolute responsibility of the authors.
ICFO-QOT group acknowledges support from:
European Research Council AdG NOQIA;
MCIN/AEI (PGC2018-0910.13039/501100011033, CEX2019-000910-S/10.13039/501100011033, Plan National FIDEUA PID2019-106901GB-I00, Plan National STAMEENA PID2022-139099NB, I00, project funded by MCIN/AEI/10.13039/501100011033 and by the “European Union NextGenerationEU/PRTR" (PRTR-C17.I1), FPI); QUANTERA DYNAMITE PCI2022-132919, QuantERA II Programme co-funded by European Union’s Horizon 2020 program under Grant Agreement No 101017733; Ministry for Digital Transformation and of Civil Service of the Spanish Government through the QUANTUM ENIA project call - Quantum Spain project, and by the European Union through the Recovery, Transformation and Resilience Plan - NextGenerationEU within the framework of the Digital Spain 2026 Agenda; Fundació Cellex; Fundació Mir-Puig; Generalitat de Catalunya (European Social Fund FEDER and CERCA program; Barcelona Supercomputing Center MareNostrum (FI-2023-3-0024); Funded by the European Union. Views and opinions expressed are however those of the author(s) only and do not necessarily reflect those of the European Union, European Commission, European Climate, Infrastructure and Environment Executive Agency (CINEA), or any other granting authority. Neither the European Union nor any granting authority can be held responsible for them (HORIZON-CL4-2022-QUANTUM-02-SGA PASQuanS2.1, 101113690, EU Horizon 2020 FET-OPEN OPTOlogic, Grant No 899794, QU-ATTO, 101168628), EU Horizon Europe Program (This project has received funding from the European Union’s Horizon Europe research and innovation program under grant agreement No 101080086 NeQSTGrant Agreement 101080086 — NeQST); ICFO Internal “QuantumGaudi” project.
B.J.-D. and A.R.-F. funding from Grant PID2023-147475NB-I00 funded by MICIU/AEI/10.13039/501100011033 and FEDER, UE, by grants 
2021SGR01095 from Generalitat de Catalunya, and by 
Project CEX2019-000918-M of ICCUB (Unidad de Excelencia María 
de Maeztu).
K.D. acknowledges the support of the "la Caixa" Foundation.
The project that gave rise to these results
received the support of a fellowship from
the ”la Caixa” Foundation (ID
100010434). The fellowship code is
LCF/BQ/DFI25/13000067.
The data that support the findings of this article are openly available \cite{dutkiewicz_2025_17831905}.


\begin{thebibliography}{144}%
\makeatletter
\providecommand \@ifxundefined [1]{%
 \@ifx{#1\undefined}
}%
\providecommand \@ifnum [1]{%
 \ifnum #1\expandafter \@firstoftwo
 \else \expandafter \@secondoftwo
 \fi
}%
\providecommand \@ifx [1]{%
 \ifx #1\expandafter \@firstoftwo
 \else \expandafter \@secondoftwo
 \fi
}%
\providecommand \natexlab [1]{#1}%
\providecommand \enquote  [1]{``#1''}%
\providecommand \bibnamefont  [1]{#1}%
\providecommand \bibfnamefont [1]{#1}%
\providecommand \citenamefont [1]{#1}%
\providecommand \href@noop [0]{\@secondoftwo}%
\providecommand \href [0]{\begingroup \@sanitize@url \@href}%
\providecommand \@href[1]{\@@startlink{#1}\@@href}%
\providecommand \@@href[1]{\endgroup#1\@@endlink}%
\providecommand \@sanitize@url [0]{\catcode `\\12\catcode `\$12\catcode `\&12\catcode `\#12\catcode `\^12\catcode `\_12\catcode `\%12\relax}%
\providecommand \@@startlink[1]{}%
\providecommand \@@endlink[0]{}%
\providecommand \url  [0]{\begingroup\@sanitize@url \@url }%
\providecommand \@url [1]{\endgroup\@href {#1}{\urlprefix }}%
\providecommand \urlprefix  [0]{URL }%
\providecommand \Eprint [0]{\href }%
\providecommand \doibase [0]{https://doi.org/}%
\providecommand \selectlanguage [0]{\@gobble}%
\providecommand \bibinfo  [0]{\@secondoftwo}%
\providecommand \bibfield  [0]{\@secondoftwo}%
\providecommand \translation [1]{[#1]}%
\providecommand \BibitemOpen [0]{}%
\providecommand \bibitemStop [0]{}%
\providecommand \bibitemNoStop [0]{.\EOS\space}%
\providecommand \EOS [0]{\spacefactor3000\relax}%
\providecommand \BibitemShut  [1]{\csname bibitem#1\endcsname}%
\let\auto@bib@innerbib\@empty
%</preamble>
\bibitem [{\citenamefont {Davis}\ \emph {et~al.}(1995)\citenamefont {Davis}, \citenamefont {Mewes}, \citenamefont {Andrews}, \citenamefont {van Druten}, \citenamefont {Durfee}, \citenamefont {Kurn},\ and\ \citenamefont {Ketterle}}]{Davis1995}%
  \BibitemOpen
  \bibfield  {author} {\bibinfo {author} {\bibfnamefont {K.~B.}\ \bibnamefont {Davis}}, \bibinfo {author} {\bibfnamefont {M.~O.}\ \bibnamefont {Mewes}}, \bibinfo {author} {\bibfnamefont {M.~R.}\ \bibnamefont {Andrews}}, \bibinfo {author} {\bibfnamefont {N.~J.}\ \bibnamefont {van Druten}}, \bibinfo {author} {\bibfnamefont {D.~S.}\ \bibnamefont {Durfee}}, \bibinfo {author} {\bibfnamefont {D.~M.}\ \bibnamefont {Kurn}},\ and\ \bibinfo {author} {\bibfnamefont {W.}~\bibnamefont {Ketterle}},\ }\bibfield  {title} {\bibinfo {title} {{Bose}-{Einstein} condensation in a gas of sodium atoms},\ }\href {https://doi.org/10.1103/PhysRevLett.75.3969} {\bibfield  {journal} {\bibinfo  {journal} {Phys. Rev. Lett.}\ }\textbf {\bibinfo {volume} {75}},\ \bibinfo {pages} {3969} (\bibinfo {year} {1995})}\BibitemShut {NoStop}%
\bibitem [{\citenamefont {Anderson}\ \emph {et~al.}(1995)\citenamefont {Anderson}, \citenamefont {Ensher}, \citenamefont {Matthews}, \citenamefont {Wieman},\ and\ \citenamefont {Cornell}}]{Anderson1995}%
  \BibitemOpen
  \bibfield  {author} {\bibinfo {author} {\bibfnamefont {M.~H.}\ \bibnamefont {Anderson}}, \bibinfo {author} {\bibfnamefont {J.~R.}\ \bibnamefont {Ensher}}, \bibinfo {author} {\bibfnamefont {M.~R.}\ \bibnamefont {Matthews}}, \bibinfo {author} {\bibfnamefont {C.~E.}\ \bibnamefont {Wieman}},\ and\ \bibinfo {author} {\bibfnamefont {E.~A.}\ \bibnamefont {Cornell}},\ }\bibfield  {title} {\bibinfo {title} {Observation of {Bose}-{Einstein} condensation in a dilute atomic vapor},\ }\href {https://doi.org/10.1126/science.269.5221.198} {\bibfield  {journal} {\bibinfo  {journal} {Science}\ }\textbf {\bibinfo {volume} {269}},\ \bibinfo {pages} {198} (\bibinfo {year} {1995})}\BibitemShut {NoStop}%
\bibitem [{\citenamefont {{{Bose}}}(1924)}]{Bose1924}%
  \BibitemOpen
  \bibfield  {author} {\bibinfo {author} {\bibnamefont {{{Bose}}}},\ }\bibfield  {title} {\bibinfo {title} {Plancks gesetz und lichtquantenhypothese},\ }\href {https://doi.org/10.1007/BF01327326} {\bibfield  {journal} {\bibinfo  {journal} {Zeitschrift f{\"u}r Physik}\ }\textbf {\bibinfo {volume} {26}},\ \bibinfo {pages} {178} (\bibinfo {year} {1924})}\BibitemShut {NoStop}%
\bibitem [{\citenamefont {{Einstein}}(2005)}]{Einstein1925}%
  \BibitemOpen
  \bibfield  {author} {\bibinfo {author} {\bibfnamefont {A.}~\bibnamefont {{Einstein}}},\ }\bibinfo {title} {Quantentheorie des einatomigen idealen gases},\ in\ \href {https://doi.org/https://doi.org/10.1002/3527608958.ch27} {\emph {\bibinfo {booktitle} {Albert {Einstein}: Akademie‐Vorträge}}}\ (\bibinfo  {publisher} {John Wiley I\& Sons, Ltd},\ \bibinfo {year} {2005})\ pp.\ \bibinfo {pages} {237--244}\BibitemShut {NoStop}%
\bibitem [{\citenamefont {Nikuni}\ \emph {et~al.}(2000)\citenamefont {Nikuni}, \citenamefont {Oshikawa}, \citenamefont {Oosawa},\ and\ \citenamefont {Tanaka}}]{Nikuni2000}%
  \BibitemOpen
  \bibfield  {author} {\bibinfo {author} {\bibfnamefont {T.}~\bibnamefont {Nikuni}}, \bibinfo {author} {\bibfnamefont {M.}~\bibnamefont {Oshikawa}}, \bibinfo {author} {\bibfnamefont {A.}~\bibnamefont {Oosawa}},\ and\ \bibinfo {author} {\bibfnamefont {H.}~\bibnamefont {Tanaka}},\ }\bibfield  {title} {\bibinfo {title} {{Bose}-{Einstein} condensation of dilute magnons in ${\mathrm{tlcucl}}_{3}$},\ }\href {https://doi.org/10.1103/PhysRevLett.84.5868} {\bibfield  {journal} {\bibinfo  {journal} {Phys. Rev. Lett.}\ }\textbf {\bibinfo {volume} {84}},\ \bibinfo {pages} {5868} (\bibinfo {year} {2000})}\BibitemShut {NoStop}%
\bibitem [{\citenamefont {Eisenstein}\ and\ \citenamefont {MacDonald}(2004)}]{Eisenstein2004}%
  \BibitemOpen
  \bibfield  {author} {\bibinfo {author} {\bibfnamefont {J.~P.}\ \bibnamefont {Eisenstein}}\ and\ \bibinfo {author} {\bibfnamefont {A.~H.}\ \bibnamefont {MacDonald}},\ }\bibfield  {title} {\bibinfo {title} {{Bose}--{Einstein} condensation of excitons in bilayer electron systems},\ }\href {https://doi.org/10.1038/nature03081} {\bibfield  {journal} {\bibinfo  {journal} {Nature}\ }\textbf {\bibinfo {volume} {432}},\ \bibinfo {pages} {691} (\bibinfo {year} {2004})}\BibitemShut {NoStop}%
\bibitem [{\citenamefont {Deng}\ \emph {et~al.}(2002)\citenamefont {Deng}, \citenamefont {Weihs}, \citenamefont {Santori}, \citenamefont {Bloch},\ and\ \citenamefont {Yamamoto}}]{Deng2002}%
  \BibitemOpen
  \bibfield  {author} {\bibinfo {author} {\bibfnamefont {H.}~\bibnamefont {Deng}}, \bibinfo {author} {\bibfnamefont {G.}~\bibnamefont {Weihs}}, \bibinfo {author} {\bibfnamefont {C.}~\bibnamefont {Santori}}, \bibinfo {author} {\bibfnamefont {J.}~\bibnamefont {Bloch}},\ and\ \bibinfo {author} {\bibfnamefont {Y.}~\bibnamefont {Yamamoto}},\ }\bibfield  {title} {\bibinfo {title} {Condensation of semiconductor microcavity exciton polaritons},\ }\href {https://doi.org/10.1126/science.1074464} {\bibfield  {journal} {\bibinfo  {journal} {Science}\ }\textbf {\bibinfo {volume} {298}},\ \bibinfo {pages} {199} (\bibinfo {year} {2002})}\BibitemShut {NoStop}%
\bibitem [{\citenamefont {Kasprzak}\ \emph {et~al.}(2006)\citenamefont {Kasprzak}, \citenamefont {Richard}, \citenamefont {Kundermann}, \citenamefont {Baas}, \citenamefont {Jeambrun}, \citenamefont {Keeling}, \citenamefont {Marchetti}, \citenamefont {Szyma{\'{n}}ska}, \citenamefont {Andr{\'e}}, \citenamefont {Staehli}, \citenamefont {Savona}, \citenamefont {Littlewood}, \citenamefont {Deveaud},\ and\ \citenamefont {Dang}}]{Kasprzak2006}%
  \BibitemOpen
  \bibfield  {author} {\bibinfo {author} {\bibfnamefont {J.}~\bibnamefont {Kasprzak}}, \bibinfo {author} {\bibfnamefont {M.}~\bibnamefont {Richard}}, \bibinfo {author} {\bibfnamefont {S.}~\bibnamefont {Kundermann}}, \bibinfo {author} {\bibfnamefont {A.}~\bibnamefont {Baas}}, \bibinfo {author} {\bibfnamefont {P.}~\bibnamefont {Jeambrun}}, \bibinfo {author} {\bibfnamefont {J.~M.~J.}\ \bibnamefont {Keeling}}, \bibinfo {author} {\bibfnamefont {F.~M.}\ \bibnamefont {Marchetti}}, \bibinfo {author} {\bibfnamefont {M.~H.}\ \bibnamefont {Szyma{\'{n}}ska}}, \bibinfo {author} {\bibfnamefont {R.}~\bibnamefont {Andr{\'e}}}, \bibinfo {author} {\bibfnamefont {J.~L.}\ \bibnamefont {Staehli}}, \bibinfo {author} {\bibfnamefont {V.}~\bibnamefont {Savona}}, \bibinfo {author} {\bibfnamefont {P.~B.}\ \bibnamefont {Littlewood}}, \bibinfo {author} {\bibfnamefont {B.}~\bibnamefont {Deveaud}},\ and\ \bibinfo {author} {\bibfnamefont {L.~S.}\ \bibnamefont {Dang}},\ }\bibfield  {title} {\bibinfo {title} {{Bose}--{Einstein} condensation
  of exciton polaritons},\ }\href {https://doi.org/10.1038/nature05131} {\bibfield  {journal} {\bibinfo  {journal} {Nature}\ }\textbf {\bibinfo {volume} {443}},\ \bibinfo {pages} {409} (\bibinfo {year} {2006})}\BibitemShut {NoStop}%
\bibitem [{\citenamefont {Klaers}\ \emph {et~al.}(2010)\citenamefont {Klaers}, \citenamefont {Schmitt}, \citenamefont {Vewinger},\ and\ \citenamefont {Weitz}}]{Klaers2010}%
  \BibitemOpen
  \bibfield  {author} {\bibinfo {author} {\bibfnamefont {J.}~\bibnamefont {Klaers}}, \bibinfo {author} {\bibfnamefont {J.}~\bibnamefont {Schmitt}}, \bibinfo {author} {\bibfnamefont {F.}~\bibnamefont {Vewinger}},\ and\ \bibinfo {author} {\bibfnamefont {M.}~\bibnamefont {Weitz}},\ }\bibfield  {title} {\bibinfo {title} {{Bose}--{Einstein} condensation of photons in an optical microcavity},\ }\href {https://doi.org/10.1038/nature09567} {\bibfield  {journal} {\bibinfo  {journal} {Nature}\ }\textbf {\bibinfo {volume} {468}},\ \bibinfo {pages} {545} (\bibinfo {year} {2010})}\BibitemShut {NoStop}%
\bibitem [{\citenamefont {Dalfovo}\ \emph {et~al.}(1999)\citenamefont {Dalfovo}, \citenamefont {Giorgini}, \citenamefont {Pitaevskii},\ and\ \citenamefont {Stringari}}]{Dalfovo1999}%
  \BibitemOpen
  \bibfield  {author} {\bibinfo {author} {\bibfnamefont {F.}~\bibnamefont {Dalfovo}}, \bibinfo {author} {\bibfnamefont {S.}~\bibnamefont {Giorgini}}, \bibinfo {author} {\bibfnamefont {L.~P.}\ \bibnamefont {Pitaevskii}},\ and\ \bibinfo {author} {\bibfnamefont {S.}~\bibnamefont {Stringari}},\ }\bibfield  {title} {\bibinfo {title} {Theory of {Bose}-{Einstein} condensation in trapped gases},\ }\href {https://doi.org/10.1103/RevModPhys.71.463} {\bibfield  {journal} {\bibinfo  {journal} {Rev. Mod. Phys.}\ }\textbf {\bibinfo {volume} {71}},\ \bibinfo {pages} {463} (\bibinfo {year} {1999})}\BibitemShut {NoStop}%
\bibitem [{\citenamefont {Matthews}\ \emph {et~al.}(1999)\citenamefont {Matthews}, \citenamefont {Anderson}, \citenamefont {Haljan}, \citenamefont {Hall}, \citenamefont {Wieman},\ and\ \citenamefont {Cornell}}]{Matthews1999}%
  \BibitemOpen
  \bibfield  {author} {\bibinfo {author} {\bibfnamefont {M.~R.}\ \bibnamefont {Matthews}}, \bibinfo {author} {\bibfnamefont {B.~P.}\ \bibnamefont {Anderson}}, \bibinfo {author} {\bibfnamefont {P.~C.}\ \bibnamefont {Haljan}}, \bibinfo {author} {\bibfnamefont {D.~S.}\ \bibnamefont {Hall}}, \bibinfo {author} {\bibfnamefont {C.~E.}\ \bibnamefont {Wieman}},\ and\ \bibinfo {author} {\bibfnamefont {E.~A.}\ \bibnamefont {Cornell}},\ }\bibfield  {title} {\bibinfo {title} {Vortices in a {Bose}-{Einstein} condensate},\ }\href {https://doi.org/10.1103/PhysRevLett.83.2498} {\bibfield  {journal} {\bibinfo  {journal} {Phys. Rev. Lett.}\ }\textbf {\bibinfo {volume} {83}},\ \bibinfo {pages} {2498} (\bibinfo {year} {1999})}\BibitemShut {NoStop}%
\bibitem [{\citenamefont {Madison}\ \emph {et~al.}(2000)\citenamefont {Madison}, \citenamefont {Chevy}, \citenamefont {Wohlleben},\ and\ \citenamefont {Dalibard}}]{Madison2000}%
  \BibitemOpen
  \bibfield  {author} {\bibinfo {author} {\bibfnamefont {K.~W.}\ \bibnamefont {Madison}}, \bibinfo {author} {\bibfnamefont {F.}~\bibnamefont {Chevy}}, \bibinfo {author} {\bibfnamefont {W.}~\bibnamefont {Wohlleben}},\ and\ \bibinfo {author} {\bibfnamefont {J.}~\bibnamefont {Dalibard}},\ }\bibfield  {title} {\bibinfo {title} {Vortex formation in a stirred {Bose}-{Einstein} condensate},\ }\href {https://doi.org/10.1103/PhysRevLett.84.806} {\bibfield  {journal} {\bibinfo  {journal} {Phys. Rev. Lett.}\ }\textbf {\bibinfo {volume} {84}},\ \bibinfo {pages} {806} (\bibinfo {year} {2000})}\BibitemShut {NoStop}%
\bibitem [{\citenamefont {Gajda}\ and\ \citenamefont {Rzążewski}(1997)}]{Gajda1997}%
  \BibitemOpen
  \bibfield  {author} {\bibinfo {author} {\bibfnamefont {M.}~\bibnamefont {Gajda}}\ and\ \bibinfo {author} {\bibfnamefont {K.}~\bibnamefont {Rzążewski}},\ }\bibfield  {title} {\bibinfo {title} {Fluctuations of {Bose}--{Einstein} condensate},\ }\href {https://doi.org/10.1103/PhysRevLett.78.2686} {\bibfield  {journal} {\bibinfo  {journal} {Physical Review Letters}\ }\textbf {\bibinfo {volume} {78}},\ \bibinfo {pages} {2686} (\bibinfo {year} {1997})}\BibitemShut {NoStop}%
\bibitem [{\citenamefont {Giorgini}\ \emph {et~al.}(1998)\citenamefont {Giorgini}, \citenamefont {Pitaevskii},\ and\ \citenamefont {Stringari}}]{Giorgini1998}%
  \BibitemOpen
  \bibfield  {author} {\bibinfo {author} {\bibfnamefont {S.}~\bibnamefont {Giorgini}}, \bibinfo {author} {\bibfnamefont {L.~P.}\ \bibnamefont {Pitaevskii}},\ and\ \bibinfo {author} {\bibfnamefont {S.}~\bibnamefont {Stringari}},\ }\bibfield  {title} {\bibinfo {title} {Anomalous fluctuations of the condensate in interacting {{Bose}} gases},\ }\href {https://doi.org/10.1103/PhysRevLett.80.5040} {\bibfield  {journal} {\bibinfo  {journal} {Physical Review Letters}\ }\textbf {\bibinfo {volume} {80}},\ \bibinfo {pages} {5040} (\bibinfo {year} {1998})}\BibitemShut {NoStop}%
\bibitem [{\citenamefont {Idziaszek}\ \emph {et~al.}(1999)\citenamefont {Idziaszek}, \citenamefont {Gajda}, \citenamefont {Navez}, \citenamefont {Wilkens},\ and\ \citenamefont {Rza{\.{z}}ewski}}]{Idziaszek1999}%
  \BibitemOpen
  \bibfield  {author} {\bibinfo {author} {\bibfnamefont {Z.}~\bibnamefont {Idziaszek}}, \bibinfo {author} {\bibfnamefont {M.}~\bibnamefont {Gajda}}, \bibinfo {author} {\bibfnamefont {P.}~\bibnamefont {Navez}}, \bibinfo {author} {\bibfnamefont {M.}~\bibnamefont {Wilkens}},\ and\ \bibinfo {author} {\bibfnamefont {K.}~\bibnamefont {Rza{\.{z}}ewski}},\ }\bibfield  {title} {\bibinfo {title} {Fluctuations of the weakly interacting {{Bose}--{Einstein}} condensate},\ }\href {https://doi.org/10.1103/PhysRevLett.82.4376} {\bibfield  {journal} {\bibinfo  {journal} {Physical Review Letters}\ }\textbf {\bibinfo {volume} {82}},\ \bibinfo {pages} {4376} (\bibinfo {year} {1999})}\BibitemShut {NoStop}%
\bibitem [{\citenamefont {Meier}\ and\ \citenamefont {Zwerger}(1999)}]{Meier1999}%
  \BibitemOpen
  \bibfield  {author} {\bibinfo {author} {\bibfnamefont {F.}~\bibnamefont {Meier}}\ and\ \bibinfo {author} {\bibfnamefont {W.}~\bibnamefont {Zwerger}},\ }\bibfield  {title} {\bibinfo {title} {Anomalous condensate fluctuations in strongly interacting superfluids},\ }\href {https://doi.org/10.1103/PhysRevA.60.5133} {\bibfield  {journal} {\bibinfo  {journal} {Physical Review A}\ }\textbf {\bibinfo {volume} {60}},\ \bibinfo {pages} {5133} (\bibinfo {year} {1999})}\BibitemShut {NoStop}%
\bibitem [{\citenamefont {Kocharovsky}\ \emph {et~al.}(2006)\citenamefont {Kocharovsky}, \citenamefont {Kocharovsky}, \citenamefont {Holthaus}, \citenamefont {Ooi}, \citenamefont {Svidzinsky}, \citenamefont {Ketterle},\ and\ \citenamefont {Scully}}]{Kocharovsky2006}%
  \BibitemOpen
  \bibfield  {author} {\bibinfo {author} {\bibfnamefont {V.~V.}\ \bibnamefont {Kocharovsky}}, \bibinfo {author} {\bibfnamefont {V.~V.}\ \bibnamefont {Kocharovsky}}, \bibinfo {author} {\bibfnamefont {M.}~\bibnamefont {Holthaus}}, \bibinfo {author} {\bibfnamefont {C.~H.~R.}\ \bibnamefont {Ooi}}, \bibinfo {author} {\bibfnamefont {A.}~\bibnamefont {Svidzinsky}}, \bibinfo {author} {\bibfnamefont {W.}~\bibnamefont {Ketterle}},\ and\ \bibinfo {author} {\bibfnamefont {M.~O.}\ \bibnamefont {Scully}},\ }\bibfield  {title} {\bibinfo {title} {Fluctuations in ideal and interacting {{Bose}--{Einstein}} condensates: From the laser phase transition analogy to squeezed states and {Bogoliubov} quasiparticles},\ }in\ \href {https://doi.org/10.1016/S1049-250X(06)53010-1} {\emph {\bibinfo {booktitle} {Advances in Atomic, Molecular, and Optical Physics}}},\ Vol.~\bibinfo {volume} {53}\ (\bibinfo  {publisher} {Elsevier},\ \bibinfo {year} {2006})\ pp.\ \bibinfo {pages} {291--411}\BibitemShut {NoStop}%
\bibitem [{\citenamefont {Idziaszek}(2005)}]{Idziaszek2005}%
  \BibitemOpen
  \bibfield  {author} {\bibinfo {author} {\bibfnamefont {Z.}~\bibnamefont {Idziaszek}},\ }\bibfield  {title} {\bibinfo {title} {Microcanonical fluctuations of the condensate in weakly interacting {{Bose}} gases},\ }\href {https://doi.org/10.1103/PhysRevA.71.053604} {\bibfield  {journal} {\bibinfo  {journal} {Physical Review A}\ }\textbf {\bibinfo {volume} {71}},\ \bibinfo {pages} {053604} (\bibinfo {year} {2005})}\BibitemShut {NoStop}%
\bibitem [{\citenamefont {Brewczyk}\ \emph {et~al.}(2007)\citenamefont {Brewczyk}, \citenamefont {Gajda},\ and\ \citenamefont {Rzążewski}}]{Brewczyk2007}%
  \BibitemOpen
  \bibfield  {author} {\bibinfo {author} {\bibfnamefont {M.}~\bibnamefont {Brewczyk}}, \bibinfo {author} {\bibfnamefont {M.}~\bibnamefont {Gajda}},\ and\ \bibinfo {author} {\bibfnamefont {K.}~\bibnamefont {Rzążewski}},\ }\bibfield  {title} {\bibinfo {title} {Classical fields approximation for bosons at nonzero temperatures},\ }\href {https://doi.org/10.1088/0953-4075/40/2/R01} {\bibfield  {journal} {\bibinfo  {journal} {Journal of Physics B: Atomic, Molecular and Optical Physics}\ }\textbf {\bibinfo {volume} {40}},\ \bibinfo {pages} {R1} (\bibinfo {year} {2007})}\BibitemShut {NoStop}%
\bibitem [{\citenamefont {Idziaszek}\ \emph {et~al.}(2009)\citenamefont {Idziaszek}, \citenamefont {Łukasz Zawitkowski}, \citenamefont {Gajda},\ and\ \citenamefont {Rzążewski}}]{Idziaszek2009}%
  \BibitemOpen
  \bibfield  {author} {\bibinfo {author} {\bibfnamefont {Z.}~\bibnamefont {Idziaszek}}, \bibinfo {author} {\bibnamefont {Łukasz Zawitkowski}}, \bibinfo {author} {\bibfnamefont {M.}~\bibnamefont {Gajda}},\ and\ \bibinfo {author} {\bibfnamefont {K.}~\bibnamefont {Rzążewski}},\ }\bibfield  {title} {\bibinfo {title} {Fluctuations of a weakly interacting {Bose}--{Einstein} condensate},\ }\href {https://doi.org/10.1209/0295-5075/86/10002} {\bibfield  {journal} {\bibinfo  {journal} {EPL (Europhysics Letters)}\ }\textbf {\bibinfo {volume} {86}},\ \bibinfo {pages} {10002} (\bibinfo {year} {2009})}\BibitemShut {NoStop}%
\bibitem [{\citenamefont {Witkowska}\ \emph {et~al.}(2009)\citenamefont {Witkowska}, \citenamefont {Gajda},\ and\ \citenamefont {Rzążewski}}]{Witkowska2009}%
  \BibitemOpen
  \bibfield  {author} {\bibinfo {author} {\bibfnamefont {E.}~\bibnamefont {Witkowska}}, \bibinfo {author} {\bibfnamefont {M.}~\bibnamefont {Gajda}},\ and\ \bibinfo {author} {\bibfnamefont {K.}~\bibnamefont {Rzążewski}},\ }\bibfield  {title} {\bibinfo {title} {{Bose} statistics and classical fields},\ }\href {https://doi.org/10.1103/PhysRevA.79.033631} {\bibfield  {journal} {\bibinfo  {journal} {Physical Review A}\ }\textbf {\bibinfo {volume} {79}},\ \bibinfo {pages} {033631} (\bibinfo {year} {2009})}\BibitemShut {NoStop}%
\bibitem [{\citenamefont {Bienias}\ \emph {et~al.}(2011{\natexlab{a}})\citenamefont {Bienias}, \citenamefont {Pawłowski}, \citenamefont {Gajda},\ and\ \citenamefont {Rzążewski}}]{Bienias2011}%
  \BibitemOpen
  \bibfield  {author} {\bibinfo {author} {\bibfnamefont {P.}~\bibnamefont {Bienias}}, \bibinfo {author} {\bibfnamefont {K.}~\bibnamefont {Pawłowski}}, \bibinfo {author} {\bibfnamefont {M.}~\bibnamefont {Gajda}},\ and\ \bibinfo {author} {\bibfnamefont {K.}~\bibnamefont {Rzążewski}},\ }\bibfield  {title} {\bibinfo {title} {Statistical properties of one-dimensional {Bose} gas},\ }\href {https://doi.org/10.1103/PhysRevA.83.033610} {\bibfield  {journal} {\bibinfo  {journal} {Physical Review A}\ }\textbf {\bibinfo {volume} {83}},\ \bibinfo {pages} {033610} (\bibinfo {year} {2011}{\natexlab{a}})}\BibitemShut {NoStop}%
\bibitem [{\citenamefont {Bienias}\ \emph {et~al.}(2011{\natexlab{b}})\citenamefont {Bienias}, \citenamefont {Pawłowski}, \citenamefont {Gajda},\ and\ \citenamefont {Rzążewski}}]{Bienias2011Attr}%
  \BibitemOpen
  \bibfield  {author} {\bibinfo {author} {\bibfnamefont {P.}~\bibnamefont {Bienias}}, \bibinfo {author} {\bibfnamefont {K.}~\bibnamefont {Pawłowski}}, \bibinfo {author} {\bibfnamefont {M.}~\bibnamefont {Gajda}},\ and\ \bibinfo {author} {\bibfnamefont {K.}~\bibnamefont {Rzążewski}},\ }\bibfield  {title} {\bibinfo {title} {Statistical properties of one-dimensional attractive {Bose} gas},\ }\href {https://doi.org/10.1209/0295-5075/96/10011} {\bibfield  {journal} {\bibinfo  {journal} {EPL (Europhysics Letters)}\ }\textbf {\bibinfo {volume} {96}},\ \bibinfo {pages} {10011} (\bibinfo {year} {2011}{\natexlab{b}})}\BibitemShut {NoStop}%
\bibitem [{\citenamefont {Kocharovsky}\ \emph {et~al.}(2016)\citenamefont {Kocharovsky}, \citenamefont {Kocharovsky},\ and\ \citenamefont {Tarasov}}]{Kocharovsky16}%
  \BibitemOpen
  \bibfield  {author} {\bibinfo {author} {\bibfnamefont {V.~V.}\ \bibnamefont {Kocharovsky}}, \bibinfo {author} {\bibfnamefont {V.~V.}\ \bibnamefont {Kocharovsky}},\ and\ \bibinfo {author} {\bibfnamefont {S.~V.}\ \bibnamefont {Tarasov}},\ }\bibfield  {title} {\bibinfo {title} {{Bose}--{Einstein} condensation in mesoscopic systems: The self-similar structure of the critical region and the nonequivalence of the canonical and grand canonical ensembles},\ }\href {https://doi.org/10.1134/S0021364016010070} {\bibfield  {journal} {\bibinfo  {journal} {JETP Letters}\ }\textbf {\bibinfo {volume} {103}},\ \bibinfo {pages} {62} (\bibinfo {year} {2016})}\BibitemShut {NoStop}%
\bibitem [{\citenamefont {Touchette}(2015)}]{Touchette15}%
  \BibitemOpen
  \bibfield  {author} {\bibinfo {author} {\bibfnamefont {H.}~\bibnamefont {Touchette}},\ }\bibfield  {title} {\bibinfo {title} {Equivalence and nonequivalence of ensembles: Thermodynamic, macrostate, and measure levels},\ }\href {https://doi.org/10.1007/s10955-015-1212-2} {\bibfield  {journal} {\bibinfo  {journal} {Journal of Statistical Physics}\ }\textbf {\bibinfo {volume} {159}},\ \bibinfo {pages} {987} (\bibinfo {year} {2015})}\BibitemShut {NoStop}%
\bibitem [{\citenamefont {Yukalov}(2024)}]{Yukalov24}%
  \BibitemOpen
  \bibfield  {author} {\bibinfo {author} {\bibfnamefont {V.~I.}\ \bibnamefont {Yukalov}},\ }\bibfield  {title} {\bibinfo {title} {Particle fluctuations in systems with {{Bose}--{Einstein}} condensate},\ }\href {https://doi.org/10.1088/1555-6611/ad8221} {\bibfield  {journal} {\bibinfo  {journal} {Laser Physics}\ }\textbf {\bibinfo {volume} {34}},\ \bibinfo {pages} {113001} (\bibinfo {year} {2024})}\BibitemShut {NoStop}%
\bibitem [{\citenamefont {Kruk}\ \emph {et~al.}(2023)\citenamefont {Kruk}, \citenamefont {Hryniuk}, \citenamefont {Kristensen}, \citenamefont {Vibel}, \citenamefont {Pawłowski}, \citenamefont {Arlt},\ and\ \citenamefont {Rzążewski}}]{Kruk2023}%
  \BibitemOpen
  \bibfield  {author} {\bibinfo {author} {\bibfnamefont {M.~B.}\ \bibnamefont {Kruk}}, \bibinfo {author} {\bibfnamefont {D.}~\bibnamefont {Hryniuk}}, \bibinfo {author} {\bibfnamefont {M.}~\bibnamefont {Kristensen}}, \bibinfo {author} {\bibfnamefont {T.}~\bibnamefont {Vibel}}, \bibinfo {author} {\bibfnamefont {K.}~\bibnamefont {Pawłowski}}, \bibinfo {author} {\bibfnamefont {J.}~\bibnamefont {Arlt}},\ and\ \bibinfo {author} {\bibfnamefont {K.}~\bibnamefont {Rzążewski}},\ }\bibfield  {title} {\bibinfo {title} {Microcanonical and canonical fluctuations in atomic {Bose}-{Einstein} condensates – fock state sampling approach},\ }\href {https://doi.org/10.21468/scipostphys.14.3.036} {\bibfield  {journal} {\bibinfo  {journal} {SciPost Physics}\ }\textbf {\bibinfo {volume} {14}},\ \bibinfo {pages} {036} (\bibinfo {year} {2023})}\BibitemShut {NoStop}%
\bibitem [{\citenamefont {Vibel}\ \emph {et~al.}(2024)\citenamefont {Vibel}, \citenamefont {Christensen}, \citenamefont {F~Andersen}, \citenamefont {Stokholm}, \citenamefont {Pawłowski}, \citenamefont {Rzążewski}, \citenamefont {Kristensen},\ and\ \citenamefont {Arlt}}]{Vibel2024}%
  \BibitemOpen
  \bibfield  {author} {\bibinfo {author} {\bibfnamefont {T.}~\bibnamefont {Vibel}}, \bibinfo {author} {\bibfnamefont {M.~B.}\ \bibnamefont {Christensen}}, \bibinfo {author} {\bibfnamefont {R.~M.}\ \bibnamefont {F~Andersen}}, \bibinfo {author} {\bibfnamefont {L.~N.}\ \bibnamefont {Stokholm}}, \bibinfo {author} {\bibfnamefont {K.}~\bibnamefont {Pawłowski}}, \bibinfo {author} {\bibfnamefont {K.}~\bibnamefont {Rzążewski}}, \bibinfo {author} {\bibfnamefont {M.~A.}\ \bibnamefont {Kristensen}},\ and\ \bibinfo {author} {\bibfnamefont {J.~J.}\ \bibnamefont {Arlt}},\ }\bibfield  {title} {\bibinfo {title} {Atom number fluctuations in {Bose} gases—statistical analysis of parameter estimation},\ }\href {https://doi.org/10.1088/1361-6455/ad7458} {\bibfield  {journal} {\bibinfo  {journal} {Journal of Physics B: Atomic, Molecular and Optical Physics}\ }\textbf {\bibinfo {volume} {57}},\ \bibinfo {pages} {195301} (\bibinfo {year} {2024})}\BibitemShut {NoStop}%
\bibitem [{\citenamefont {Chuu}\ \emph {et~al.}(2005)\citenamefont {Chuu}, \citenamefont {Schreck}, \citenamefont {Meyrath}, \citenamefont {Hanssen}, \citenamefont {Price},\ and\ \citenamefont {Raizen}}]{Chuu2005}%
  \BibitemOpen
  \bibfield  {author} {\bibinfo {author} {\bibfnamefont {C.-S.}\ \bibnamefont {Chuu}}, \bibinfo {author} {\bibfnamefont {F.}~\bibnamefont {Schreck}}, \bibinfo {author} {\bibfnamefont {T.~P.}\ \bibnamefont {Meyrath}}, \bibinfo {author} {\bibfnamefont {J.~L.}\ \bibnamefont {Hanssen}}, \bibinfo {author} {\bibfnamefont {G.~N.}\ \bibnamefont {Price}},\ and\ \bibinfo {author} {\bibfnamefont {M.~G.}\ \bibnamefont {Raizen}},\ }\bibfield  {title} {\bibinfo {title} {Direct observation of sub-{Poissonian} number statistics in a degenerate {Bose} gas},\ }\href {https://doi.org/10.1103/PhysRevLett.95.260403} {\bibfield  {journal} {\bibinfo  {journal} {Phys. Rev. Lett.}\ }\textbf {\bibinfo {volume} {95}},\ \bibinfo {pages} {260403} (\bibinfo {year} {2005})}\BibitemShut {NoStop}%
\bibitem [{\citenamefont {Esteve}\ \emph {et~al.}(2006)\citenamefont {Esteve}, \citenamefont {Trebbia}, \citenamefont {Schumm}, \citenamefont {Aspect}, \citenamefont {Westbrook},\ and\ \citenamefont {Bouchoule}}]{Esteve2006}%
  \BibitemOpen
  \bibfield  {author} {\bibinfo {author} {\bibfnamefont {J.}~\bibnamefont {Esteve}}, \bibinfo {author} {\bibfnamefont {J.-B.}\ \bibnamefont {Trebbia}}, \bibinfo {author} {\bibfnamefont {T.}~\bibnamefont {Schumm}}, \bibinfo {author} {\bibfnamefont {A.}~\bibnamefont {Aspect}}, \bibinfo {author} {\bibfnamefont {C.~I.}\ \bibnamefont {Westbrook}},\ and\ \bibinfo {author} {\bibfnamefont {I.}~\bibnamefont {Bouchoule}},\ }\bibfield  {title} {\bibinfo {title} {Observations of density fluctuations in an elongated {Bose} gas: Ideal-gas and quasi-condensate regimes},\ }\href {https://doi.org/10.1103/PhysRevLett.96.130403} {\bibfield  {journal} {\bibinfo  {journal} {Phys. Rev. Lett.}\ }\textbf {\bibinfo {volume} {96}},\ \bibinfo {pages} {130403} (\bibinfo {year} {2006})}\BibitemShut {NoStop}%
\bibitem [{\citenamefont {Maussang}\ \emph {et~al.}(2010)\citenamefont {Maussang}, \citenamefont {Marti}, \citenamefont {Schneider}, \citenamefont {Treutlein}, \citenamefont {Li}, \citenamefont {Sinatra}, \citenamefont {Long}, \citenamefont {Estève},\ and\ \citenamefont {Reichel}}]{Maussang2010}%
  \BibitemOpen
  \bibfield  {author} {\bibinfo {author} {\bibfnamefont {K.}~\bibnamefont {Maussang}}, \bibinfo {author} {\bibfnamefont {G.~E.}\ \bibnamefont {Marti}}, \bibinfo {author} {\bibfnamefont {T.~M.}\ \bibnamefont {Schneider}}, \bibinfo {author} {\bibfnamefont {P.}~\bibnamefont {Treutlein}}, \bibinfo {author} {\bibfnamefont {Y.}~\bibnamefont {Li}}, \bibinfo {author} {\bibfnamefont {A.}~\bibnamefont {Sinatra}}, \bibinfo {author} {\bibfnamefont {R.}~\bibnamefont {Long}}, \bibinfo {author} {\bibfnamefont {J.}~\bibnamefont {Estève}},\ and\ \bibinfo {author} {\bibfnamefont {J.}~\bibnamefont {Reichel}},\ }\bibfield  {title} {\bibinfo {title} {Enhanced and reduced atom number fluctuations in a {BEC} splitter},\ }\href {https://doi.org/10.1103/PhysRevLett.105.080403} {\bibfield  {journal} {\bibinfo  {journal} {Phys. Rev. Lett.}\ }\textbf {\bibinfo {volume} {105}},\ \bibinfo {pages} {080403} (\bibinfo {year} {2010})}\BibitemShut {NoStop}%
\bibitem [{\citenamefont {Armijo}\ \emph {et~al.}(2010)\citenamefont {Armijo}, \citenamefont {Jacqmin}, \citenamefont {Kheruntsyan},\ and\ \citenamefont {Bouchoule}}]{Armijo2010}%
  \BibitemOpen
  \bibfield  {author} {\bibinfo {author} {\bibfnamefont {J.}~\bibnamefont {Armijo}}, \bibinfo {author} {\bibfnamefont {T.}~\bibnamefont {Jacqmin}}, \bibinfo {author} {\bibfnamefont {K.~V.}\ \bibnamefont {Kheruntsyan}},\ and\ \bibinfo {author} {\bibfnamefont {I.}~\bibnamefont {Bouchoule}},\ }\bibfield  {title} {\bibinfo {title} {Probing three-body correlations in a quantum gas using the measurement of the third moment of density fluctuations},\ }\href {https://doi.org/10.1103/PhysRevLett.105.230402} {\bibfield  {journal} {\bibinfo  {journal} {Phys. Rev. Lett.}\ }\textbf {\bibinfo {volume} {105}},\ \bibinfo {pages} {230402} (\bibinfo {year} {2010})}\BibitemShut {NoStop}%
\bibitem [{\citenamefont {Jacqmin}\ \emph {et~al.}(2011)\citenamefont {Jacqmin}, \citenamefont {Armijo}, \citenamefont {Berrada}, \citenamefont {Kheruntsyan},\ and\ \citenamefont {Bouchoule}}]{Jacqmin2011}%
  \BibitemOpen
  \bibfield  {author} {\bibinfo {author} {\bibfnamefont {T.}~\bibnamefont {Jacqmin}}, \bibinfo {author} {\bibfnamefont {J.}~\bibnamefont {Armijo}}, \bibinfo {author} {\bibfnamefont {T.}~\bibnamefont {Berrada}}, \bibinfo {author} {\bibfnamefont {K.~V.}\ \bibnamefont {Kheruntsyan}},\ and\ \bibinfo {author} {\bibfnamefont {I.}~\bibnamefont {Bouchoule}},\ }\bibfield  {title} {\bibinfo {title} {Sub-{Poissonian} fluctuations in a 1d {Bose} gas: From the quantum quasicondensate to the strongly interacting regime},\ }\href {https://doi.org/10.1103/PhysRevLett.106.230405} {\bibfield  {journal} {\bibinfo  {journal} {Phys. Rev. Lett.}\ }\textbf {\bibinfo {volume} {106}},\ \bibinfo {pages} {230405} (\bibinfo {year} {2011})}\BibitemShut {NoStop}%
\bibitem [{\citenamefont {Chomaz}\ \emph {et~al.}(2015)\citenamefont {Chomaz}, \citenamefont {Corman}, \citenamefont {Bienaim}, \citenamefont {Desbuquois}, \citenamefont {Weitenberg}, \citenamefont {Nascimb{\`e}ne}, \citenamefont {Beugnon},\ and\ \citenamefont {Dalibard}}]{Chomaz2015}%
  \BibitemOpen
  \bibfield  {author} {\bibinfo {author} {\bibfnamefont {L.}~\bibnamefont {Chomaz}}, \bibinfo {author} {\bibfnamefont {L.}~\bibnamefont {Corman}}, \bibinfo {author} {\bibfnamefont {T.}~\bibnamefont {Bienaim}}, \bibinfo {author} {\bibfnamefont {R.}~\bibnamefont {Desbuquois}}, \bibinfo {author} {\bibfnamefont {C.}~\bibnamefont {Weitenberg}}, \bibinfo {author} {\bibfnamefont {S.}~\bibnamefont {Nascimb{\`e}ne}}, \bibinfo {author} {\bibfnamefont {J.}~\bibnamefont {Beugnon}},\ and\ \bibinfo {author} {\bibfnamefont {J.}~\bibnamefont {Dalibard}},\ }\bibfield  {title} {\bibinfo {title} {Emergence of coherence via transverse condensation in a uniform quasi-two-dimensional {Bose} gas},\ }\href {https://doi.org/10.1038/ncomms7162} {\bibfield  {journal} {\bibinfo  {journal} {Nature Communications}\ }\textbf {\bibinfo {volume} {6}},\ \bibinfo {pages} {6162} (\bibinfo {year} {2015})}\BibitemShut {NoStop}%
\bibitem [{\citenamefont {Gajdacz}\ \emph {et~al.}(2016)\citenamefont {Gajdacz}, \citenamefont {Hilliard}, \citenamefont {Kristensen}, \citenamefont {Pedersen}, \citenamefont {Klempt}, \citenamefont {Arlt},\ and\ \citenamefont {Sherson}}]{Gajdacz2016}%
  \BibitemOpen
  \bibfield  {author} {\bibinfo {author} {\bibfnamefont {M.}~\bibnamefont {Gajdacz}}, \bibinfo {author} {\bibfnamefont {A.~J.}\ \bibnamefont {Hilliard}}, \bibinfo {author} {\bibfnamefont {M.~A.}\ \bibnamefont {Kristensen}}, \bibinfo {author} {\bibfnamefont {P.~L.}\ \bibnamefont {Pedersen}}, \bibinfo {author} {\bibfnamefont {C.}~\bibnamefont {Klempt}}, \bibinfo {author} {\bibfnamefont {J.~J.}\ \bibnamefont {Arlt}},\ and\ \bibinfo {author} {\bibfnamefont {J.~F.}\ \bibnamefont {Sherson}},\ }\bibfield  {title} {\bibinfo {title} {Preparation of ultracold atom clouds at the shot noise level},\ }\href {https://doi.org/10.1103/PhysRevLett.117.073604} {\bibfield  {journal} {\bibinfo  {journal} {Physical Review Letters}\ }\textbf {\bibinfo {volume} {117}},\ \bibinfo {pages} {073604} (\bibinfo {year} {2016})}\BibitemShut {NoStop}%
\bibitem [{\citenamefont {Kristensen}\ \emph {et~al.}(2019)\citenamefont {Kristensen}, \citenamefont {Christensen}, \citenamefont {Gajdacz}, \citenamefont {Iglicki}, \citenamefont {Paw{\l}owski}, \citenamefont {Klempt}, \citenamefont {Sherson}, \citenamefont {Rzążewski}, \citenamefont {Hilliard},\ and\ \citenamefont {Arlt}}]{Kristensen2019}%
  \BibitemOpen
  \bibfield  {author} {\bibinfo {author} {\bibfnamefont {M.~A.}\ \bibnamefont {Kristensen}}, \bibinfo {author} {\bibfnamefont {M.~B.}\ \bibnamefont {Christensen}}, \bibinfo {author} {\bibfnamefont {M.}~\bibnamefont {Gajdacz}}, \bibinfo {author} {\bibfnamefont {M.}~\bibnamefont {Iglicki}}, \bibinfo {author} {\bibfnamefont {K.}~\bibnamefont {Paw{\l}owski}}, \bibinfo {author} {\bibfnamefont {C.}~\bibnamefont {Klempt}}, \bibinfo {author} {\bibfnamefont {J.~F.}\ \bibnamefont {Sherson}}, \bibinfo {author} {\bibfnamefont {K.}~\bibnamefont {Rzążewski}}, \bibinfo {author} {\bibfnamefont {A.~J.}\ \bibnamefont {Hilliard}},\ and\ \bibinfo {author} {\bibfnamefont {J.~J.}\ \bibnamefont {Arlt}},\ }\bibfield  {title} {\bibinfo {title} {Observation of atom number fluctuations in a {{Bose}--{Einstein}} condensate},\ }\href {https://doi.org/10.1103/PhysRevLett.122.163601} {\bibfield  {journal} {\bibinfo  {journal} {Physical Review Letters}\ }\textbf {\bibinfo {volume} {122}},\ \bibinfo {pages} {163601} (\bibinfo {year}
  {2019})}\BibitemShut {NoStop}%
\bibitem [{\citenamefont {Asteria}\ \emph {et~al.}(2021)\citenamefont {Asteria}, \citenamefont {Zahn}, \citenamefont {Kosch}, \citenamefont {Sengstock},\ and\ \citenamefont {Weitenberg}}]{Asteria_2021}%
  \BibitemOpen
  \bibfield  {author} {\bibinfo {author} {\bibfnamefont {L.}~\bibnamefont {Asteria}}, \bibinfo {author} {\bibfnamefont {H.~P.}\ \bibnamefont {Zahn}}, \bibinfo {author} {\bibfnamefont {M.~N.}\ \bibnamefont {Kosch}}, \bibinfo {author} {\bibfnamefont {K.}~\bibnamefont {Sengstock}},\ and\ \bibinfo {author} {\bibfnamefont {C.}~\bibnamefont {Weitenberg}},\ }\bibfield  {title} {\bibinfo {title} {Quantum gas magnifier for sub-lattice-resolved imaging of 3d quantum systems},\ }\href {https://doi.org/10.1038/s41586-021-04011-2} {\bibfield  {journal} {\bibinfo  {journal} {Nature}\ }\textbf {\bibinfo {volume} {599}},\ \bibinfo {pages} {571–575} (\bibinfo {year} {2021})}\BibitemShut {NoStop}%
\bibitem [{\citenamefont {Christensen}\ \emph {et~al.}(2021)\citenamefont {Christensen}, \citenamefont {Vibel}, \citenamefont {Hilliard}, \citenamefont {Kruk}, \citenamefont {Paw{\l}owski}, \citenamefont {Hryniuk}, \citenamefont {Rzążewski}, \citenamefont {Kristensen},\ and\ \citenamefont {Arlt}}]{Christensen2021}%
  \BibitemOpen
  \bibfield  {author} {\bibinfo {author} {\bibfnamefont {M.~B.}\ \bibnamefont {Christensen}}, \bibinfo {author} {\bibfnamefont {T.}~\bibnamefont {Vibel}}, \bibinfo {author} {\bibfnamefont {A.~J.}\ \bibnamefont {Hilliard}}, \bibinfo {author} {\bibfnamefont {M.~B.}\ \bibnamefont {Kruk}}, \bibinfo {author} {\bibfnamefont {K.}~\bibnamefont {Paw{\l}owski}}, \bibinfo {author} {\bibfnamefont {D.}~\bibnamefont {Hryniuk}}, \bibinfo {author} {\bibfnamefont {K.}~\bibnamefont {Rzążewski}}, \bibinfo {author} {\bibfnamefont {M.~A.}\ \bibnamefont {Kristensen}},\ and\ \bibinfo {author} {\bibfnamefont {J.~J.}\ \bibnamefont {Arlt}},\ }\bibfield  {title} {\bibinfo {title} {Observation of microcanonical atom number fluctuations in a {{Bose}--{Einstein}} condensate},\ }\href {https://doi.org/10.1103/PhysRevLett.126.153601} {\bibfield  {journal} {\bibinfo  {journal} {Physical Review Letters}\ }\textbf {\bibinfo {volume} {126}},\ \bibinfo {pages} {153601} (\bibinfo {year} {2021})}\BibitemShut {NoStop}%
\bibitem [{\citenamefont {Kruk}\ \emph {et~al.}(2025)\citenamefont {Kruk}, \citenamefont {Kulik}, \citenamefont {Andersen}, \citenamefont {Deuar}, \citenamefont {Gajda}, \citenamefont {Pawłowski}, \citenamefont {Witkowska}, \citenamefont {Arlt},\ and\ \citenamefont {Rzążewski}}]{kruk2025}%
  \BibitemOpen
  \bibfield  {author} {\bibinfo {author} {\bibfnamefont {M.~B.}\ \bibnamefont {Kruk}}, \bibinfo {author} {\bibfnamefont {P.}~\bibnamefont {Kulik}}, \bibinfo {author} {\bibfnamefont {M.~F.}\ \bibnamefont {Andersen}}, \bibinfo {author} {\bibfnamefont {P.}~\bibnamefont {Deuar}}, \bibinfo {author} {\bibfnamefont {M.}~\bibnamefont {Gajda}}, \bibinfo {author} {\bibfnamefont {K.}~\bibnamefont {Pawłowski}}, \bibinfo {author} {\bibfnamefont {E.}~\bibnamefont {Witkowska}}, \bibinfo {author} {\bibfnamefont {J.~J.}\ \bibnamefont {Arlt}},\ and\ \bibinfo {author} {\bibfnamefont {K.}~\bibnamefont {Rzążewski}},\ }\bibfield  {title} {\bibinfo {title} {On the fluctuations of the number of atoms in the condensate},\ }\href {https://doi.org/10.1088/1361-6633/ae0e33} {\bibfield  {journal} {\bibinfo  {journal} {Reports on Progress in Physics}\ }\textbf {\bibinfo {volume} {88}},\ \bibinfo {pages} {106401} (\bibinfo {year} {2025})}\BibitemShut {NoStop}%
\bibitem [{\citenamefont {Grass}\ \emph {et~al.}(2025)\citenamefont {Grass}, \citenamefont {Bercioux}, \citenamefont {Bhattacharya}, \citenamefont {Lewenstein}, \citenamefont {Nguyen},\ and\ \citenamefont {Weitenberg}}]{Grass2024}%
  \BibitemOpen
  \bibfield  {author} {\bibinfo {author} {\bibfnamefont {T.}~\bibnamefont {Grass}}, \bibinfo {author} {\bibfnamefont {D.}~\bibnamefont {Bercioux}}, \bibinfo {author} {\bibfnamefont {U.}~\bibnamefont {Bhattacharya}}, \bibinfo {author} {\bibfnamefont {M.}~\bibnamefont {Lewenstein}}, \bibinfo {author} {\bibfnamefont {H.~S.}\ \bibnamefont {Nguyen}},\ and\ \bibinfo {author} {\bibfnamefont {C.}~\bibnamefont {Weitenberg}},\ }\bibfield  {title} {\bibinfo {title} {Colloquium: Synthetic quantum matter in nonstandard geometries},\ }\href {https://doi.org/10.1103/RevModPhys.97.011001} {\bibfield  {journal} {\bibinfo  {journal} {Rev. Mod. Phys.}\ }\textbf {\bibinfo {volume} {97}},\ \bibinfo {pages} {011001} (\bibinfo {year} {2025})}\BibitemShut {NoStop}%
\bibitem [{\citenamefont {Xu}\ \emph {et~al.}(2021)\citenamefont {Xu}, \citenamefont {Wang}, \citenamefont {Chen}, \citenamefont {Smith},\ and\ \citenamefont {Jin}}]{Xu2021}%
  \BibitemOpen
  \bibfield  {author} {\bibinfo {author} {\bibfnamefont {X.-Y.}\ \bibnamefont {Xu}}, \bibinfo {author} {\bibfnamefont {X.-W.}\ \bibnamefont {Wang}}, \bibinfo {author} {\bibfnamefont {D.-Y.}\ \bibnamefont {Chen}}, \bibinfo {author} {\bibfnamefont {C.~M.}\ \bibnamefont {Smith}},\ and\ \bibinfo {author} {\bibfnamefont {X.-M.}\ \bibnamefont {Jin}},\ }\bibfield  {title} {\bibinfo {title} {Quantum transport in fractal networks},\ }\href {https://doi.org/10.1038/s41566-021-00845-4} {\bibfield  {journal} {\bibinfo  {journal} {Nat. Photonics}\ }\textbf {\bibinfo {volume} {15}},\ \bibinfo {pages} {703} (\bibinfo {year} {2021})}\BibitemShut {NoStop}%
\bibitem [{\citenamefont {Biesenthal}\ \emph {et~al.}(2022)\citenamefont {Biesenthal}, \citenamefont {Maczewsky}, \citenamefont {Yang}, \citenamefont {Kremer}, \citenamefont {Segev}, \citenamefont {Szameit},\ and\ \citenamefont {Heinrich}}]{Biesenthal2022}%
  \BibitemOpen
  \bibfield  {author} {\bibinfo {author} {\bibfnamefont {T.}~\bibnamefont {Biesenthal}}, \bibinfo {author} {\bibfnamefont {L.~J.}\ \bibnamefont {Maczewsky}}, \bibinfo {author} {\bibfnamefont {Z.}~\bibnamefont {Yang}}, \bibinfo {author} {\bibfnamefont {M.}~\bibnamefont {Kremer}}, \bibinfo {author} {\bibfnamefont {M.}~\bibnamefont {Segev}}, \bibinfo {author} {\bibfnamefont {A.}~\bibnamefont {Szameit}},\ and\ \bibinfo {author} {\bibfnamefont {M.}~\bibnamefont {Heinrich}},\ }\bibfield  {title} {\bibinfo {title} {Fractal photonic topological insulators},\ }\href {https://doi.org/10.1126/science.abm2842} {\bibfield  {journal} {\bibinfo  {journal} {Science}\ }\textbf {\bibinfo {volume} {376}},\ \bibinfo {pages} {1114} (\bibinfo {year} {2022})}\BibitemShut {NoStop}%
\bibitem [{\citenamefont {Shang}\ \emph {et~al.}(2015)\citenamefont {Shang}, \citenamefont {Wang}, \citenamefont {Chen}, \citenamefont {Dai}, \citenamefont {Zhou}, \citenamefont {Kuttner}, \citenamefont {Hilt}, \citenamefont {Shao}, \citenamefont {Gottfried},\ and\ \citenamefont {Wu}}]{Shang2015}%
  \BibitemOpen
  \bibfield  {author} {\bibinfo {author} {\bibfnamefont {J.}~\bibnamefont {Shang}}, \bibinfo {author} {\bibfnamefont {Y.}~\bibnamefont {Wang}}, \bibinfo {author} {\bibfnamefont {M.}~\bibnamefont {Chen}}, \bibinfo {author} {\bibfnamefont {J.}~\bibnamefont {Dai}}, \bibinfo {author} {\bibfnamefont {X.}~\bibnamefont {Zhou}}, \bibinfo {author} {\bibfnamefont {J.}~\bibnamefont {Kuttner}}, \bibinfo {author} {\bibfnamefont {G.}~\bibnamefont {Hilt}}, \bibinfo {author} {\bibfnamefont {X.}~\bibnamefont {Shao}}, \bibinfo {author} {\bibfnamefont {J.~M.}\ \bibnamefont {Gottfried}},\ and\ \bibinfo {author} {\bibfnamefont {K.}~\bibnamefont {Wu}},\ }\bibfield  {title} {\bibinfo {title} {{Assembling molecular Sierpi{\'{n}}ski triangle fractals}},\ }\href {https://doi.org/10.1038/nchem.2211} {\bibfield  {journal} {\bibinfo  {journal} {Nature Chemistry}\ }\textbf {\bibinfo {volume} {7}},\ \bibinfo {pages} {389} (\bibinfo {year} {2015})}\BibitemShut {NoStop}%
\bibitem [{\citenamefont {Kempkes}\ \emph {et~al.}(2019)\citenamefont {Kempkes}, \citenamefont {Slot}, \citenamefont {Freeney}, \citenamefont {Zevenhuizen}, \citenamefont {Vanmaekelbergh}, \citenamefont {Swart},\ and\ \citenamefont {Smith}}]{Kempkes_2019_b}%
  \BibitemOpen
  \bibfield  {author} {\bibinfo {author} {\bibfnamefont {S.~N.}\ \bibnamefont {Kempkes}}, \bibinfo {author} {\bibfnamefont {M.~R.}\ \bibnamefont {Slot}}, \bibinfo {author} {\bibfnamefont {S.~E.}\ \bibnamefont {Freeney}}, \bibinfo {author} {\bibfnamefont {S.~J.~M.}\ \bibnamefont {Zevenhuizen}}, \bibinfo {author} {\bibfnamefont {D.}~\bibnamefont {Vanmaekelbergh}}, \bibinfo {author} {\bibfnamefont {I.}~\bibnamefont {Swart}},\ and\ \bibinfo {author} {\bibfnamefont {C.~M.}\ \bibnamefont {Smith}},\ }\bibfield  {title} {\bibinfo {title} {Design and characterization of electrons in a fractal geometry},\ }\href {https://doi.org/10.1038/s41567-018-0328-0} {\bibfield  {journal} {\bibinfo  {journal} {Nat. Phys.}\ }\textbf {\bibinfo {volume} {15}},\ \bibinfo {pages} {127} (\bibinfo {year} {2019})}\BibitemShut {NoStop}%
\bibitem [{\citenamefont {Tian}\ \emph {et~al.}(2023)\citenamefont {Tian}, \citenamefont {Wee}, \citenamefont {Qu}, \citenamefont {Lim}, \citenamefont {Datla}, \citenamefont {Koh},\ and\ \citenamefont {Loh}}]{Tian2023}%
  \BibitemOpen
  \bibfield  {author} {\bibinfo {author} {\bibfnamefont {W.}~\bibnamefont {Tian}}, \bibinfo {author} {\bibfnamefont {W.~J.}\ \bibnamefont {Wee}}, \bibinfo {author} {\bibfnamefont {A.}~\bibnamefont {Qu}}, \bibinfo {author} {\bibfnamefont {B.~J.~M.}\ \bibnamefont {Lim}}, \bibinfo {author} {\bibfnamefont {P.~R.}\ \bibnamefont {Datla}}, \bibinfo {author} {\bibfnamefont {V.~P.~W.}\ \bibnamefont {Koh}},\ and\ \bibinfo {author} {\bibfnamefont {H.}~\bibnamefont {Loh}},\ }\bibfield  {title} {\bibinfo {title} {Parallel assembly of arbitrary defect-free atom arrays with a multitweezer algorithm},\ }\href {https://doi.org/10.1103/PhysRevApplied.19.034048} {\bibfield  {journal} {\bibinfo  {journal} {Phys. Rev. Appl.}\ }\textbf {\bibinfo {volume} {19}},\ \bibinfo {pages} {034048} (\bibinfo {year} {2023})}\BibitemShut {NoStop}%
\bibitem [{\citenamefont {Koll{\'a}r}\ \emph {et~al.}(2019)\citenamefont {Koll{\'a}r}, \citenamefont {Fitzpatrick},\ and\ \citenamefont {Houck}}]{Kollar2019}%
  \BibitemOpen
  \bibfield  {author} {\bibinfo {author} {\bibfnamefont {A.~J.}\ \bibnamefont {Koll{\'a}r}}, \bibinfo {author} {\bibfnamefont {M.}~\bibnamefont {Fitzpatrick}},\ and\ \bibinfo {author} {\bibfnamefont {A.~A.}\ \bibnamefont {Houck}},\ }\bibfield  {title} {\bibinfo {title} {Hyperbolic lattices in circuit quantum electrodynamics},\ }\href {https://doi.org/10.1038/s41586-019-1348-3} {\bibfield  {journal} {\bibinfo  {journal} {Nature}\ }\textbf {\bibinfo {volume} {571}},\ \bibinfo {pages} {45} (\bibinfo {year} {2019})}\BibitemShut {NoStop}%
\bibitem [{\citenamefont {Pavlyshynets}\ \emph {et~al.}(2025)\citenamefont {Pavlyshynets}, \citenamefont {Salasnich}, \citenamefont {Malomed},\ and\ \citenamefont {Yakimenko}}]{Pavlyshynets2025}%
  \BibitemOpen
  \bibfield  {author} {\bibinfo {author} {\bibfnamefont {E.}~\bibnamefont {Pavlyshynets}}, \bibinfo {author} {\bibfnamefont {L.}~\bibnamefont {Salasnich}}, \bibinfo {author} {\bibfnamefont {B.~A.}\ \bibnamefont {Malomed}},\ and\ \bibinfo {author} {\bibfnamefont {A.}~\bibnamefont {Yakimenko}},\ }\bibfield  {title} {\bibinfo {title} {Solitons in quasiperiodic lattices with fractional diffraction},\ }\href {https://doi.org/10.1103/physreve.111.044206} {\bibfield  {journal} {\bibinfo  {journal} {Physical Review E}\ }\textbf {\bibinfo {volume} {111}},\ \bibinfo {pages} {044206} (\bibinfo {year} {2025})}\BibitemShut {NoStop}%
\bibitem [{\citenamefont {Aubry}\ and\ \citenamefont {Andr{\'e}}(1980)}]{Aubry_1980}%
  \BibitemOpen
  \bibfield  {author} {\bibinfo {author} {\bibfnamefont {S.}~\bibnamefont {Aubry}}\ and\ \bibinfo {author} {\bibfnamefont {G.}~\bibnamefont {Andr{\'e}}},\ }\bibfield  {title} {\bibinfo {title} {Analyticity breaking and anderson localization in incommensurate lattices},\ }\href@noop {} {\bibfield  {journal} {\bibinfo  {journal} {Ann. Israel Phys. Soc}\ }\textbf {\bibinfo {volume} {3}},\ \bibinfo {pages} {18} (\bibinfo {year} {1980})}\BibitemShut {NoStop}%
\bibitem [{\citenamefont {Kohmoto}\ \emph {et~al.}(1983)\citenamefont {Kohmoto}, \citenamefont {Kadanoff},\ and\ \citenamefont {Tang}}]{Kohmoto_1983}%
  \BibitemOpen
  \bibfield  {author} {\bibinfo {author} {\bibfnamefont {M.}~\bibnamefont {Kohmoto}}, \bibinfo {author} {\bibfnamefont {L.~P.}\ \bibnamefont {Kadanoff}},\ and\ \bibinfo {author} {\bibfnamefont {C.}~\bibnamefont {Tang}},\ }\bibfield  {title} {\bibinfo {title} {Localization problem in one dimension: Mapping and escape},\ }\href {https://doi.org/10.1103/PhysRevLett.50.1870} {\bibfield  {journal} {\bibinfo  {journal} {Phys. Rev. Lett.}\ }\textbf {\bibinfo {volume} {50}},\ \bibinfo {pages} {1870} (\bibinfo {year} {1983})}\BibitemShut {NoStop}%
\bibitem [{\citenamefont {Domany}\ \emph {et~al.}(1983)\citenamefont {Domany}, \citenamefont {Alexander}, \citenamefont {Bensimon},\ and\ \citenamefont {Kadanoff}}]{Domany_1983}%
  \BibitemOpen
  \bibfield  {author} {\bibinfo {author} {\bibfnamefont {E.}~\bibnamefont {Domany}}, \bibinfo {author} {\bibfnamefont {S.}~\bibnamefont {Alexander}}, \bibinfo {author} {\bibfnamefont {D.}~\bibnamefont {Bensimon}},\ and\ \bibinfo {author} {\bibfnamefont {L.~P.}\ \bibnamefont {Kadanoff}},\ }\bibfield  {title} {\bibinfo {title} {Solutions to the {Schr\"odinger} equation on some fractal lattices},\ }\href {https://doi.org/10.1103/PhysRevB.28.3110} {\bibfield  {journal} {\bibinfo  {journal} {Phys. Rev. B}\ }\textbf {\bibinfo {volume} {28}},\ \bibinfo {pages} {3110} (\bibinfo {year} {1983})}\BibitemShut {NoStop}%
\bibitem [{\citenamefont {Wang}(1995)}]{Wang_1995}%
  \BibitemOpen
  \bibfield  {author} {\bibinfo {author} {\bibfnamefont {X.~R.}\ \bibnamefont {Wang}},\ }\bibfield  {title} {\bibinfo {title} {Localization in fractal spaces: Exact results on the {Sierpinski} gasket},\ }\href {https://doi.org/10.1103/PhysRevB.51.9310} {\bibfield  {journal} {\bibinfo  {journal} {Phys. Rev. B}\ }\textbf {\bibinfo {volume} {51}},\ \bibinfo {pages} {9310} (\bibinfo {year} {1995})}\BibitemShut {NoStop}%
\bibitem [{\citenamefont {Manna}\ and\ \citenamefont {Roy}(2023)}]{Manna2023}%
  \BibitemOpen
  \bibfield  {author} {\bibinfo {author} {\bibfnamefont {S.}~\bibnamefont {Manna}}\ and\ \bibinfo {author} {\bibfnamefont {B.}~\bibnamefont {Roy}},\ }\bibfield  {title} {\bibinfo {title} {Inner skin effects on non-hermitian topological fractals},\ }\href {https://doi.org/10.1038/s42005-023-01130-2} {\bibfield  {journal} {\bibinfo  {journal} {Communications Physics}\ }\textbf {\bibinfo {volume} {6}},\ \bibinfo {pages} {10} (\bibinfo {year} {2023})}\BibitemShut {NoStop}%
\bibitem [{\citenamefont {Manna}\ \emph {et~al.}(2024)\citenamefont {Manna}, \citenamefont {Jaworowski},\ and\ \citenamefont {Nielsen}}]{Manna_2024}%
  \BibitemOpen
  \bibfield  {author} {\bibinfo {author} {\bibfnamefont {S.}~\bibnamefont {Manna}}, \bibinfo {author} {\bibfnamefont {B.}~\bibnamefont {Jaworowski}},\ and\ \bibinfo {author} {\bibfnamefont {A.~E.~B.}\ \bibnamefont {Nielsen}},\ }\bibfield  {title} {\bibinfo {title} {Many-body localization on finite generation fractal lattices},\ }\href {https://doi.org/10.1088/1742-5468/ad4538} {\bibfield  {journal} {\bibinfo  {journal} {Journal of Statistical Mechanics: Theory and Experiment}\ }\textbf {\bibinfo {volume} {2024}},\ \bibinfo {pages} {053301} (\bibinfo {year} {2024})}\BibitemShut {NoStop}%
\bibitem [{\citenamefont {Gefen}\ \emph {et~al.}(1981)\citenamefont {Gefen}, \citenamefont {Aharony}, \citenamefont {Mandelbrot},\ and\ \citenamefont {Kirkpatrick}}]{Gefen1981}%
  \BibitemOpen
  \bibfield  {author} {\bibinfo {author} {\bibfnamefont {Y.}~\bibnamefont {Gefen}}, \bibinfo {author} {\bibfnamefont {A.}~\bibnamefont {Aharony}}, \bibinfo {author} {\bibfnamefont {B.~B.}\ \bibnamefont {Mandelbrot}},\ and\ \bibinfo {author} {\bibfnamefont {S.}~\bibnamefont {Kirkpatrick}},\ }\bibfield  {title} {\bibinfo {title} {Solvable fractal family, and its possible relation to the backbone at percolation},\ }\href {https://doi.org/10.1103/PhysRevLett.47.1771} {\bibfield  {journal} {\bibinfo  {journal} {Phys. Rev. Lett.}\ }\textbf {\bibinfo {volume} {47}},\ \bibinfo {pages} {1771} (\bibinfo {year} {1981})}\BibitemShut {NoStop}%
\bibitem [{\citenamefont {Alexander}\ and\ \citenamefont {Orbach}(1982)}]{Alexander_1982}%
  \BibitemOpen
  \bibfield  {author} {\bibinfo {author} {\bibfnamefont {S.}~\bibnamefont {Alexander}}\ and\ \bibinfo {author} {\bibfnamefont {R.}~\bibnamefont {Orbach}},\ }\bibfield  {title} {\bibinfo {title} {Density of states on fractals : fractons},\ }\href {https://doi.org/10.1051/jphyslet:019820043017062500} {\bibfield  {journal} {\bibinfo  {journal} {J. Physique Lett.}\ }\textbf {\bibinfo {volume} {43}},\ \bibinfo {pages} {625} (\bibinfo {year} {1982})}\BibitemShut {NoStop}%
\bibitem [{\citenamefont {Rammal}\ and\ \citenamefont {Toulouse}(1982)}]{Rammal_1983}%
  \BibitemOpen
  \bibfield  {author} {\bibinfo {author} {\bibfnamefont {R.}~\bibnamefont {Rammal}}\ and\ \bibinfo {author} {\bibfnamefont {G.}~\bibnamefont {Toulouse}},\ }\bibfield  {title} {\bibinfo {title} {Random walks on fractal structures and percolation clusters},\ }\href {https://doi.org/10.1051/jphyslet:0198300440101300} {\bibfield  {journal} {\bibinfo  {journal} {J. Physique Lett.}\ }\textbf {\bibinfo {volume} {44}},\ \bibinfo {pages} {13} (\bibinfo {year} {1982})}\BibitemShut {NoStop}%
\bibitem [{\citenamefont {Gefen}\ \emph {et~al.}(1983)\citenamefont {Gefen}, \citenamefont {Aharony},\ and\ \citenamefont {Alexander}}]{Gefen1983}%
  \BibitemOpen
  \bibfield  {author} {\bibinfo {author} {\bibfnamefont {Y.}~\bibnamefont {Gefen}}, \bibinfo {author} {\bibfnamefont {A.}~\bibnamefont {Aharony}},\ and\ \bibinfo {author} {\bibfnamefont {S.}~\bibnamefont {Alexander}},\ }\bibfield  {title} {\bibinfo {title} {Anomalous diffusion on percolating clusters},\ }\href {https://doi.org/10.1103/PhysRevLett.50.77} {\bibfield  {journal} {\bibinfo  {journal} {Phys. Rev. Lett.}\ }\textbf {\bibinfo {volume} {50}},\ \bibinfo {pages} {77} (\bibinfo {year} {1983})}\BibitemShut {NoStop}%
\bibitem [{\citenamefont {Havlin}\ and\ \citenamefont {Ben-Avraham}(1987)}]{Havlin_1987}%
  \BibitemOpen
  \bibfield  {author} {\bibinfo {author} {\bibfnamefont {S.}~\bibnamefont {Havlin}}\ and\ \bibinfo {author} {\bibfnamefont {D.}~\bibnamefont {Ben-Avraham}},\ }\bibfield  {title} {\bibinfo {title} {Diffusion in disordered media},\ }\href {https://doi.org/10.1080/00018738700101072} {\bibfield  {journal} {\bibinfo  {journal} {Advances in Physics}\ }\textbf {\bibinfo {volume} {36}},\ \bibinfo {pages} {695} (\bibinfo {year} {1987})}\BibitemShut {NoStop}%
\bibitem [{\citenamefont {Dar\'azs}\ \emph {et~al.}(2014)\citenamefont {Dar\'azs}, \citenamefont {Anishchenko}, \citenamefont {Kiss}, \citenamefont {Blumen},\ and\ \citenamefont {M\"ulken}}]{Darazs_2014}%
  \BibitemOpen
  \bibfield  {author} {\bibinfo {author} {\bibfnamefont {Z.}~\bibnamefont {Dar\'azs}}, \bibinfo {author} {\bibfnamefont {A.}~\bibnamefont {Anishchenko}}, \bibinfo {author} {\bibfnamefont {T.}~\bibnamefont {Kiss}}, \bibinfo {author} {\bibfnamefont {A.}~\bibnamefont {Blumen}},\ and\ \bibinfo {author} {\bibfnamefont {O.}~\bibnamefont {M\"ulken}},\ }\bibfield  {title} {\bibinfo {title} {Transport properties of continuous-time quantum walks on {Sierpinski} fractals},\ }\href {https://doi.org/10.1103/PhysRevE.90.032113} {\bibfield  {journal} {\bibinfo  {journal} {Phys. Rev. E}\ }\textbf {\bibinfo {volume} {90}},\ \bibinfo {pages} {032113} (\bibinfo {year} {2014})}\BibitemShut {NoStop}%
\bibitem [{\citenamefont {Kosior}\ and\ \citenamefont {Sacha}(2017)}]{Kosior2017}%
  \BibitemOpen
  \bibfield  {author} {\bibinfo {author} {\bibfnamefont {A.}~\bibnamefont {Kosior}}\ and\ \bibinfo {author} {\bibfnamefont {K.}~\bibnamefont {Sacha}},\ }\bibfield  {title} {\bibinfo {title} {Localization in random fractal lattices},\ }\href {https://doi.org/10.1103/PhysRevB.95.104206} {\bibfield  {journal} {\bibinfo  {journal} {Phys. Rev. B}\ }\textbf {\bibinfo {volume} {95}},\ \bibinfo {pages} {104206} (\bibinfo {year} {2017})}\BibitemShut {NoStop}%
\bibitem [{\citenamefont {van Veen}\ \emph {et~al.}(2016)\citenamefont {van Veen}, \citenamefont {Yuan}, \citenamefont {Katsnelson}, \citenamefont {Polini},\ and\ \citenamefont {Tomadin}}]{vanVeen_2016}%
  \BibitemOpen
  \bibfield  {author} {\bibinfo {author} {\bibfnamefont {E.}~\bibnamefont {van Veen}}, \bibinfo {author} {\bibfnamefont {S.}~\bibnamefont {Yuan}}, \bibinfo {author} {\bibfnamefont {M.~I.}\ \bibnamefont {Katsnelson}}, \bibinfo {author} {\bibfnamefont {M.}~\bibnamefont {Polini}},\ and\ \bibinfo {author} {\bibfnamefont {A.}~\bibnamefont {Tomadin}},\ }\bibfield  {title} {\bibinfo {title} {Quantum transport in {Sierpinski} carpets},\ }\href {https://doi.org/10.1103/PhysRevB.93.115428} {\bibfield  {journal} {\bibinfo  {journal} {Phys. Rev. B}\ }\textbf {\bibinfo {volume} {93}},\ \bibinfo {pages} {115428} (\bibinfo {year} {2016})}\BibitemShut {NoStop}%
\bibitem [{\citenamefont {Rojo-Franc{\`a}s}\ \emph {et~al.}(2024)\citenamefont {Rojo-Franc{\`a}s}, \citenamefont {Pansari}, \citenamefont {Bhattacharya}, \citenamefont {Juli{\'a}-D{\'i}az},\ and\ \citenamefont {Grass}}]{Rojo-Francas2024}%
  \BibitemOpen
  \bibfield  {author} {\bibinfo {author} {\bibfnamefont {A.}~\bibnamefont {Rojo-Franc{\`a}s}}, \bibinfo {author} {\bibfnamefont {P.}~\bibnamefont {Pansari}}, \bibinfo {author} {\bibfnamefont {U.}~\bibnamefont {Bhattacharya}}, \bibinfo {author} {\bibfnamefont {B.}~\bibnamefont {Juli{\'a}-D{\'i}az}},\ and\ \bibinfo {author} {\bibfnamefont {T.}~\bibnamefont {Grass}},\ }\bibfield  {title} {\bibinfo {title} {Anomalous quantum transport in fractal lattices},\ }\href {https://doi.org/10.1038/s42005-024-01747-x} {\bibfield  {journal} {\bibinfo  {journal} {Communications Physics}\ }\textbf {\bibinfo {volume} {7}},\ \bibinfo {pages} {259} (\bibinfo {year} {2024})}\BibitemShut {NoStop}%
\bibitem [{\citenamefont {Salvati}\ \emph {et~al.}(2024)\citenamefont {Salvati}, \citenamefont {Katsnelson},\ and\ \citenamefont {Bagrov}}]{Salvati2024}%
  \BibitemOpen
  \bibfield  {author} {\bibinfo {author} {\bibfnamefont {F.}~\bibnamefont {Salvati}}, \bibinfo {author} {\bibfnamefont {M.~I.}\ \bibnamefont {Katsnelson}},\ and\ \bibinfo {author} {\bibfnamefont {A.~A.}\ \bibnamefont {Bagrov}},\ }\href {https://arxiv.org/abs/2410.18559} {\bibinfo {title} {Emergence of non-ergodic multifractal quantum states in geometrical fractals}} (\bibinfo {year} {2024}),\ \Eprint {https://arxiv.org/abs/2410.18559} {arXiv:2410.18559 [cond-mat.dis-nn]} \BibitemShut {NoStop}%
\bibitem [{\citenamefont {Brzezi\ifmmode~\acute{n}\else \'{n}\fi{}ska}\ \emph {et~al.}(2018)\citenamefont {Brzezi\ifmmode~\acute{n}\else \'{n}\fi{}ska}, \citenamefont {Cook},\ and\ \citenamefont {Neupert}}]{Brzezinska2018}%
  \BibitemOpen
  \bibfield  {author} {\bibinfo {author} {\bibfnamefont {M.}~\bibnamefont {Brzezi\ifmmode~\acute{n}\else \'{n}\fi{}ska}}, \bibinfo {author} {\bibfnamefont {A.~M.}\ \bibnamefont {Cook}},\ and\ \bibinfo {author} {\bibfnamefont {T.}~\bibnamefont {Neupert}},\ }\bibfield  {title} {\bibinfo {title} {{Topology in the Sierpi\ifmmode \acute{n}\else \'{n}\fi{}ski-Hofstadter problem}},\ }\href {https://doi.org/10.1103/PhysRevB.98.205116} {\bibfield  {journal} {\bibinfo  {journal} {Phys. Rev. B}\ }\textbf {\bibinfo {volume} {98}},\ \bibinfo {pages} {205116} (\bibinfo {year} {2018})}\BibitemShut {NoStop}%
\bibitem [{\citenamefont {Pai}\ and\ \citenamefont {Prem}(2019)}]{Pai2019}%
  \BibitemOpen
  \bibfield  {author} {\bibinfo {author} {\bibfnamefont {S.}~\bibnamefont {Pai}}\ and\ \bibinfo {author} {\bibfnamefont {A.}~\bibnamefont {Prem}},\ }\bibfield  {title} {\bibinfo {title} {Topological states on fractal lattices},\ }\href {https://doi.org/10.1103/PhysRevB.100.155135} {\bibfield  {journal} {\bibinfo  {journal} {Phys. Rev. B}\ }\textbf {\bibinfo {volume} {100}},\ \bibinfo {pages} {155135} (\bibinfo {year} {2019})}\BibitemShut {NoStop}%
\bibitem [{\citenamefont {Iliasov}\ \emph {et~al.}(2020)\citenamefont {Iliasov}, \citenamefont {Katsnelson},\ and\ \citenamefont {Yuan}}]{Iliasov2020}%
  \BibitemOpen
  \bibfield  {author} {\bibinfo {author} {\bibfnamefont {A.~A.}\ \bibnamefont {Iliasov}}, \bibinfo {author} {\bibfnamefont {M.~I.}\ \bibnamefont {Katsnelson}},\ and\ \bibinfo {author} {\bibfnamefont {S.}~\bibnamefont {Yuan}},\ }\bibfield  {title} {\bibinfo {title} {{Hall conductivity of a Sierpi\ifmmode \acute{n}\else \'{n}\fi{}ski carpet}},\ }\href {https://doi.org/10.1103/PhysRevB.101.045413} {\bibfield  {journal} {\bibinfo  {journal} {Phys. Rev. B}\ }\textbf {\bibinfo {volume} {101}},\ \bibinfo {pages} {045413} (\bibinfo {year} {2020})}\BibitemShut {NoStop}%
\bibitem [{\citenamefont {Fremling}\ \emph {et~al.}(2020)\citenamefont {Fremling}, \citenamefont {van Hooft}, \citenamefont {Smith},\ and\ \citenamefont {Fritz}}]{Fremling2020}%
  \BibitemOpen
  \bibfield  {author} {\bibinfo {author} {\bibfnamefont {M.}~\bibnamefont {Fremling}}, \bibinfo {author} {\bibfnamefont {M.}~\bibnamefont {van Hooft}}, \bibinfo {author} {\bibfnamefont {C.~M.}\ \bibnamefont {Smith}},\ and\ \bibinfo {author} {\bibfnamefont {L.}~\bibnamefont {Fritz}},\ }\bibfield  {title} {\bibinfo {title} {{Existence of robust edge currents in Sierpi\ifmmode \acute{n}\else \'{n}\fi{}ski fractals}},\ }\href {https://doi.org/10.1103/PhysRevResearch.2.013044} {\bibfield  {journal} {\bibinfo  {journal} {Phys. Rev. Res.}\ }\textbf {\bibinfo {volume} {2}},\ \bibinfo {pages} {013044} (\bibinfo {year} {2020})}\BibitemShut {NoStop}%
\bibitem [{\citenamefont {Moustaj}\ \emph {et~al.}(2021)\citenamefont {Moustaj}, \citenamefont {Kempkes},\ and\ \citenamefont {Smith}}]{Moustaj2021}%
  \BibitemOpen
  \bibfield  {author} {\bibinfo {author} {\bibfnamefont {A.}~\bibnamefont {Moustaj}}, \bibinfo {author} {\bibfnamefont {S.}~\bibnamefont {Kempkes}},\ and\ \bibinfo {author} {\bibfnamefont {C.~M.}\ \bibnamefont {Smith}},\ }\bibfield  {title} {\bibinfo {title} {Effects of disorder in the fibonacci quasicrystal},\ }\href {https://doi.org/10.1103/physrevb.104.144201} {\bibfield  {journal} {\bibinfo  {journal} {Physical Review B}\ }\textbf {\bibinfo {volume} {104}},\ \bibinfo {pages} {144201} (\bibinfo {year} {2021})}\BibitemShut {NoStop}%
\bibitem [{\citenamefont {Manna}\ \emph {et~al.}(2022{\natexlab{a}})\citenamefont {Manna}, \citenamefont {Nandy},\ and\ \citenamefont {Roy}}]{Manna2022b}%
  \BibitemOpen
  \bibfield  {author} {\bibinfo {author} {\bibfnamefont {S.}~\bibnamefont {Manna}}, \bibinfo {author} {\bibfnamefont {S.}~\bibnamefont {Nandy}},\ and\ \bibinfo {author} {\bibfnamefont {B.}~\bibnamefont {Roy}},\ }\bibfield  {title} {\bibinfo {title} {Higher-order topological phases on fractal lattices},\ }\href {https://doi.org/10.1103/PhysRevB.105.L201301} {\bibfield  {journal} {\bibinfo  {journal} {Phys. Rev. B}\ }\textbf {\bibinfo {volume} {105}},\ \bibinfo {pages} {L201301} (\bibinfo {year} {2022}{\natexlab{a}})}\BibitemShut {NoStop}%
\bibitem [{\citenamefont {Ivaki}\ \emph {et~al.}(2022)\citenamefont {Ivaki}, \citenamefont {Sahlberg}, \citenamefont {P{\"o}yh{\"o}nen},\ and\ \citenamefont {Ojanen}}]{Ivaki2022}%
  \BibitemOpen
  \bibfield  {author} {\bibinfo {author} {\bibfnamefont {M.~N.}\ \bibnamefont {Ivaki}}, \bibinfo {author} {\bibfnamefont {I.}~\bibnamefont {Sahlberg}}, \bibinfo {author} {\bibfnamefont {K.}~\bibnamefont {P{\"o}yh{\"o}nen}},\ and\ \bibinfo {author} {\bibfnamefont {T.}~\bibnamefont {Ojanen}},\ }\bibfield  {title} {\bibinfo {title} {Topological random fractals},\ }\href {https://doi.org/10.1038/s42005-022-01101-z} {\bibfield  {journal} {\bibinfo  {journal} {Communications Physics}\ }\textbf {\bibinfo {volume} {5}},\ \bibinfo {pages} {327} (\bibinfo {year} {2022})}\BibitemShut {NoStop}%
\bibitem [{\citenamefont {Li}\ \emph {et~al.}(2023)\citenamefont {Li}, \citenamefont {Sun}, \citenamefont {Mo}, \citenamefont {Ruan},\ and\ \citenamefont {Yang}}]{Li2023}%
  \BibitemOpen
  \bibfield  {author} {\bibinfo {author} {\bibfnamefont {J.}~\bibnamefont {Li}}, \bibinfo {author} {\bibfnamefont {Y.}~\bibnamefont {Sun}}, \bibinfo {author} {\bibfnamefont {Q.}~\bibnamefont {Mo}}, \bibinfo {author} {\bibfnamefont {Z.}~\bibnamefont {Ruan}},\ and\ \bibinfo {author} {\bibfnamefont {Z.}~\bibnamefont {Yang}},\ }\bibfield  {title} {\bibinfo {title} {Fractality-induced topological phase squeezing and devil’s staircase},\ }\href {https://doi.org/10.1103/physrevresearch.5.023189} {\bibfield  {journal} {\bibinfo  {journal} {Phys. Rev. Res.}\ }\textbf {\bibinfo {volume} {5}},\ \bibinfo {pages} {23189} (\bibinfo {year} {2023})}\BibitemShut {NoStop}%
\bibitem [{\citenamefont {Krebbekx}\ \emph {et~al.}(2023)\citenamefont {Krebbekx}, \citenamefont {Moustaj}, \citenamefont {Dajani},\ and\ \citenamefont {Morais~Smith}}]{Krebbekx2023}%
  \BibitemOpen
  \bibfield  {author} {\bibinfo {author} {\bibfnamefont {J.~P.~J.}\ \bibnamefont {Krebbekx}}, \bibinfo {author} {\bibfnamefont {A.}~\bibnamefont {Moustaj}}, \bibinfo {author} {\bibfnamefont {K.}~\bibnamefont {Dajani}},\ and\ \bibinfo {author} {\bibfnamefont {C.}~\bibnamefont {Morais~Smith}},\ }\bibfield  {title} {\bibinfo {title} {Multifractal properties of tribonacci chains},\ }\href {https://doi.org/10.1103/PhysRevB.108.104204} {\bibfield  {journal} {\bibinfo  {journal} {Phys. Rev. B}\ }\textbf {\bibinfo {volume} {108}},\ \bibinfo {pages} {104204} (\bibinfo {year} {2023})}\BibitemShut {NoStop}%
\bibitem [{\citenamefont {Moustaj}\ \emph {et~al.}(2023)\citenamefont {Moustaj}, \citenamefont {R\"{o}ntgen}, \citenamefont {Morfonios}, \citenamefont {Schmelcher},\ and\ \citenamefont {Morais~Smith}}]{Moustaj2023}%
  \BibitemOpen
  \bibfield  {author} {\bibinfo {author} {\bibfnamefont {A.}~\bibnamefont {Moustaj}}, \bibinfo {author} {\bibfnamefont {M.}~\bibnamefont {R\"{o}ntgen}}, \bibinfo {author} {\bibfnamefont {C.~V.}\ \bibnamefont {Morfonios}}, \bibinfo {author} {\bibfnamefont {P.}~\bibnamefont {Schmelcher}},\ and\ \bibinfo {author} {\bibfnamefont {C.}~\bibnamefont {Morais~Smith}},\ }\bibfield  {title} {\bibinfo {title} {Spectral properties of two coupled fibonacci chains},\ }\href {https://doi.org/10.1088/1367-2630/acf0e0} {\bibfield  {journal} {\bibinfo  {journal} {New Journal of Physics}\ }\textbf {\bibinfo {volume} {25}},\ \bibinfo {pages} {093019} (\bibinfo {year} {2023})}\BibitemShut {NoStop}%
\bibitem [{\citenamefont {Stålhammar}\ and\ \citenamefont {Morais~Smith}(2023)}]{Stlhammar2023}%
  \BibitemOpen
  \bibfield  {author} {\bibinfo {author} {\bibfnamefont {M.}~\bibnamefont {Stålhammar}}\ and\ \bibinfo {author} {\bibfnamefont {C.}~\bibnamefont {Morais~Smith}},\ }\bibfield  {title} {\bibinfo {title} {Fractal nodal band structures},\ }\href {https://doi.org/10.1103/physrevresearch.5.043043} {\bibfield  {journal} {\bibinfo  {journal} {Physical Review Research}\ }\textbf {\bibinfo {volume} {5}},\ \bibinfo {pages} {043043} (\bibinfo {year} {2023})}\BibitemShut {NoStop}%
\bibitem [{\citenamefont {Eek}\ \emph {et~al.}(2024)\citenamefont {Eek}, \citenamefont {Osseweijer},\ and\ \citenamefont {Smith}}]{Eek2024}%
  \BibitemOpen
  \bibfield  {author} {\bibinfo {author} {\bibfnamefont {L.}~\bibnamefont {Eek}}, \bibinfo {author} {\bibfnamefont {Z.~F.}\ \bibnamefont {Osseweijer}},\ and\ \bibinfo {author} {\bibfnamefont {C.~M.}\ \bibnamefont {Smith}},\ }\href {https://doi.org/10.48550/ARXIV.2411.12341} {\bibinfo {title} {Fractality-induced topology}} (\bibinfo {year} {2024})\BibitemShut {NoStop}%
\bibitem [{\citenamefont {Osseweijer}\ \emph {et~al.}(2024)\citenamefont {Osseweijer}, \citenamefont {Eek}, \citenamefont {Moustaj}, \citenamefont {Fremling},\ and\ \citenamefont {Morais~Smith}}]{Osseweijer2024}%
  \BibitemOpen
  \bibfield  {author} {\bibinfo {author} {\bibfnamefont {Z.~F.}\ \bibnamefont {Osseweijer}}, \bibinfo {author} {\bibfnamefont {L.}~\bibnamefont {Eek}}, \bibinfo {author} {\bibfnamefont {A.}~\bibnamefont {Moustaj}}, \bibinfo {author} {\bibfnamefont {M.}~\bibnamefont {Fremling}},\ and\ \bibinfo {author} {\bibfnamefont {C.}~\bibnamefont {Morais~Smith}},\ }\bibfield  {title} {\bibinfo {title} {{Haldane} model on the {S}ierpi\ifmmode \acute{n}\else \'{n}\fi{}ski gasket},\ }\href {https://doi.org/10.1103/PhysRevB.110.245405} {\bibfield  {journal} {\bibinfo  {journal} {Phys. Rev. B}\ }\textbf {\bibinfo {volume} {110}},\ \bibinfo {pages} {245405} (\bibinfo {year} {2024})}\BibitemShut {NoStop}%
\bibitem [{\citenamefont {Canyellas}\ \emph {et~al.}(2024)\citenamefont {Canyellas}, \citenamefont {Liu}, \citenamefont {Arouca}, \citenamefont {Eek}, \citenamefont {Wang}, \citenamefont {Yin}, \citenamefont {Guan}, \citenamefont {Li}, \citenamefont {Wang}, \citenamefont {Zheng}, \citenamefont {Liu}, \citenamefont {Jia},\ and\ \citenamefont {Morais~Smith}}]{Canyellas2024}%
  \BibitemOpen
  \bibfield  {author} {\bibinfo {author} {\bibfnamefont {R.}~\bibnamefont {Canyellas}}, \bibinfo {author} {\bibfnamefont {C.}~\bibnamefont {Liu}}, \bibinfo {author} {\bibfnamefont {R.}~\bibnamefont {Arouca}}, \bibinfo {author} {\bibfnamefont {L.}~\bibnamefont {Eek}}, \bibinfo {author} {\bibfnamefont {G.}~\bibnamefont {Wang}}, \bibinfo {author} {\bibfnamefont {Y.}~\bibnamefont {Yin}}, \bibinfo {author} {\bibfnamefont {D.}~\bibnamefont {Guan}}, \bibinfo {author} {\bibfnamefont {Y.}~\bibnamefont {Li}}, \bibinfo {author} {\bibfnamefont {S.}~\bibnamefont {Wang}}, \bibinfo {author} {\bibfnamefont {H.}~\bibnamefont {Zheng}}, \bibinfo {author} {\bibfnamefont {C.}~\bibnamefont {Liu}}, \bibinfo {author} {\bibfnamefont {J.}~\bibnamefont {Jia}},\ and\ \bibinfo {author} {\bibfnamefont {C.}~\bibnamefont {Morais~Smith}},\ }\bibfield  {title} {\bibinfo {title} {Topological edge and corner states in bismuth fractal nanostructures},\ }\href {https://doi.org/10.1038/s41567-024-02551-8} {\bibfield  {journal} {\bibinfo  {journal}
  {Nature Physics}\ }\textbf {\bibinfo {volume} {20}},\ \bibinfo {pages} {1421–1428} (\bibinfo {year} {2024})}\BibitemShut {NoStop}%
\bibitem [{\citenamefont {Manna}\ \emph {et~al.}(2020)\citenamefont {Manna}, \citenamefont {Pal}, \citenamefont {Wang},\ and\ \citenamefont {Nielsen}}]{PhysRevResearch.2.023401}%
  \BibitemOpen
  \bibfield  {author} {\bibinfo {author} {\bibfnamefont {S.}~\bibnamefont {Manna}}, \bibinfo {author} {\bibfnamefont {B.}~\bibnamefont {Pal}}, \bibinfo {author} {\bibfnamefont {W.}~\bibnamefont {Wang}},\ and\ \bibinfo {author} {\bibfnamefont {A.~E.~B.}\ \bibnamefont {Nielsen}},\ }\bibfield  {title} {\bibinfo {title} {Anyons and fractional quantum hall effect in fractal dimensions},\ }\href {https://doi.org/10.1103/PhysRevResearch.2.023401} {\bibfield  {journal} {\bibinfo  {journal} {Phys. Rev. Res.}\ }\textbf {\bibinfo {volume} {2}},\ \bibinfo {pages} {023401} (\bibinfo {year} {2020})}\BibitemShut {NoStop}%
\bibitem [{\citenamefont {Manna}\ \emph {et~al.}(2022{\natexlab{b}})\citenamefont {Manna}, \citenamefont {Duncan}, \citenamefont {Weidner}, \citenamefont {Sherson},\ and\ \citenamefont {Nielsen}}]{PhysRevA.105.L021302}%
  \BibitemOpen
  \bibfield  {author} {\bibinfo {author} {\bibfnamefont {S.}~\bibnamefont {Manna}}, \bibinfo {author} {\bibfnamefont {C.~W.}\ \bibnamefont {Duncan}}, \bibinfo {author} {\bibfnamefont {C.~A.}\ \bibnamefont {Weidner}}, \bibinfo {author} {\bibfnamefont {J.~F.}\ \bibnamefont {Sherson}},\ and\ \bibinfo {author} {\bibfnamefont {A.~E.~B.}\ \bibnamefont {Nielsen}},\ }\bibfield  {title} {\bibinfo {title} {Anyon braiding on a fractal lattice with a local {Hamiltonian}},\ }\href {https://doi.org/10.1103/PhysRevA.105.L021302} {\bibfield  {journal} {\bibinfo  {journal} {Phys. Rev. A}\ }\textbf {\bibinfo {volume} {105}},\ \bibinfo {pages} {L021302} (\bibinfo {year} {2022}{\natexlab{b}})}\BibitemShut {NoStop}%
\bibitem [{\citenamefont {Iliasov}\ \emph {et~al.}(2024)\citenamefont {Iliasov}, \citenamefont {Canyellas}, \citenamefont {Katsnelson},\ and\ \citenamefont {Bagrov}}]{Iliasov2024}%
  \BibitemOpen
  \bibfield  {author} {\bibinfo {author} {\bibfnamefont {A.~A.}\ \bibnamefont {Iliasov}}, \bibinfo {author} {\bibfnamefont {R.}~\bibnamefont {Canyellas}}, \bibinfo {author} {\bibfnamefont {M.~I.}\ \bibnamefont {Katsnelson}},\ and\ \bibinfo {author} {\bibfnamefont {A.~A.}\ \bibnamefont {Bagrov}},\ }\href {https://arxiv.org/abs/2310.11497} {\bibinfo {title} {Strong enhancement of superconductivity on finitely ramified fractal lattices}} (\bibinfo {year} {2024}),\ \Eprint {https://arxiv.org/abs/2310.11497} {arXiv:2310.11497 [cond-mat.supr-con]} \BibitemShut {NoStop}%
\bibitem [{\citenamefont {Caracanhas}\ \emph {et~al.}(2017)\citenamefont {Caracanhas}, \citenamefont {Schreck},\ and\ \citenamefont {Smith}}]{Caracanhas2017}%
  \BibitemOpen
  \bibfield  {author} {\bibinfo {author} {\bibfnamefont {M.~A.}\ \bibnamefont {Caracanhas}}, \bibinfo {author} {\bibfnamefont {F.}~\bibnamefont {Schreck}},\ and\ \bibinfo {author} {\bibfnamefont {C.~M.}\ \bibnamefont {Smith}},\ }\bibfield  {title} {\bibinfo {title} {Fermi–{Bose} mixture in mixed dimensions},\ }\href {https://doi.org/10.1088/1367-2630/aa8e56} {\bibfield  {journal} {\bibinfo  {journal} {New Journal of Physics}\ }\textbf {\bibinfo {volume} {19}},\ \bibinfo {pages} {115011} (\bibinfo {year} {2017})}\BibitemShut {NoStop}%
\bibitem [{\citenamefont {Conte}\ \emph {et~al.}(2024)\citenamefont {Conte}, \citenamefont {Zampronio}, \citenamefont {R\"{o}ntgen},\ and\ \citenamefont {Smith}}]{Conte2024}%
  \BibitemOpen
  \bibfield  {author} {\bibinfo {author} {\bibfnamefont {M.}~\bibnamefont {Conte}}, \bibinfo {author} {\bibfnamefont {V.}~\bibnamefont {Zampronio}}, \bibinfo {author} {\bibfnamefont {M.}~\bibnamefont {R\"{o}ntgen}},\ and\ \bibinfo {author} {\bibfnamefont {C.~M.}\ \bibnamefont {Smith}},\ }\bibfield  {title} {\bibinfo {title} {The fractal-lattice {Hubbard} model},\ }\href {https://doi.org/10.22331/q-2024-09-11-1469} {\bibfield  {journal} {\bibinfo  {journal} {Quantum}\ }\textbf {\bibinfo {volume} {8}},\ \bibinfo {pages} {1469} (\bibinfo {year} {2024})}\BibitemShut {NoStop}%
\bibitem [{\citenamefont {Koch}\ and\ \citenamefont {Posazhennikova}(2024)}]{Koch2024}%
  \BibitemOpen
  \bibfield  {author} {\bibinfo {author} {\bibfnamefont {G.}~\bibnamefont {Koch}}\ and\ \bibinfo {author} {\bibfnamefont {A.}~\bibnamefont {Posazhennikova}},\ }\bibfield  {title} {\bibinfo {title} {Loop current states and their stability in small fractal lattices of {Bose}-{Einstein} condensates},\ }\href {https://doi.org/10.1103/PhysRevA.110.033301} {\bibfield  {journal} {\bibinfo  {journal} {Phys. Rev. A}\ }\textbf {\bibinfo {volume} {110}},\ \bibinfo {pages} {033301} (\bibinfo {year} {2024})}\BibitemShut {NoStop}%
\bibitem [{\citenamefont {Zhang}\ \emph {et~al.}(2022)\citenamefont {Zhang}, \citenamefont {Yuan}, \citenamefont {Sun}, \citenamefont {Sun},\ and\ \citenamefont {Zhang}}]{Zhang_2022}%
  \BibitemOpen
  \bibfield  {author} {\bibinfo {author} {\bibfnamefont {W.}~\bibnamefont {Zhang}}, \bibinfo {author} {\bibfnamefont {H.}~\bibnamefont {Yuan}}, \bibinfo {author} {\bibfnamefont {N.}~\bibnamefont {Sun}}, \bibinfo {author} {\bibfnamefont {H.}~\bibnamefont {Sun}},\ and\ \bibinfo {author} {\bibfnamefont {X.}~\bibnamefont {Zhang}},\ }\bibfield  {title} {\bibinfo {title} {Observation of novel topological states in hyperbolic lattices},\ }\href {https://doi.org/10.1038/s41467-022-30631-x} {\bibfield  {journal} {\bibinfo  {journal} {Nat. Commun.}\ }\textbf {\bibinfo {volume} {13}},\ \bibinfo {pages} {2937} (\bibinfo {year} {2022})}\BibitemShut {NoStop}%
\bibitem [{\citenamefont {Yu}\ \emph {et~al.}(2020)\citenamefont {Yu}, \citenamefont {Piao},\ and\ \citenamefont {Park}}]{Yu_2020}%
  \BibitemOpen
  \bibfield  {author} {\bibinfo {author} {\bibfnamefont {S.}~\bibnamefont {Yu}}, \bibinfo {author} {\bibfnamefont {X.}~\bibnamefont {Piao}},\ and\ \bibinfo {author} {\bibfnamefont {N.}~\bibnamefont {Park}},\ }\bibfield  {title} {\bibinfo {title} {Topological hyperbolic lattices},\ }\href {https://doi.org/10.1103/physrevlett.125.053901} {\bibfield  {journal} {\bibinfo  {journal} {Physical Review Letters}\ }\textbf {\bibinfo {volume} {125}},\ \bibinfo {pages} {53901} (\bibinfo {year} {2020})}\BibitemShut {NoStop}%
\bibitem [{\citenamefont {Chen}\ \emph {et~al.}(2023)\citenamefont {Chen}, \citenamefont {Guan}, \citenamefont {Lenggenhager}, \citenamefont {Maciejko}, \citenamefont {Boettcher},\ and\ \citenamefont {Bzdušek}}]{Chen_2023}%
  \BibitemOpen
  \bibfield  {author} {\bibinfo {author} {\bibfnamefont {A.}~\bibnamefont {Chen}}, \bibinfo {author} {\bibfnamefont {Y.}~\bibnamefont {Guan}}, \bibinfo {author} {\bibfnamefont {P.~M.}\ \bibnamefont {Lenggenhager}}, \bibinfo {author} {\bibfnamefont {J.}~\bibnamefont {Maciejko}}, \bibinfo {author} {\bibfnamefont {I.}~\bibnamefont {Boettcher}},\ and\ \bibinfo {author} {\bibfnamefont {T.}~\bibnamefont {Bzdušek}},\ }\bibfield  {title} {\bibinfo {title} {Symmetry and topology of hyperbolic {Haldane} models},\ }\href {https://doi.org/10.1103/physrevb.108.085114} {\bibfield  {journal} {\bibinfo  {journal} {Physical Review B}\ }\textbf {\bibinfo {volume} {108}},\ \bibinfo {pages} {85114} (\bibinfo {year} {2023})}\BibitemShut {NoStop}%
\bibitem [{\citenamefont {Tummuru}\ \emph {et~al.}(2024)\citenamefont {Tummuru}, \citenamefont {Chen}, \citenamefont {Lenggenhager}, \citenamefont {Neupert}, \citenamefont {Maciejko},\ and\ \citenamefont {Bzdušek}}]{Tummuru_2024}%
  \BibitemOpen
  \bibfield  {author} {\bibinfo {author} {\bibfnamefont {T.}~\bibnamefont {Tummuru}}, \bibinfo {author} {\bibfnamefont {A.}~\bibnamefont {Chen}}, \bibinfo {author} {\bibfnamefont {P.~M.}\ \bibnamefont {Lenggenhager}}, \bibinfo {author} {\bibfnamefont {T.}~\bibnamefont {Neupert}}, \bibinfo {author} {\bibfnamefont {J.}~\bibnamefont {Maciejko}},\ and\ \bibinfo {author} {\bibfnamefont {T.}~\bibnamefont {Bzdušek}},\ }\bibfield  {title} {\bibinfo {title} {Hyperbolic non-abelian semimetal},\ }\href {https://doi.org/10.1103/physrevlett.132.206601} {\bibfield  {journal} {\bibinfo  {journal} {Physical Review Letters}\ }\textbf {\bibinfo {volume} {132}},\ \bibinfo {pages} {206601} (\bibinfo {year} {2024})}\BibitemShut {NoStop}%
\bibitem [{\citenamefont {Sun}\ \emph {et~al.}(2024)\citenamefont {Sun}, \citenamefont {Chen}, \citenamefont {Bzdusek},\ and\ \citenamefont {Maciejko}}]{Sun_2024}%
  \BibitemOpen
  \bibfield  {author} {\bibinfo {author} {\bibfnamefont {C.}~\bibnamefont {Sun}}, \bibinfo {author} {\bibfnamefont {A.}~\bibnamefont {Chen}}, \bibinfo {author} {\bibfnamefont {T.}~\bibnamefont {Bzdusek}},\ and\ \bibinfo {author} {\bibfnamefont {J.}~\bibnamefont {Maciejko}},\ }\bibfield  {title} {\bibinfo {title} {Topological linear response of hyperbolic chern insulators},\ }\href {https://doi.org/10.21468/scipostphys.17.5.124} {\bibfield  {journal} {\bibinfo  {journal} {SciPost Physics}\ }\textbf {\bibinfo {volume} {17}},\ \bibinfo {pages} {124} (\bibinfo {year} {2024})}\BibitemShut {NoStop}%
\bibitem [{\citenamefont {Götz}\ \emph {et~al.}(2024)\citenamefont {Götz}, \citenamefont {Rein}, \citenamefont {Inácio},\ and\ \citenamefont {Assaad}}]{G_tz_2024}%
  \BibitemOpen
  \bibfield  {author} {\bibinfo {author} {\bibfnamefont {A.}~\bibnamefont {Götz}}, \bibinfo {author} {\bibfnamefont {G.}~\bibnamefont {Rein}}, \bibinfo {author} {\bibfnamefont {J.~C.}\ \bibnamefont {Inácio}},\ and\ \bibinfo {author} {\bibfnamefont {F.~F.}\ \bibnamefont {Assaad}},\ }\bibfield  {title} {\bibinfo {title} {{Hubbard} and heisenberg models on hyperbolic lattices: Metal-insulator transitions, global antiferromagnetism, and enhanced boundary fluctuations},\ }\href {https://doi.org/10.1103/physrevb.110.235105} {\bibfield  {journal} {\bibinfo  {journal} {Physical Review B}\ }\textbf {\bibinfo {volume} {110}},\ \bibinfo {pages} {235105} (\bibinfo {year} {2024})}\BibitemShut {NoStop}%
\bibitem [{\citenamefont {Gluscevich}\ \emph {et~al.}(2025)\citenamefont {Gluscevich}, \citenamefont {Samanta}, \citenamefont {Manna},\ and\ \citenamefont {Roy}}]{PhysRevB.111.L121108}%
  \BibitemOpen
  \bibfield  {author} {\bibinfo {author} {\bibfnamefont {N.}~\bibnamefont {Gluscevich}}, \bibinfo {author} {\bibfnamefont {A.}~\bibnamefont {Samanta}}, \bibinfo {author} {\bibfnamefont {S.}~\bibnamefont {Manna}},\ and\ \bibinfo {author} {\bibfnamefont {B.}~\bibnamefont {Roy}},\ }\bibfield  {title} {\bibinfo {title} {Dynamic mass generation on two-dimensional electronic hyperbolic lattices},\ }\href {https://doi.org/10.1103/PhysRevB.111.L121108} {\bibfield  {journal} {\bibinfo  {journal} {Phys. Rev. B}\ }\textbf {\bibinfo {volume} {111}},\ \bibinfo {pages} {L121108} (\bibinfo {year} {2025})}\BibitemShut {NoStop}%
\bibitem [{\citenamefont {Lenggenhager}\ \emph {et~al.}(2024)\citenamefont {Lenggenhager}, \citenamefont {Dey}, \citenamefont {Bzdušek},\ and\ \citenamefont {Maciejko}}]{lenggenhager2024hyperbolicspinliquids}%
  \BibitemOpen
  \bibfield  {author} {\bibinfo {author} {\bibfnamefont {P.~M.}\ \bibnamefont {Lenggenhager}}, \bibinfo {author} {\bibfnamefont {S.}~\bibnamefont {Dey}}, \bibinfo {author} {\bibfnamefont {T.}~\bibnamefont {Bzdušek}},\ and\ \bibinfo {author} {\bibfnamefont {J.}~\bibnamefont {Maciejko}},\ }\href {https://arxiv.org/abs/2407.09601} {\bibinfo {title} {Hyperbolic spin liquids}} (\bibinfo {year} {2024}),\ \Eprint {https://arxiv.org/abs/2407.09601} {arXiv:2407.09601 [cond-mat.str-el]} \BibitemShut {NoStop}%
\bibitem [{\citenamefont {Zhu}\ \emph {et~al.}(2021)\citenamefont {Zhu}, \citenamefont {Guo}, \citenamefont {Breuckmann}, \citenamefont {Guo},\ and\ \citenamefont {Feng}}]{Zhu_2021}%
  \BibitemOpen
  \bibfield  {author} {\bibinfo {author} {\bibfnamefont {X.}~\bibnamefont {Zhu}}, \bibinfo {author} {\bibfnamefont {J.}~\bibnamefont {Guo}}, \bibinfo {author} {\bibfnamefont {N.~P.}\ \bibnamefont {Breuckmann}}, \bibinfo {author} {\bibfnamefont {H.}~\bibnamefont {Guo}},\ and\ \bibinfo {author} {\bibfnamefont {S.}~\bibnamefont {Feng}},\ }\bibfield  {title} {\bibinfo {title} {Quantum phase transitions of interacting bosons on hyperbolic lattices},\ }\href {https://doi.org/10.1088/1361-648x/ac0a1a} {\bibfield  {journal} {\bibinfo  {journal} {Journal of Physics: Condensed Matter}\ }\textbf {\bibinfo {volume} {33}},\ \bibinfo {pages} {335602} (\bibinfo {year} {2021})}\BibitemShut {NoStop}%
\bibitem [{\citenamefont {Lemm}\ and\ \citenamefont {Siebert}(2022)}]{Lemm_2022}%
  \BibitemOpen
  \bibfield  {author} {\bibinfo {author} {\bibfnamefont {M.}~\bibnamefont {Lemm}}\ and\ \bibinfo {author} {\bibfnamefont {O.}~\bibnamefont {Siebert}},\ }\bibfield  {title} {\bibinfo {title} {{Bose}–{Einstein} condensation on hyperbolic spaces},\ }\href {https://doi.org/10.1063/5.0088383} {\bibfield  {journal} {\bibinfo  {journal} {Journal of Mathematical Physics}\ }\textbf {\bibinfo {volume} {63}},\ \bibinfo {pages} {081903} (\bibinfo {year} {2022})}\BibitemShut {NoStop}%
\bibitem [{\citenamefont {Tononi}\ and\ \citenamefont {Salasnich}(2019)}]{Tononi_2019}%
  \BibitemOpen
  \bibfield  {author} {\bibinfo {author} {\bibfnamefont {A.}~\bibnamefont {Tononi}}\ and\ \bibinfo {author} {\bibfnamefont {L.}~\bibnamefont {Salasnich}},\ }\bibfield  {title} {\bibinfo {title} {{Bose}-{Einstein} condensation on the surface of a sphere},\ }\href {https://doi.org/10.1103/physrevlett.123.160403} {\bibfield  {journal} {\bibinfo  {journal} {Physical Review Letters}\ }\textbf {\bibinfo {volume} {123}},\ \bibinfo {pages} {160403} (\bibinfo {year} {2019})}\BibitemShut {NoStop}%
\bibitem [{\citenamefont {Diniz}\ \emph {et~al.}(2020)\citenamefont {Diniz}, \citenamefont {Oliveira}, \citenamefont {Lima},\ and\ \citenamefont {Henn}}]{Diniz2020}%
  \BibitemOpen
  \bibfield  {author} {\bibinfo {author} {\bibfnamefont {P.~C.}\ \bibnamefont {Diniz}}, \bibinfo {author} {\bibfnamefont {E.~A.~B.}\ \bibnamefont {Oliveira}}, \bibinfo {author} {\bibfnamefont {A.~R.~P.}\ \bibnamefont {Lima}},\ and\ \bibinfo {author} {\bibfnamefont {E.~A.~L.}\ \bibnamefont {Henn}},\ }\bibfield  {title} {\bibinfo {title} {Ground state and collective excitations of a dipolar {Bose}-{Einstein} condensate in a bubble trap},\ }\href {https://doi.org/10.1038/s41598-020-61657-0} {\bibfield  {journal} {\bibinfo  {journal} {Scientific Reports}\ }\textbf {\bibinfo {volume} {10}},\ \bibinfo {pages} {4831} (\bibinfo {year} {2020})}\BibitemShut {NoStop}%
\bibitem [{\citenamefont {Tononi}\ \emph {et~al.}(2020)\citenamefont {Tononi}, \citenamefont {Cinti},\ and\ \citenamefont {Salasnich}}]{Tononi2020}%
  \BibitemOpen
  \bibfield  {author} {\bibinfo {author} {\bibfnamefont {A.}~\bibnamefont {Tononi}}, \bibinfo {author} {\bibfnamefont {F.}~\bibnamefont {Cinti}},\ and\ \bibinfo {author} {\bibfnamefont {L.}~\bibnamefont {Salasnich}},\ }\bibfield  {title} {\bibinfo {title} {Quantum bubbles in microgravity},\ }\href {https://doi.org/10.1103/physrevlett.125.010402} {\bibfield  {journal} {\bibinfo  {journal} {Physical Review Letters}\ }\textbf {\bibinfo {volume} {125}},\ \bibinfo {pages} {010402} (\bibinfo {year} {2020})}\BibitemShut {NoStop}%
\bibitem [{\citenamefont {Salasnich}(2022)}]{Salasnich2022}%
  \BibitemOpen
  \bibfield  {author} {\bibinfo {author} {\bibfnamefont {L.}~\bibnamefont {Salasnich}},\ }\bibfield  {title} {\bibinfo {title} {{Bose}-{Einstein} condensate in an elliptical waveguide},\ }\href {https://doi.org/10.21468/scipostphyscore.5.1.015} {\bibfield  {journal} {\bibinfo  {journal} {SciPost Physics Core}\ }\textbf {\bibinfo {volume} {5}},\ \bibinfo {pages} {015} (\bibinfo {year} {2022})}\BibitemShut {NoStop}%
\bibitem [{\citenamefont {Furutani}\ and\ \citenamefont {Salasnich}(2022)}]{Furutani2022}%
  \BibitemOpen
  \bibfield  {author} {\bibinfo {author} {\bibfnamefont {K.}~\bibnamefont {Furutani}}\ and\ \bibinfo {author} {\bibfnamefont {L.}~\bibnamefont {Salasnich}},\ }\bibfield  {title} {\bibinfo {title} {Superfluid properties of bright solitons in a ring},\ }\href {https://doi.org/10.1103/physreva.105.033320} {\bibfield  {journal} {\bibinfo  {journal} {Physical Review A}\ }\textbf {\bibinfo {volume} {105}},\ \bibinfo {pages} {033320} (\bibinfo {year} {2022})}\BibitemShut {NoStop}%
\bibitem [{\citenamefont {Tononi}\ \emph {et~al.}(2022)\citenamefont {Tononi}, \citenamefont {Pelster},\ and\ \citenamefont {Salasnich}}]{Tononi2022}%
  \BibitemOpen
  \bibfield  {author} {\bibinfo {author} {\bibfnamefont {A.}~\bibnamefont {Tononi}}, \bibinfo {author} {\bibfnamefont {A.}~\bibnamefont {Pelster}},\ and\ \bibinfo {author} {\bibfnamefont {L.}~\bibnamefont {Salasnich}},\ }\bibfield  {title} {\bibinfo {title} {Topological superfluid transition in bubble-trapped condensates},\ }\href {https://doi.org/10.1103/physrevresearch.4.013122} {\bibfield  {journal} {\bibinfo  {journal} {Physical Review Research}\ }\textbf {\bibinfo {volume} {4}},\ \bibinfo {pages} {013122} (\bibinfo {year} {2022})}\BibitemShut {NoStop}%
\bibitem [{\citenamefont {de~Oliveira}\ and\ \citenamefont {Móller}(2025)}]{sheilla2025}%
  \BibitemOpen
  \bibfield  {author} {\bibinfo {author} {\bibfnamefont {S.~M.}\ \bibnamefont {de~Oliveira}}\ and\ \bibinfo {author} {\bibfnamefont {N.~S.}\ \bibnamefont {Móller}},\ }\href {https://arxiv.org/abs/2504.20751} {\bibinfo {title} {Geometric potential for a {Bose}-{Einstein} condensate on a curved surface}} (\bibinfo {year} {2025}),\ \Eprint {https://arxiv.org/abs/2504.20751} {arXiv:2504.20751 [cond-mat.quant-gas]} \BibitemShut {NoStop}%
\bibitem [{\citenamefont {Tononi}\ and\ \citenamefont {Salasnich}(2023)}]{Tononi_2023}%
  \BibitemOpen
  \bibfield  {author} {\bibinfo {author} {\bibfnamefont {A.}~\bibnamefont {Tononi}}\ and\ \bibinfo {author} {\bibfnamefont {L.}~\bibnamefont {Salasnich}},\ }\bibfield  {title} {\bibinfo {title} {Low-dimensional quantum gases in curved geometries},\ }\href {https://doi.org/10.1038/s42254-023-00591-2} {\bibfield  {journal} {\bibinfo  {journal} {Nature Reviews Physics}\ }\textbf {\bibinfo {volume} {5}},\ \bibinfo {pages} {398–406} (\bibinfo {year} {2023})}\BibitemShut {NoStop}%
\bibitem [{\citenamefont {Nikolaieva}\ \emph {et~al.}(2023)\citenamefont {Nikolaieva}, \citenamefont {Salasnich},\ and\ \citenamefont {Yakimenko}}]{Nikolaieva2023}%
  \BibitemOpen
  \bibfield  {author} {\bibinfo {author} {\bibfnamefont {Y.}~\bibnamefont {Nikolaieva}}, \bibinfo {author} {\bibfnamefont {L.}~\bibnamefont {Salasnich}},\ and\ \bibinfo {author} {\bibfnamefont {A.}~\bibnamefont {Yakimenko}},\ }\bibfield  {title} {\bibinfo {title} {Engineering phase and density of {Bose}–{Einstein} condensates in curved waveguides with toroidal topology},\ }\href {https://doi.org/10.1088/1367-2630/acf783} {\bibfield  {journal} {\bibinfo  {journal} {New Journal of Physics}\ }\textbf {\bibinfo {volume} {25}},\ \bibinfo {pages} {103003} (\bibinfo {year} {2023})}\BibitemShut {NoStop}%
\bibitem [{\citenamefont {Brito}\ \emph {et~al.}(2023)\citenamefont {Brito}, \citenamefont {Tomio},\ and\ \citenamefont {Gammal}}]{Brito2023}%
  \BibitemOpen
  \bibfield  {author} {\bibinfo {author} {\bibfnamefont {L.}~\bibnamefont {Brito}}, \bibinfo {author} {\bibfnamefont {L.}~\bibnamefont {Tomio}},\ and\ \bibinfo {author} {\bibfnamefont {A.}~\bibnamefont {Gammal}},\ }\bibfield  {title} {\bibinfo {title} {Faraday waves on a bubble-trapped {Bose}-{Einstein}-condensed binary mixture},\ }\href {https://doi.org/10.1103/PhysRevA.108.053315} {\bibfield  {journal} {\bibinfo  {journal} {Phys. Rev. A}\ }\textbf {\bibinfo {volume} {108}},\ \bibinfo {pages} {053315} (\bibinfo {year} {2023})}\BibitemShut {NoStop}%
\bibitem [{\citenamefont {Biral}\ \emph {et~al.}(2024)\citenamefont {Biral}, \citenamefont {Móller}, \citenamefont {Pelster},\ and\ \citenamefont {dos Santos}}]{Biral2024}%
  \BibitemOpen
  \bibfield  {author} {\bibinfo {author} {\bibfnamefont {E.~J.~P.}\ \bibnamefont {Biral}}, \bibinfo {author} {\bibfnamefont {N.~S.}\ \bibnamefont {Móller}}, \bibinfo {author} {\bibfnamefont {A.}~\bibnamefont {Pelster}},\ and\ \bibinfo {author} {\bibfnamefont {F.~E.~A.}\ \bibnamefont {dos Santos}},\ }\bibfield  {title} {\bibinfo {title} {{Bose}–{Einstein} condensates and the thin-shell limit in anisotropic bubble traps},\ }\href {https://doi.org/10.1088/1367-2630/ad1a29} {\bibfield  {journal} {\bibinfo  {journal} {New Journal of Physics}\ }\textbf {\bibinfo {volume} {26}},\ \bibinfo {pages} {013035} (\bibinfo {year} {2024})}\BibitemShut {NoStop}%
\bibitem [{\citenamefont {Tononi}\ \emph {et~al.}(2024{\natexlab{a}})\citenamefont {Tononi}, \citenamefont {Salasnich},\ and\ \citenamefont {Yakimenko}}]{Tononi2024quantumvortices}%
  \BibitemOpen
  \bibfield  {author} {\bibinfo {author} {\bibfnamefont {A.}~\bibnamefont {Tononi}}, \bibinfo {author} {\bibfnamefont {L.}~\bibnamefont {Salasnich}},\ and\ \bibinfo {author} {\bibfnamefont {A.}~\bibnamefont {Yakimenko}},\ }\bibfield  {title} {\bibinfo {title} {Quantum vortices in curved geometries},\ }\href {https://doi.org/10.1116/5.0211426} {\bibfield  {journal} {\bibinfo  {journal} {AVS Quantum Science}\ }\textbf {\bibinfo {volume} {6}},\ \bibinfo {pages} {030502} (\bibinfo {year} {2024}{\natexlab{a}})}\BibitemShut {NoStop}%
\bibitem [{\citenamefont {Tononi}\ and\ \citenamefont {Salasnich}(2024)}]{Tononi2024shellshaped}%
  \BibitemOpen
  \bibfield  {author} {\bibinfo {author} {\bibfnamefont {A.}~\bibnamefont {Tononi}}\ and\ \bibinfo {author} {\bibfnamefont {L.}~\bibnamefont {Salasnich}},\ }\bibfield  {title} {\bibinfo {title} {Shell-shaped atomic gases},\ }\href {https://doi.org/10.1016/j.physrep.2024.04.004} {\bibfield  {journal} {\bibinfo  {journal} {Physics Reports}\ }\textbf {\bibinfo {volume} {1072}},\ \bibinfo {pages} {1–48} (\bibinfo {year} {2024})}\BibitemShut {NoStop}%
\bibitem [{\citenamefont {Tononi}\ \emph {et~al.}(2024{\natexlab{b}})\citenamefont {Tononi}, \citenamefont {Astrakharchik},\ and\ \citenamefont {Petrov}}]{Tononi_2024gastosoliton}%
  \BibitemOpen
  \bibfield  {author} {\bibinfo {author} {\bibfnamefont {A.}~\bibnamefont {Tononi}}, \bibinfo {author} {\bibfnamefont {G.~E.}\ \bibnamefont {Astrakharchik}},\ and\ \bibinfo {author} {\bibfnamefont {D.~S.}\ \bibnamefont {Petrov}},\ }\bibfield  {title} {\bibinfo {title} {Gas-to-soliton transition of attractive bosons on a spherical surface},\ }\href {https://doi.org/10.1116/5.0190767} {\bibfield  {journal} {\bibinfo  {journal} {AVS Quantum Science}\ }\textbf {\bibinfo {volume} {6}},\ \bibinfo {pages} {023201} (\bibinfo {year} {2024}{\natexlab{b}})}\BibitemShut {NoStop}%
\bibitem [{\citenamefont {Tononi}\ and\ \citenamefont {Astrakharchik}(2025)}]{tononi2025boseeinsteincondensationaxiallysymmetricsurfaces}%
  \BibitemOpen
  \bibfield  {author} {\bibinfo {author} {\bibfnamefont {A.}~\bibnamefont {Tononi}}\ and\ \bibinfo {author} {\bibfnamefont {G.~E.}\ \bibnamefont {Astrakharchik}},\ }\href {https://arxiv.org/abs/2503.17835} {\bibinfo {title} {{Bose}-{Einstein} condensation on axially-symmetric surfaces}} (\bibinfo {year} {2025}),\ \Eprint {https://arxiv.org/abs/2503.17835} {arXiv:2503.17835 [cond-mat.quant-gas]} \BibitemShut {NoStop}%
\bibitem [{\citenamefont {Tononi}\ \emph {et~al.}(2025)\citenamefont {Tononi}, \citenamefont {Petrov},\ and\ \citenamefont {Lewenstein}}]{tononi2025dimerproblemsphericalsurface}%
  \BibitemOpen
  \bibfield  {author} {\bibinfo {author} {\bibfnamefont {A.}~\bibnamefont {Tononi}}, \bibinfo {author} {\bibfnamefont {D.~S.}\ \bibnamefont {Petrov}},\ and\ \bibinfo {author} {\bibfnamefont {M.}~\bibnamefont {Lewenstein}},\ }\href {https://arxiv.org/abs/2502.06724} {\bibinfo {title} {Dimer problem on a spherical surface}} (\bibinfo {year} {2025}),\ \Eprint {https://arxiv.org/abs/2502.06724} {arXiv:2502.06724 [cond-mat.quant-gas]} \BibitemShut {NoStop}%
\bibitem [{\citenamefont {Mosseri}\ and\ \citenamefont {Vidal}(2023)}]{Mosseri_2023}%
  \BibitemOpen
  \bibfield  {author} {\bibinfo {author} {\bibfnamefont {R.}~\bibnamefont {Mosseri}}\ and\ \bibinfo {author} {\bibfnamefont {J.}~\bibnamefont {Vidal}},\ }\bibfield  {title} {\bibinfo {title} {Density of states of tight-binding models in the hyperbolic plane},\ }\href {https://doi.org/10.1103/physrevb.108.035154} {\bibfield  {journal} {\bibinfo  {journal} {Physical Review B}\ }\textbf {\bibinfo {volume} {108}},\ \bibinfo {pages} {035154} (\bibinfo {year} {2023})}\BibitemShut {NoStop}%
\bibitem [{\citenamefont {Boettcher}\ \emph {et~al.}(2022)\citenamefont {Boettcher}, \citenamefont {Gorshkov}, \citenamefont {Koll\'ar}, \citenamefont {Maciejko}, \citenamefont {Rayan},\ and\ \citenamefont {Thomale}}]{PhysRevB.105.125118}%
  \BibitemOpen
  \bibfield  {author} {\bibinfo {author} {\bibfnamefont {I.}~\bibnamefont {Boettcher}}, \bibinfo {author} {\bibfnamefont {A.~V.}\ \bibnamefont {Gorshkov}}, \bibinfo {author} {\bibfnamefont {A.~J.}\ \bibnamefont {Koll\'ar}}, \bibinfo {author} {\bibfnamefont {J.}~\bibnamefont {Maciejko}}, \bibinfo {author} {\bibfnamefont {S.}~\bibnamefont {Rayan}},\ and\ \bibinfo {author} {\bibfnamefont {R.}~\bibnamefont {Thomale}},\ }\bibfield  {title} {\bibinfo {title} {Crystallography of hyperbolic lattices},\ }\href {https://doi.org/10.1103/PhysRevB.105.125118} {\bibfield  {journal} {\bibinfo  {journal} {Phys. Rev. B}\ }\textbf {\bibinfo {volume} {105}},\ \bibinfo {pages} {125118} (\bibinfo {year} {2022})}\BibitemShut {NoStop}%
\bibitem [{\citenamefont {Lenggenhager}\ \emph {et~al.}(2023)\citenamefont {Lenggenhager}, \citenamefont {Maciejko},\ and\ \citenamefont {Bzdušek}}]{Lenggenhager_2023}%
  \BibitemOpen
  \bibfield  {author} {\bibinfo {author} {\bibfnamefont {P.~M.}\ \bibnamefont {Lenggenhager}}, \bibinfo {author} {\bibfnamefont {J.}~\bibnamefont {Maciejko}},\ and\ \bibinfo {author} {\bibfnamefont {T.}~\bibnamefont {Bzdušek}},\ }\bibfield  {title} {\bibinfo {title} {Non-abelian hyperbolic band theory from supercells},\ }\href {https://doi.org/10.1103/physrevlett.131.226401} {\bibfield  {journal} {\bibinfo  {journal} {Physical Review Letters}\ }\textbf {\bibinfo {volume} {131}},\ \bibinfo {pages} {226401} (\bibinfo {year} {2023})}\BibitemShut {NoStop}%
\bibitem [{\citenamefont {Ramakumar}\ and\ \citenamefont {Das}(2005)}]{Ramakumar_2005}%
  \BibitemOpen
  \bibfield  {author} {\bibinfo {author} {\bibfnamefont {R.}~\bibnamefont {Ramakumar}}\ and\ \bibinfo {author} {\bibfnamefont {A.~N.}\ \bibnamefont {Das}},\ }\bibfield  {title} {\bibinfo {title} {{Bose}-{Einstein} condensation in tight-binding bands},\ }\href {https://doi.org/10.1103/physrevb.72.094301} {\bibfield  {journal} {\bibinfo  {journal} {Physical Review B}\ }\textbf {\bibinfo {volume} {72}},\ \bibinfo {pages} {094301} (\bibinfo {year} {2005})}\BibitemShut {NoStop}%
\bibitem [{\citenamefont {Burioni}\ \emph {et~al.}(2001)\citenamefont {Burioni}, \citenamefont {Cassi}, \citenamefont {Rasetti}, \citenamefont {Sodano},\ and\ \citenamefont {Vezzani}}]{Burioni_2001}%
  \BibitemOpen
  \bibfield  {author} {\bibinfo {author} {\bibfnamefont {R.}~\bibnamefont {Burioni}}, \bibinfo {author} {\bibfnamefont {D.}~\bibnamefont {Cassi}}, \bibinfo {author} {\bibfnamefont {M.}~\bibnamefont {Rasetti}}, \bibinfo {author} {\bibfnamefont {P.}~\bibnamefont {Sodano}},\ and\ \bibinfo {author} {\bibfnamefont {A.}~\bibnamefont {Vezzani}},\ }\bibfield  {title} {\bibinfo {title} {{Bose}-{Einstein} condensation on inhomogeneous complex networks},\ }\href {https://doi.org/10.1088/0953-4075/34/23/314} {\bibfield  {journal} {\bibinfo  {journal} {Journal of Physics B: Atomic, Molecular and Optical Physics}\ }\textbf {\bibinfo {volume} {34}},\ \bibinfo {pages} {4697} (\bibinfo {year} {2001})}\BibitemShut {NoStop}%
\bibitem [{\citenamefont {Buonsante}\ \emph {et~al.}(2002)\citenamefont {Buonsante}, \citenamefont {Burioni}, \citenamefont {Cassi},\ and\ \citenamefont {Vezzani}}]{PhysRevB.66.094207}%
  \BibitemOpen
  \bibfield  {author} {\bibinfo {author} {\bibfnamefont {P.}~\bibnamefont {Buonsante}}, \bibinfo {author} {\bibfnamefont {R.}~\bibnamefont {Burioni}}, \bibinfo {author} {\bibfnamefont {D.}~\bibnamefont {Cassi}},\ and\ \bibinfo {author} {\bibfnamefont {A.}~\bibnamefont {Vezzani}},\ }\bibfield  {title} {\bibinfo {title} {{Bose}-{Einstein} condensation on inhomogeneous networks: Mesoscopic aspects versus thermodynamic limit},\ }\href {https://doi.org/10.1103/PhysRevB.66.094207} {\bibfield  {journal} {\bibinfo  {journal} {Phys. Rev. B}\ }\textbf {\bibinfo {volume} {66}},\ \bibinfo {pages} {094207} (\bibinfo {year} {2002})}\BibitemShut {NoStop}%
\bibitem [{\citenamefont {Burioni}\ \emph {et~al.}(2000)\citenamefont {Burioni}, \citenamefont {Cassi}, \citenamefont {Meccoli}, \citenamefont {Rasetti}, \citenamefont {Regina}, \citenamefont {Sodano},\ and\ \citenamefont {Vezzani}}]{Burioni_2000}%
  \BibitemOpen
  \bibfield  {author} {\bibinfo {author} {\bibfnamefont {R.}~\bibnamefont {Burioni}}, \bibinfo {author} {\bibfnamefont {D.}~\bibnamefont {Cassi}}, \bibinfo {author} {\bibfnamefont {I.}~\bibnamefont {Meccoli}}, \bibinfo {author} {\bibfnamefont {M.}~\bibnamefont {Rasetti}}, \bibinfo {author} {\bibfnamefont {S.}~\bibnamefont {Regina}}, \bibinfo {author} {\bibfnamefont {P.}~\bibnamefont {Sodano}},\ and\ \bibinfo {author} {\bibfnamefont {A.}~\bibnamefont {Vezzani}},\ }\bibfield  {title} {\bibinfo {title} {{Bose}-{Einstein} condensation in inhomogeneous josephson arrays},\ }\href {https://doi.org/10.1209/epl/i2000-00431-5} {\bibfield  {journal} {\bibinfo  {journal} {Europhysics Letters}\ }\textbf {\bibinfo {volume} {52}},\ \bibinfo {pages} {251} (\bibinfo {year} {2000})}\BibitemShut {NoStop}%
\bibitem [{\citenamefont {Halu}\ \emph {et~al.}(2012)\citenamefont {Halu}, \citenamefont {Ferretti}, \citenamefont {Vezzani},\ and\ \citenamefont {Bianconi}}]{Halu_2012}%
  \BibitemOpen
  \bibfield  {author} {\bibinfo {author} {\bibfnamefont {A.}~\bibnamefont {Halu}}, \bibinfo {author} {\bibfnamefont {L.}~\bibnamefont {Ferretti}}, \bibinfo {author} {\bibfnamefont {A.}~\bibnamefont {Vezzani}},\ and\ \bibinfo {author} {\bibfnamefont {G.}~\bibnamefont {Bianconi}},\ }\bibfield  {title} {\bibinfo {title} {Phase diagram of the {Bose}-{Hubbard} model on complex networks},\ }\href {https://doi.org/10.1209/0295-5075/99/18001} {\bibfield  {journal} {\bibinfo  {journal} {Europhysics Letters}\ }\textbf {\bibinfo {volume} {99}},\ \bibinfo {pages} {18001} (\bibinfo {year} {2012})}\BibitemShut {NoStop}%
\bibitem [{\citenamefont {Buonsante}\ \emph {et~al.}(2004{\natexlab{a}})\citenamefont {Buonsante}, \citenamefont {Penna},\ and\ \citenamefont {Vezzani}}]{PhysRevB.70.184520}%
  \BibitemOpen
  \bibfield  {author} {\bibinfo {author} {\bibfnamefont {P.}~\bibnamefont {Buonsante}}, \bibinfo {author} {\bibfnamefont {V.}~\bibnamefont {Penna}},\ and\ \bibinfo {author} {\bibfnamefont {A.}~\bibnamefont {Vezzani}},\ }\bibfield  {title} {\bibinfo {title} {Strong-coupling expansions for the topologically inhomogeneous {Bose}-{Hubbard} model},\ }\href {https://doi.org/10.1103/PhysRevB.70.184520} {\bibfield  {journal} {\bibinfo  {journal} {Phys. Rev. B}\ }\textbf {\bibinfo {volume} {70}},\ \bibinfo {pages} {184520} (\bibinfo {year} {2004}{\natexlab{a}})}\BibitemShut {NoStop}%
\bibitem [{\citenamefont {Buonsante}\ \emph {et~al.}(2004{\natexlab{b}})\citenamefont {Buonsante}, \citenamefont {Burioni}, \citenamefont {Cassi}, \citenamefont {Penna},\ and\ \citenamefont {Vezzani}}]{PhysRevB.70.224510}%
  \BibitemOpen
  \bibfield  {author} {\bibinfo {author} {\bibfnamefont {P.}~\bibnamefont {Buonsante}}, \bibinfo {author} {\bibfnamefont {R.}~\bibnamefont {Burioni}}, \bibinfo {author} {\bibfnamefont {D.}~\bibnamefont {Cassi}}, \bibinfo {author} {\bibfnamefont {V.}~\bibnamefont {Penna}},\ and\ \bibinfo {author} {\bibfnamefont {A.}~\bibnamefont {Vezzani}},\ }\bibfield  {title} {\bibinfo {title} {Topology-induced confined superfluidity in inhomogeneous arrays},\ }\href {https://doi.org/10.1103/PhysRevB.70.224510} {\bibfield  {journal} {\bibinfo  {journal} {Phys. Rev. B}\ }\textbf {\bibinfo {volume} {70}},\ \bibinfo {pages} {224510} (\bibinfo {year} {2004}{\natexlab{b}})}\BibitemShut {NoStop}%
\bibitem [{\citenamefont {Cheng}\ \emph {et~al.}(2021)\citenamefont {Cheng}, \citenamefont {Wang}, \citenamefont {Wang},\ and\ \citenamefont {Zong}}]{sym13020300}%
  \BibitemOpen
  \bibfield  {author} {\bibinfo {author} {\bibfnamefont {R.}~\bibnamefont {Cheng}}, \bibinfo {author} {\bibfnamefont {Q.-Y.}\ \bibnamefont {Wang}}, \bibinfo {author} {\bibfnamefont {Y.-L.}\ \bibnamefont {Wang}},\ and\ \bibinfo {author} {\bibfnamefont {H.-S.}\ \bibnamefont {Zong}},\ }\bibfield  {title} {\bibinfo {title} {Finite-size effects with boundary conditions on {Bose}-{Einstein} condensation},\ }\href {https://doi.org/10.3390/sym13020300} {\bibfield  {journal} {\bibinfo  {journal} {Symmetry}\ }\textbf {\bibinfo {volume} {13}},\ \bibinfo {pages} {300} (\bibinfo {year} {2021})}\BibitemShut {NoStop}%
\bibitem [{\citenamefont {Greenspoon}\ and\ \citenamefont {Pathria}(1974)}]{PhysRevA.9.2103}%
  \BibitemOpen
  \bibfield  {author} {\bibinfo {author} {\bibfnamefont {S.}~\bibnamefont {Greenspoon}}\ and\ \bibinfo {author} {\bibfnamefont {R.~K.}\ \bibnamefont {Pathria}},\ }\bibfield  {title} {\bibinfo {title} {{Bose}-{Einstein} condensation in finite noninteracting systems: A new law of corresponding states},\ }\href {https://doi.org/10.1103/PhysRevA.9.2103} {\bibfield  {journal} {\bibinfo  {journal} {Phys. Rev. A}\ }\textbf {\bibinfo {volume} {9}},\ \bibinfo {pages} {2103} (\bibinfo {year} {1974})}\BibitemShut {NoStop}%
\bibitem [{\citenamefont {Akaturk}\ and\ \citenamefont {Hen}(2024)}]{Akaturk}%
  \BibitemOpen
  \bibfield  {author} {\bibinfo {author} {\bibfnamefont {E.}~\bibnamefont {Akaturk}}\ and\ \bibinfo {author} {\bibfnamefont {I.}~\bibnamefont {Hen}},\ }\bibfield  {title} {\bibinfo {title} {Quantum monte carlo algorithm for {Bose}-{Hubbard} models on arbitrary graphs},\ }\href {https://doi.org/10.1103/PhysRevB.109.134519} {\bibfield  {journal} {\bibinfo  {journal} {Phys. Rev. B}\ }\textbf {\bibinfo {volume} {109}},\ \bibinfo {pages} {134519} (\bibinfo {year} {2024})}\BibitemShut {NoStop}%
\bibitem [{\citenamefont {Ghadimi}\ \emph {et~al.}(2020)\citenamefont {Ghadimi}, \citenamefont {Sugimoto},\ and\ \citenamefont {Tohyama}}]{Ghadimi}%
  \BibitemOpen
  \bibfield  {author} {\bibinfo {author} {\bibfnamefont {R.}~\bibnamefont {Ghadimi}}, \bibinfo {author} {\bibfnamefont {T.}~\bibnamefont {Sugimoto}},\ and\ \bibinfo {author} {\bibfnamefont {T.}~\bibnamefont {Tohyama}},\ }\bibfield  {title} {\bibinfo {title} {Mean-field study of the {Bose}-{Hubbard} model in the {Penrose} lattice},\ }\href {https://doi.org/10.1103/PhysRevB.102.224201} {\bibfield  {journal} {\bibinfo  {journal} {Phys. Rev. B}\ }\textbf {\bibinfo {volume} {102}},\ \bibinfo {pages} {224201} (\bibinfo {year} {2020})}\BibitemShut {NoStop}%
\bibitem [{\citenamefont {van Oosten}\ \emph {et~al.}(2001)\citenamefont {van Oosten}, \citenamefont {van~der Straten},\ and\ \citenamefont {Stoof}}]{van_Oosten_2001}%
  \BibitemOpen
  \bibfield  {author} {\bibinfo {author} {\bibfnamefont {D.}~\bibnamefont {van Oosten}}, \bibinfo {author} {\bibfnamefont {P.}~\bibnamefont {van~der Straten}},\ and\ \bibinfo {author} {\bibfnamefont {H.~T.~C.}\ \bibnamefont {Stoof}},\ }\bibfield  {title} {\bibinfo {title} {Quantum phases in an optical lattice},\ }\href {https://doi.org/10.1103/PhysRevA.63.053601} {\bibfield  {journal} {\bibinfo  {journal} {Phys. Rev. A}\ }\textbf {\bibinfo {volume} {63}},\ \bibinfo {pages} {053601} (\bibinfo {year} {2001})}\BibitemShut {NoStop}%
\bibitem [{\citenamefont {dos Santos}\ and\ \citenamefont {Pelster}(2009)}]{dosSantos2009}%
  \BibitemOpen
  \bibfield  {author} {\bibinfo {author} {\bibfnamefont {F.~E.~A.}\ \bibnamefont {dos Santos}}\ and\ \bibinfo {author} {\bibfnamefont {A.}~\bibnamefont {Pelster}},\ }\bibfield  {title} {\bibinfo {title} {Quantum phase diagram of bosons in optical lattices},\ }\href {https://doi.org/10.1103/PhysRevA.79.013614} {\bibfield  {journal} {\bibinfo  {journal} {Phys. Rev. A}\ }\textbf {\bibinfo {volume} {79}},\ \bibinfo {pages} {013614} (\bibinfo {year} {2009})}\BibitemShut {NoStop}%
\bibitem [{\citenamefont {Bradlyn}\ \emph {et~al.}(2009)\citenamefont {Bradlyn}, \citenamefont {dos Santos},\ and\ \citenamefont {Pelster}}]{Bradlyn2009}%
  \BibitemOpen
  \bibfield  {author} {\bibinfo {author} {\bibfnamefont {B.}~\bibnamefont {Bradlyn}}, \bibinfo {author} {\bibfnamefont {F.~E.~A.}\ \bibnamefont {dos Santos}},\ and\ \bibinfo {author} {\bibfnamefont {A.}~\bibnamefont {Pelster}},\ }\bibfield  {title} {\bibinfo {title} {Effective action approach for quantum phase transitions in bosonic lattices},\ }\href {https://doi.org/10.1103/PhysRevA.79.013615} {\bibfield  {journal} {\bibinfo  {journal} {Phys. Rev. A}\ }\textbf {\bibinfo {volume} {79}},\ \bibinfo {pages} {013615} (\bibinfo {year} {2009})}\BibitemShut {NoStop}%
\bibitem [{\citenamefont {Gra\ss{}}\ \emph {et~al.}(2011)\citenamefont {Gra\ss{}}, \citenamefont {dos Santos},\ and\ \citenamefont {Pelster}}]{Grass2011}%
  \BibitemOpen
  \bibfield  {author} {\bibinfo {author} {\bibfnamefont {T.~D.}\ \bibnamefont {Gra\ss{}}}, \bibinfo {author} {\bibfnamefont {F.~E.~A.}\ \bibnamefont {dos Santos}},\ and\ \bibinfo {author} {\bibfnamefont {A.}~\bibnamefont {Pelster}},\ }\bibfield  {title} {\bibinfo {title} {Excitation spectra of bosons in optical lattices from the schwinger-keldysh calculation},\ }\href {https://doi.org/10.1103/PhysRevA.84.013613} {\bibfield  {journal} {\bibinfo  {journal} {Phys. Rev. A}\ }\textbf {\bibinfo {volume} {84}},\ \bibinfo {pages} {013613} (\bibinfo {year} {2011})}\BibitemShut {NoStop}%
\bibitem [{\citenamefont {McIntosh}\ \emph {et~al.}(2012)\citenamefont {McIntosh}, \citenamefont {Pisarski}, \citenamefont {Gooding},\ and\ \citenamefont {Zaremba}}]{McIntosh_2012}%
  \BibitemOpen
  \bibfield  {author} {\bibinfo {author} {\bibfnamefont {T.}~\bibnamefont {McIntosh}}, \bibinfo {author} {\bibfnamefont {P.}~\bibnamefont {Pisarski}}, \bibinfo {author} {\bibfnamefont {R.~J.}\ \bibnamefont {Gooding}},\ and\ \bibinfo {author} {\bibfnamefont {E.}~\bibnamefont {Zaremba}},\ }\bibfield  {title} {\bibinfo {title} {Multisite mean-field theory for cold bosonic atoms in optical lattices},\ }\href {https://doi.org/10.1103/physreva.86.013623} {\bibfield  {journal} {\bibinfo  {journal} {Physical Review A}\ }\textbf {\bibinfo {volume} {86}},\ \bibinfo {pages} {013623} (\bibinfo {year} {2012})}\BibitemShut {NoStop}%
\bibitem [{\citenamefont {Pisarski}\ \emph {et~al.}(2011)\citenamefont {Pisarski}, \citenamefont {Jones},\ and\ \citenamefont {Gooding}}]{PhysRevA.83.053608}%
  \BibitemOpen
  \bibfield  {author} {\bibinfo {author} {\bibfnamefont {P.}~\bibnamefont {Pisarski}}, \bibinfo {author} {\bibfnamefont {R.~M.}\ \bibnamefont {Jones}},\ and\ \bibinfo {author} {\bibfnamefont {R.~J.}\ \bibnamefont {Gooding}},\ }\bibfield  {title} {\bibinfo {title} {Application of a multisite mean-field theory to the disordered {Bose}-{Hubbard} model},\ }\href {https://doi.org/10.1103/PhysRevA.83.053608} {\bibfield  {journal} {\bibinfo  {journal} {Phys. Rev. A}\ }\textbf {\bibinfo {volume} {83}},\ \bibinfo {pages} {053608} (\bibinfo {year} {2011})}\BibitemShut {NoStop}%
\bibitem [{\citenamefont {Lühmann}(2013)}]{L_hmann_2013}%
  \BibitemOpen
  \bibfield  {author} {\bibinfo {author} {\bibfnamefont {D.-S.}\ \bibnamefont {Lühmann}},\ }\bibfield  {title} {\bibinfo {title} {Cluster {Gutzwiller} method for bosonic lattice systems},\ }\href {https://doi.org/10.1103/physreva.87.043619} {\bibfield  {journal} {\bibinfo  {journal} {Physical Review A}\ }\textbf {\bibinfo {volume} {87}},\ \bibinfo {pages} {043619} (\bibinfo {year} {2013})}\BibitemShut {NoStop}%
\bibitem [{\citenamefont {Lühmann}(2016)}]{luhmann2016notesclustergutzwillermethod}%
  \BibitemOpen
  \bibfield  {author} {\bibinfo {author} {\bibfnamefont {D.-S.}\ \bibnamefont {Lühmann}},\ }\href {https://arxiv.org/abs/1608.02905} {\bibinfo {title} {Notes on the cluster {Gutzwiller} method: Inhomogeneous lattices, excitations, and cluster time evolution}} (\bibinfo {year} {2016}),\ \Eprint {https://arxiv.org/abs/1608.02905} {arXiv:1608.02905 [cond-mat.quant-gas]} \BibitemShut {NoStop}%
\bibitem [{\citenamefont {Fischer}\ and\ \citenamefont {Xiong}(2011)}]{Fischer2011}%
  \BibitemOpen
  \bibfield  {author} {\bibinfo {author} {\bibfnamefont {U.~R.}\ \bibnamefont {Fischer}}\ and\ \bibinfo {author} {\bibfnamefont {B.}~\bibnamefont {Xiong}},\ }\bibfield  {title} {\bibinfo {title} {Many-site coherence revivals in the extended {Bose}-{Hubbard} model and the {Gutzwiller} approximation},\ }\href {https://doi.org/10.1103/PhysRevA.84.063635} {\bibfield  {journal} {\bibinfo  {journal} {Phys. Rev. A}\ }\textbf {\bibinfo {volume} {84}},\ \bibinfo {pages} {063635} (\bibinfo {year} {2011})}\BibitemShut {NoStop}%
\bibitem [{\citenamefont {Kashurnikov}\ \emph {et~al.}(1996)\citenamefont {Kashurnikov}, \citenamefont {Krasavin},\ and\ \citenamefont {Svistunov}}]{Kashurnikov1996}%
  \BibitemOpen
  \bibfield  {author} {\bibinfo {author} {\bibfnamefont {V.~A.}\ \bibnamefont {Kashurnikov}}, \bibinfo {author} {\bibfnamefont {A.~V.}\ \bibnamefont {Krasavin}},\ and\ \bibinfo {author} {\bibfnamefont {B.~V.}\ \bibnamefont {Svistunov}},\ }\bibfield  {title} {\bibinfo {title} {Mott-insulator-superfluid-liquid transition in a one-dimensional bosonic {Hubbard} model: Quantum monte carlo method},\ }\href {https://doi.org/10.1134/1.567139} {\bibfield  {journal} {\bibinfo  {journal} {Journal of Experimental and Theoretical Physics Letters}\ }\textbf {\bibinfo {volume} {64}},\ \bibinfo {pages} {99–104} (\bibinfo {year} {1996})}\BibitemShut {NoStop}%
\bibitem [{\citenamefont {K\"{u}hner}\ and\ \citenamefont {Monien}(1998)}]{Khner1998}%
  \BibitemOpen
  \bibfield  {author} {\bibinfo {author} {\bibfnamefont {T.~D.}\ \bibnamefont {K\"{u}hner}}\ and\ \bibinfo {author} {\bibfnamefont {H.}~\bibnamefont {Monien}},\ }\bibfield  {title} {\bibinfo {title} {Phases of the one-dimensional {Bose}-{Hubbard} model},\ }\href {https://doi.org/10.1103/physrevb.58.r14741} {\bibfield  {journal} {\bibinfo  {journal} {Physical Review B}\ }\textbf {\bibinfo {volume} {58}},\ \bibinfo {pages} {R14741–R14744} (\bibinfo {year} {1998})}\BibitemShut {NoStop}%
\bibitem [{\citenamefont {Ohgoe}\ \emph {et~al.}(2012)\citenamefont {Ohgoe}, \citenamefont {Suzuki},\ and\ \citenamefont {Kawashima}}]{Ohgoe2012}%
  \BibitemOpen
  \bibfield  {author} {\bibinfo {author} {\bibfnamefont {T.}~\bibnamefont {Ohgoe}}, \bibinfo {author} {\bibfnamefont {T.}~\bibnamefont {Suzuki}},\ and\ \bibinfo {author} {\bibfnamefont {N.}~\bibnamefont {Kawashima}},\ }\bibfield  {title} {\bibinfo {title} {Ground-state phase diagram of the two-dimensional extended {Bose}-{Hubbard} model},\ }\href {https://doi.org/10.1103/PhysRevB.86.054520} {\bibfield  {journal} {\bibinfo  {journal} {Phys. Rev. B}\ }\textbf {\bibinfo {volume} {86}},\ \bibinfo {pages} {054520} (\bibinfo {year} {2012})}\BibitemShut {NoStop}%
\bibitem [{\citenamefont {Mandelbrot}(1967)}]{Mandelbrot_1967}%
  \BibitemOpen
  \bibfield  {author} {\bibinfo {author} {\bibfnamefont {B.}~\bibnamefont {Mandelbrot}},\ }\bibfield  {title} {\bibinfo {title} {How long is the coast of {Britain}? statistical self-similarity and fractional dimension},\ }\href {https://doi.org/10.1126/science.156.3775.636} {\bibfield  {journal} {\bibinfo  {journal} {Science}\ }\textbf {\bibinfo {volume} {156}},\ \bibinfo {pages} {636} (\bibinfo {year} {1967})}\BibitemShut {NoStop}%
\bibitem [{\citenamefont {Coxeter}(1973)}]{coxeter1973}%
  \BibitemOpen
  \bibfield  {author} {\bibinfo {author} {\bibfnamefont {H.~S.~M.}\ \bibnamefont {Coxeter}},\ }\href@noop {} {\emph {\bibinfo {title} {Regular polytopes}}},\ \bibinfo {edition} {3rd}\ ed.\ (\bibinfo  {publisher} {Dover Publications Inc.},\ \bibinfo {address} {New York},\ \bibinfo {year} {1973})\ pp.\ \bibinfo {pages} {xiv+321}\BibitemShut {NoStop}%
\bibitem [{\citenamefont {Edmonds}\ \emph {et~al.}(1982)\citenamefont {Edmonds}, \citenamefont {Ewing},\ and\ \citenamefont {Kulkarni}}]{edmonds1982}%
  \BibitemOpen
  \bibfield  {author} {\bibinfo {author} {\bibfnamefont {A.~L.}\ \bibnamefont {Edmonds}}, \bibinfo {author} {\bibfnamefont {J.~H.}\ \bibnamefont {Ewing}},\ and\ \bibinfo {author} {\bibfnamefont {R.~S.}\ \bibnamefont {Kulkarni}},\ }\bibfield  {title} {\bibinfo {title} {Regular tessellations of surfaces and (p, q, 2)-triangle groups},\ }\href {http://www.jstor.org/stable/2007049} {\bibfield  {journal} {\bibinfo  {journal} {Annals of Mathematics}\ }\textbf {\bibinfo {volume} {116}},\ \bibinfo {pages} {113} (\bibinfo {year} {1982})}\BibitemShut {NoStop}%
\bibitem [{\citenamefont {Pitaevskii}\ and\ \citenamefont {Stringari}(2003)}]{Pitaevskii-book}%
  \BibitemOpen
  \bibfield  {author} {\bibinfo {author} {\bibfnamefont {L.}~\bibnamefont {Pitaevskii}}\ and\ \bibinfo {author} {\bibfnamefont {S.}~\bibnamefont {Stringari}},\ }\href@noop {} {\emph {\bibinfo {title} {{Bose}-{Einstein} condensation}}},\ International series of monographs on physics; v. 116\ (\bibinfo  {publisher} {Clarendon Press},\ \bibinfo {address} {Oxford},\ \bibinfo {year} {2003})\BibitemShut {NoStop}%
\bibitem [{\citenamefont {Hohenberg}(1967)}]{Hohenberg1967}%
  \BibitemOpen
  \bibfield  {author} {\bibinfo {author} {\bibfnamefont {P.~C.}\ \bibnamefont {Hohenberg}},\ }\bibfield  {title} {\bibinfo {title} {Existence of long-range order in one and two dimensions},\ }\href {https://doi.org/10.1103/PhysRev.158.383} {\bibfield  {journal} {\bibinfo  {journal} {Phys. Rev.}\ }\textbf {\bibinfo {volume} {158}},\ \bibinfo {pages} {383} (\bibinfo {year} {1967})}\BibitemShut {NoStop}%
\bibitem [{\citenamefont {Bagnato}\ and\ \citenamefont {Kleppner}(1991)}]{Bagnato1991}%
  \BibitemOpen
  \bibfield  {author} {\bibinfo {author} {\bibfnamefont {V.}~\bibnamefont {Bagnato}}\ and\ \bibinfo {author} {\bibfnamefont {D.}~\bibnamefont {Kleppner}},\ }\bibfield  {title} {\bibinfo {title} {{Bose}-{Einstein} condensation in low-dimensional traps},\ }\href {https://doi.org/10.1103/PhysRevA.44.7439} {\bibfield  {journal} {\bibinfo  {journal} {Phys. Rev. A}\ }\textbf {\bibinfo {volume} {44}},\ \bibinfo {pages} {7439} (\bibinfo {year} {1991})}\BibitemShut {NoStop}%
\bibitem [{\citenamefont {Jelitto}(1969)}]{JELITTO1969609}%
  \BibitemOpen
  \bibfield  {author} {\bibinfo {author} {\bibfnamefont {R.~J.}\ \bibnamefont {Jelitto}},\ }\bibfield  {title} {\bibinfo {title} {The density of states of some simple excitations in solids},\ }\href {https://doi.org/https://doi.org/10.1016/0022-3697(69)90016-X} {\bibfield  {journal} {\bibinfo  {journal} {Journal of Physics and Chemistry of Solids}\ }\textbf {\bibinfo {volume} {30}},\ \bibinfo {pages} {609} (\bibinfo {year} {1969})}\BibitemShut {NoStop}%
\bibitem [{\citenamefont {Freericks}\ and\ \citenamefont {Monien}(1996)}]{PhysRevB.53.2691}%
  \BibitemOpen
  \bibfield  {author} {\bibinfo {author} {\bibfnamefont {J.~K.}\ \bibnamefont {Freericks}}\ and\ \bibinfo {author} {\bibfnamefont {H.}~\bibnamefont {Monien}},\ }\bibfield  {title} {\bibinfo {title} {Strong-coupling expansions for the pure and disordered {Bose}-{Hubbard} model},\ }\href {https://doi.org/10.1103/PhysRevB.53.2691} {\bibfield  {journal} {\bibinfo  {journal} {Phys. Rev. B}\ }\textbf {\bibinfo {volume} {53}},\ \bibinfo {pages} {2691} (\bibinfo {year} {1996})}\BibitemShut {NoStop}%
\bibitem [{\citenamefont {Dutkiewicz}\ \emph {et~al.}(2025)\citenamefont {Dutkiewicz}, \citenamefont {Płodzień}, \citenamefont {Rojo-Francàs}, \citenamefont {Julia~Diaz}, \citenamefont {Lewenstein},\ and\ \citenamefont {Grass}}]{dutkiewicz_2025_17831905}%
  \BibitemOpen
  \bibfield  {author} {\bibinfo {author} {\bibfnamefont {K.}~\bibnamefont {Dutkiewicz}}, \bibinfo {author} {\bibfnamefont {M.}~\bibnamefont {Płodzień}}, \bibinfo {author} {\bibfnamefont {A.}~\bibnamefont {Rojo-Francàs}}, \bibinfo {author} {\bibfnamefont {B.}~\bibnamefont {Julia~Diaz}}, \bibinfo {author} {\bibfnamefont {M.}~\bibnamefont {Lewenstein}},\ and\ \bibinfo {author} {\bibfnamefont {T.}~\bibnamefont {Grass}},\ }\href {https://doi.org/10.5281/zenodo.17831905} {\bibinfo {title} {Bose-einstein condensation in exotic lattice geometries: Code and data}} (\bibinfo {year} {2025})\BibitemShut {NoStop}%
\end{thebibliography}
\end{document}